\documentclass[a4paper,11pt]{article}
\pdfoutput=1
\usepackage{jheppub}
\usepackage[T1]{fontenc}
\usepackage{amsmath}
\usepackage{amssymb}
\usepackage{mathrsfs}
\usepackage{graphicx}
\usepackage{mathtools}
\usepackage{subcaption}
\usepackage{xcolor}
\usepackage{changes}
\definechangesauthor[name={Per cusse}, color=orange]{per}

\def\be{\begin{equation}}
\def\ee{\end{equation}}
\def\e{\text{e}}
\def\i{\text{i}}

\def\dd{\text{d}}
\def\sH{\text{H}}

\def\max{\text{max}}
\def\reg{\text{reg}}
\def\R{\mathbb{R}}

\def\AdS{\text{AdS}}

\begin{document}
\title{Black Hole Interior and Time-like Entanglement Entropy}

\author{Zi-Hao Li, Run-Qiu Yang}
\affiliation{Center for Joint Quantum Studies and Department of Physics, School of Science, Tianjin University, Yaguan Road 135, Jinnan District, 300350 Tianjin, P.~R.~China}
\emailAdd{lieaction.lzh@gmail.com}
\emailAdd{aqiu@tju.edu.cn}
\abstract{We investigate the possibility of using holographic time-like entanglement entropy (TEE) to probe the black hole interior. We choose the complex-valued extremal surface mixing spatial and temporal segments as one concrete example from various holographic candidates. In Schwarzschild-AdS black holes, we demonstrate that the TEE of time-like strips in such a holographic proposal exhibits linear growth at large temporal widths, possesses a physically significant imaginary part, and obeys a time-like entanglement first-law-like relation for its real part.
By analyzing charged, scalar-hairy black holes, we find this geometric TEE detects a hidden ``causal phase transition'' separating Type-I and Type-II interiors. We identify a critical temporal width $\tau_c$ serving as a bulk diagnostic, potentially acting as a boundary order parameter: for strips narrower than $\tau_c$, the system enters a distinct phase dominated purely by time-like contributions up to an emergent effective scale; conversely, for strips wider than $\tau_c$, space-like entanglement re-emerges. 
Notably, a Cauchy horizon drives $\tau_c$ to infinity, leading to a pure time-like entanglement phase. These results suggest that the TEE may supply a novel boundary quantum-information measure to detect structure hidden inside the black hole and suggests a deep connection between TEE and cosmic censorship.
}

\maketitle
\flushbottom

\section{Introduction}\label{sec:intro}

The interior of a black hole presents a fundamental challenge to our understanding of gravity and spacetime. Unlike the exterior, the interior has unique features, notably a singularity where classical gravity breaks down.
Many black hole solutions, such as the Reissner-Nordstr\"{o}m (RN) and Kerr spacetimes, also have an inner Cauchy horizon behind the event horizon. Efforts to understand the interior have yielded important insights.
For example, the work of Belinski, Khalatnikov, and Lifshitz (BKL) shows that the approach to certain space-like singularities can exhibit chaotic behavior~\cite{Lifshitz:1963ps,Belinsky:1970ew,Belinski:1973zz}.
More recent studies have demonstrated that the Cauchy horizon in certain charged black holes becomes non-linearly unstable, particularly in the presence of scalar hair~\cite{Hartnoll:2020fhc,Cai:2020wrp}.

Despite these theoretical advances, the relationship between the black hole interior and its outside is not fully understood. However, the holographic duality offers a powerful new toolkit. The anti-de Sitter/conformal field theory correspondence (AdS/CFT) claims that the physics of a bulk spacetime in an asymptotically anti-de Sitter (AdS) spacetime is equivalent to a conformal field theory (CFT) on the spacetime boundary~\cite{Maldacena:1997re,Gubser:1998bc,Witten:1998qj}.
Specifically, the duality between an eternal AdS black hole and two entangled boundary theories provides a concrete framework where the black hole interior is encoded in the entanglement structure of two boundaries~\cite{Maldacena:2001kr, Maldacena:2013xja}. Within this framework, a hierarchy of probes has emerged, yet each comes with limitations.
On one hand, ``weak'' probes can capture data behind the event horizon but are fundamentally incapable of reaching the singularity. Standard Holographic Entanglement Entropy (HEE), governed by the Ryu-Takayanagi prescription~\cite{Ryu:2006bv, Hubeny:2007xt}, and the Complexity=Volume (CV) conjecture~\cite{Stanford:2014jda} fall into this category.
Their geometric duals, space-like minimal surfaces and maximal volumes, respectively, cannot reach the space-like singularity in general dimensions~\cite{Hartman:2013qma, Caceres:2022smh}, thus providing only a partial picture of the interior geometry\footnote{Only in lower-dimensional cases, such as $d=1$ or $d=2$ (BTZ black hole case), where the surfaces reduce into geodesics, can HEE and CV probes access the full interior up to the singularity. However, these are exceptional cases and do not represent the generic behavior in higher dimensions, which is the focus of our investigation.}.
On the other hand, ``strong'' probes do exist that can reach the singularity, and they have yielded profound insights into the black hole interior. For instance, the geodesic approximation to two-sided correlation functions has been used to probe the near-singularity geometry~\cite{Fidkowski:2003nf, Caceres:2022smh}. The Complexity = Action (CA) conjecture, through the evaluation of the action on the Wheeler- DeWitt patch, provides another powerful tool to track the geometry deep inside the horizon~\cite{Brown:2015bva, Lehner:2016vdi}. However, the probes capable of reaching the singularity typically rely on correlation data from two disconnected boundaries. This leaves us with a critical dilemma: the tools that are single-sided (such as HEE) cannot probe the singularity, and the tools that can probe the singularity (such as CA conjecture) are not single-sided.

Particularly, focusing on entanglement entropy, this fundamental limitation raises a natural question: can this barrier be overcome by considering a different class of entanglement? Such a candidate has recently emerged in the form of Time-like Entanglement Entropy (TEE)~\cite{Doi:2022iyj}, a generalization of HEE to \textit{time-like} separated boundary regions, which belongs to the class of \textit{single-sided} observables.
The concept of entanglement entropy for time-like subregion was originally proposed in early holographic studies~\cite{Wang:2018jva}, where it was proposed to resolve the asymmetry between finite-size and finite-temperature corrections in $T\bar{T}$ deformed CFTs\footnote{The asymmetry is further examined in this specific context in~\cite{Jiang:2023ffu}.}.
TEE is particularly sensitive to the deep interior structure of a black hole. This is because its geometric dual, unlike that of HEE, can directly reach the singularity~\cite{Narayan:2022afv,Li:2022tsv,Doi:2023zaf,Narayan:2023zen}.
This property makes TEE a rare boundary observable that is able to directly probe the near-singularity structure of a black hole. Consequently, TEE opens a new window to investigate the physics of the black hole interior and its singularity through the behavior of a boundary quantity.

The study of TEE is a rapidly developing field, with several computational frameworks proposed in recent years. On the boundary field theory side, several conceptually distinct approaches have been developed, including the operator-algebraic definition of TEE as a real-valued quantity derived from space-like entanglement~\cite{Jiang:2025pen}, boundary field theory techniques utilizing the real-time Schwinger-Keldysh and spacetime-density-matrix constructions~\cite{Milekhin:2025ycm, Gong:2025pnu, Guo:2025ase}, Euclidean replica methods and analytic continuations of replica-twist correlators to time-ordered Lorentzian correlators~\cite{Guo:2024lrr,Xu:2024yvf,Bernamonti:2026pxo}, and non-Hermitian temporal-density-matrix formulations in quantum many-body systems~\cite{Harper:2025lav}.
It is crucial to note that while the field-theoretic calculation of TEE in two-dimensional ($d=2$) conformal field theories is essentially well-defined, its exact microscopic computation in higher dimensions ($d>2$) remains an outstanding challenge, despite a recent field-theoretic attempt utilizing disjoint entangling regions on a solid torus to evaluate HEE and TEE for arbitrary $d \geqslant 2$ CFTs~\cite{Jiang:2025jnk}. One of our core motivations for performing this holographic analysis is precisely to explore bulk geometric prescriptions that can provide concrete expectations and geometric benchmarks to guide future field-theoretic developments for higher-dimensional ($d>2$) theories. 

Complementary to these boundary perspectives, various bulk constructions have been proposed, including top-down holographic models derived from string backgrounds~\cite{Nunez:2025gxq,Nunez:2025puk}, a framework deriving the TEE from the Rindler method and modular flow~\cite{He:2023ubi,Wen:2024yny}, equations of motion for extremal surfaces in non-conformal and non-relativistic theories~\cite{Afrasiar:2024lsi,Afrasiar:2024ldn,Afrasiar:2025eam}, a geodesic-based extremization procedure in $\AdS_3$ avoiding bulk analytic continuation~\cite{Bohra:2025mhb}, and calculations involving complex bulk geometries of extremal surfaces~\cite{Heller:2024whi,Guo:2025pru, Heller:2025kvp}. 
Among these diverse candidates, in this work we choose to employ the Complex-valued Weak Extremal Surface (CWES) prescription~\cite{Doi:2022iyj, Li:2022tsv, Doi:2023zaf, Bohra:2025mhb} to explore the black hole interior. Although recent AdS$_3$/CFT$_2$ results favor single complex geodesics over piece-wise curves for the replica-twist-correlator definition of TEE~\cite{Bernamonti:2026pxo}, an equally controlled boundary derivation is not yet available in higher dimensions.  We emphasize that this selection is motivated not by a claim of its absolute superiority over alternative proposals, but rather because it provides a highly tractable and geometrically intuitive framework to investigate how time-like entanglement responds to complex interior structures. Probing the black hole interior through other holographic TEE frameworks remains an intriguing open question, which we leave for future investigations. 

Equipped with these geometric tools, TEE has been applied to diverse holographic settings, including the investigation of phase structures in AdS/BCFT~\cite{Chu:2023zah} and the detection of gravitational anomalies~\cite{Chu:2025sjv}. In the context of black hole physics, pioneering studies have begun to explore the interiors. For instance, properties of TEE, captured by space-like and time-like geodesics, have been used to gain insights into the interiors of lower-dimensional spacetimes~\cite{Guo:2024lrr,Guo:2025mwp}. Similarly, in $2+1$-dimensional black holes, TEE has been shown to serve as a preliminary probe of the singularity~\cite{Anegawa:2024kdj}.
However, these initial applications have largely focused on simpler models. The relationship between the interior geometry of higher-dimensional black holes with complex internal structures and the boundary field theory remains a crucial and largely unexplored question.

In this paper, we employ the CWES/TEE framework~\cite{Doi:2022iyj, Li:2022tsv, Doi:2023zaf, Bohra:2025mhb} to investigate that time-like entanglement entropy (TEE) provides the possibility of probing a crucial dichotomy in the causal nature of singularities, one that earlier studies found using two-sided probes requiring entanglement data from two boundaries~\cite{An:2022lvo, Caceres:2022smh, Auzzi:2022bfd}.
We frame this classification in terms of two classes, which we call \textit{Type-I} and \textit{Type-II}, distinguished by their interior causal structure. This geometric distinction is rooted in the causal shape of the interior in the Penrose diagram, with Type-I spacetimes featuring ``concave'' or ``flat'' space-like singularities and Type-II spacetimes possessing ``convex'' ones, leading to markedly different causal behavior inside the black hole.
Standard holographic observables (for example, holographic entanglement entropy and thermodynamic observables) are insensitive to this distinction because they probe the geometry only down to the event horizon. By contrast, TEE can see deep interior changes, thus revealing a dramatic transition: the TEE of boundary time-like strips in Type-II interiors exhibits what we term a ``time-like entanglement phase''.
When the temporal width is sufficiently narrow, the entanglement is dominated by time-like contributions and the space-like segments of extremal surface are reduced to the UV regulator contributions. This regulator effect stems from the existence of both a spatial and a temporal minimal resolution scale, which arises in the limit where a space-like extremal surface approaches a null hypersurface.
We identify the upper bound of this time-like dominated regime, the critical temporal width $\tau_c$, as a bulk diagnostic that potentially acts as a physically intuitive boundary order parameter. Only the Type-II spacetime admits a nonzero $\tau_c$. The emergence of a non-zero ``time-like entanglement gap'' ($\tau_c > 0$) thus provides a clear signature of hidden interior transitions, effectively ``hearing'' the shape of the singularity from the outside.
While other two-sided probes (e.g., Complexity=Action) may also partly respond to this phenomenon~\cite{An:2022lvo, Caceres:2022smh, Auzzi:2022bfd}, our central contribution is to establish TEE as a novel, single-sided diagnostic of the hidden causal phase. Furthermore, previous studies have established that the dynamics of boundary CFT entanglement entropy can be formulated as a thermodynamic-like relation, known as the entanglement first law~\cite{Bhattacharya:2012mi, Blanco:2013joa}. Inspired by recent literature utilizing double Wick rotations to study the time-like entanglement first law for complex bulk extremal surfaces~\cite{Li:2025tud, Xiao:2026nvu}, we investigate this thermodynamic-like property within the CWES proposal. We discover that the real part of the TEE obeys a time-like entanglement first-law-like relation,  and the imaginary part plays a distinctive role, characterizing the non-analytic transition of the global causal structure from a vacuum state to a thermal excitation. Translating these geometric findings into a boundary modular-Hamiltonian formulation in Lorentzian spacetime remains an intriguing open challenge. 

The organization of this paper is as follows. In Sec.~\ref{sec:SAdS} we present a detailed calculation for the Schwarzschild-AdS black hole. This not only validates our methodology and establishes the benchmark Type-I behavior, where the order parameter $\tau_c$ vanishes identically, but also reveals that the linear growth in the limit of large temporal width and the imaginary part of TEE cannot be absorbed into the usual UV spatial cutoff and therefore carries physical information. In Sec.~\ref{sec:Type-II} we construct Type-II interiors using a backreacting holographic superconductor, trace the transition from Type-II to Type-I as temperature varies, and show that a nonzero $\tau_c$ serves as a geometric indicator for this transition, exploring the rich phenomenology associated with its emergence. Notably, we show that the existence of a Cauchy horizon drives $\tau_c$ to infinity, resulting in a state of pure time-like entanglement. Sec.~\ref{sec:TEEfirstlaw} is dedicated to the explicit holographic verification of a time-like entanglement first-law-like relation, showing that the real part of the time-like strip system obeys an inverse-size scaling behavior, and the imaginary part characterizes the non-analytic transition of the global causal structure from a vacuum state to a thermal excitation. Finally, in Sec.~\ref{sec:discussion}, we summarize our findings, discuss their implications and outline open questions.

\section{Time-like Entanglement Entropy in Schwarzschild-AdS Black Hole}\label{sec:SAdS}

Before exploring the rich interior structures of complex black holes, it is instructive to first apply the TEE framework to the simplest, non-trivial example: the Schwarzschild-AdS black hole (SAdS). This black hole, with its simple interior lacking inner horizons or hair, offers a controlled setting to develop our methods. It thereby establishes a baseline against which the novel phenomena in more complex spacetimes (Sec.~\ref{sec:Type-II}) can be clearly contrasted and understood.

Our starting point is the metric for a $D=(d+1)$-dimensional SAdS black hole, which in Schwarzschild coordinates is given by\footnote{Here we have assumed the spatial dimension of bulk spacetime satisfies $d>2$. The case of BTZ black hole with $d=2$ is reduced to the situation discussed in Ref.~\cite{Li:2022tsv}. Our result about the imaginary part of TEE~\eqref{eq:BTZ} for $d=2$ in this paper is consistent with Ref.~\cite{Li:2022tsv}. The real part follows directly from Eq.~\eqref{eq:Aofd=2} and Eq.~\eqref{eq:tau_H_final} for $d=2$; we therefore omit its detailed derivation in this paper.}:
\be\label{eq:AdSmetric}
\begin{aligned}
    \dd s^2 &=-f\left( r \right) \dd t^2+\frac{\dd r^2}{f\left( r \right)}+r^{2}\dd \Sigma^{2}_{k,d-1}\ ,\\
    f(r) &= k-\frac{\omega^{d-2}}{r^{d-2}}+\frac{r^2}{\ell^2_{\AdS}} \ .
\end{aligned}\ee
Here, $\ell_{\AdS}$ denotes the AdS radius, while $k \in \{+1, 0, -1\}$ describes the curvature of the ($d-1$)-dimensional spatial section $\dd \Sigma^{2}_{k,d-1}$, which is given by
\be\label{eq:AdSMetric}
\dd \Sigma^{2}_{k,d-1} =\begin{dcases}
 \dd\theta^2+\sin^2\theta\,\dd\Omega^2_{d-2}\ ,  & \text{for}\, k=+1\ ; \\
  \dd x^2+\dd \boldsymbol{y}_{d-2}^{2}\ , & \text{for}\, k=0\ ;\\
  \dd\sigma^2+\sinh^2\sigma\,\dd\Omega^2_{d-2}\ ,& \text{for}\, k=-1\ .
 \end{dcases}
\ee
This spacetime is static, admitting the time-like Killing vector field $\xi^\mu= \partial_t$. For simplicity, and to connect with holographic applications, we will focus on the planar case throughout this paper, i.e., we set $k=0$ in Eqs.~\eqref{eq:AdSmetric} and~\eqref{eq:AdSMetric}. The horizon radius $r_h$ is defined as the largest real root of the equation $f(r_h)=0$, which relates it to the `mass' parameter $\omega$
\begin{equation}\label{eq:horizon}
\begin{aligned}
k-\frac{\omega^{d-2}}{r^{d-2}_h}+\frac{r^2_h}{\ell^2_\AdS}=0\ .
\end{aligned}
\end{equation}

\subsection{Setup}

Having defined the spacetime geometry, our goal is to compute the TEE for a specific boundary subregion within this background. To do so, we will employ the Complex-valued Weak Extremal Surface (CWES) prescription introduced in Ref.~\cite{Doi:2022iyj, Li:2022tsv, Doi:2023zaf, Bohra:2025mhb}. This prescription instructs us to find a special, piece-wise smooth, codimension-2 surface $\Gamma_{\mathcal{T}}$ in the bulk. The surface $\Gamma_{\mathcal{T}}$ is anchored on the boundary subregion of interest, $\mathcal{T}$, and its area holographically computes the TEE as follows
\be\label{eq:CWESofTEE}
S_\mathcal{T}=\text{Min}\left \{\mathop{\text{Ext}}\limits_{\partial \Gamma_\mathcal{T}}\frac{\mathscr{A}(\Gamma_\mathcal{T})}{4G_\text{N}^{(d+1)}}\right \}\ .
\ee
Here, $\mathscr{A}(\Gamma_{\mathcal{T}})$ corresponds to the complex-valued area of the piece-wise smooth surface $\Gamma_{\mathcal{T}}$. The operations ``Ext'' and  ``Min'' signify that we must find a surface that is both extremal (its area is stationary under local deformations) and minimal among all such valid candidates, according to the ordering relation ``$\prec$'' (or ``$\succ$'')  defined by~\cite{Li:2022tsv}.
The power of this prescription lies in the unique structure of $\Gamma_{\mathcal{T}}$: it is composed of both space-like and time-like extremal segments, pieced together smoothly, which allows the surface to penetrate the deep interior and reach the singularity, which is impossible for the purely space-like subregions used to compute standard HEE.

\begin{figure}[htbp]
 \begin{center}
   \includegraphics[width=0.5\textwidth]{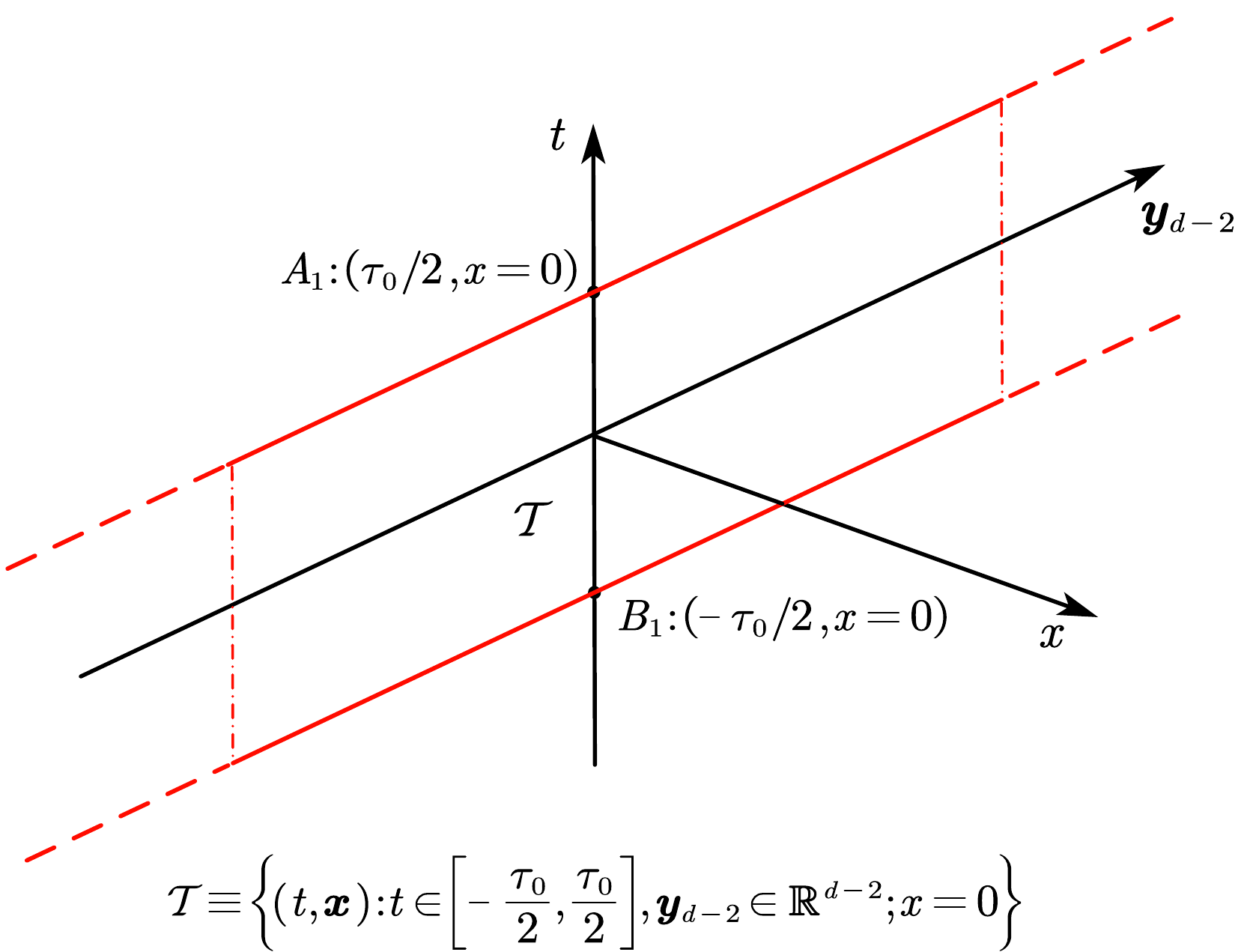}
\end{center}
\caption{A schematic diagram of time-like strip (depicted in red) on the boundary. Here the boundary is $d$-dimensional and spanned by coordinates $\{t,x,\boldsymbol{y}_{d-2}\}$ with
$\boldsymbol{y}_{d-2}\in \mathbb{R}^{d-2}$. This strip-configuration serves as the geometric generalization of a time-like interval in higher dimensions.} \label{fig:time-like strip}
\end{figure}

To apply this prescription, we must first specify the boundary subregion $\mathcal{T}$. In higher dimensions, we consider a time-like ``strip'' of temporal width $\tau_0$, extending infinitely in $d-2$ spatial directions,
\be\label{eq:time-like strip}
\mathcal{T} \equiv \left\{ (t,\boldsymbol{x}):t\in \left[ -\frac{\tau_0}{2},\frac{\tau_0}{2} \right] ,\boldsymbol{y}_{d-2}\in \mathbb{R}^{d-2};x=0 \right\}\ ,
\ee
as depicted in Fig.~\ref{fig:time-like strip}.
The endpoints of time-like strip $\mathcal{T}$ on boundary CFT$_d$ are denoted by $A_1$ and $B_1$ with the time coordinate $t = \pm \frac{\tau_0}{2}$ (note that for $d=2$,  this strip-configuration reduces to a time-like interval usually considered in the literature, such as~\cite{Li:2022tsv,Anegawa:2024kdj,Guo:2025mwp}). Our task is now clear: find the minimal CWES anchored on this strip $\mathcal{T}$ and evaluate its area.

\begin{figure}[htbp]
 \begin{center}
   \includegraphics[width=0.8\textwidth]{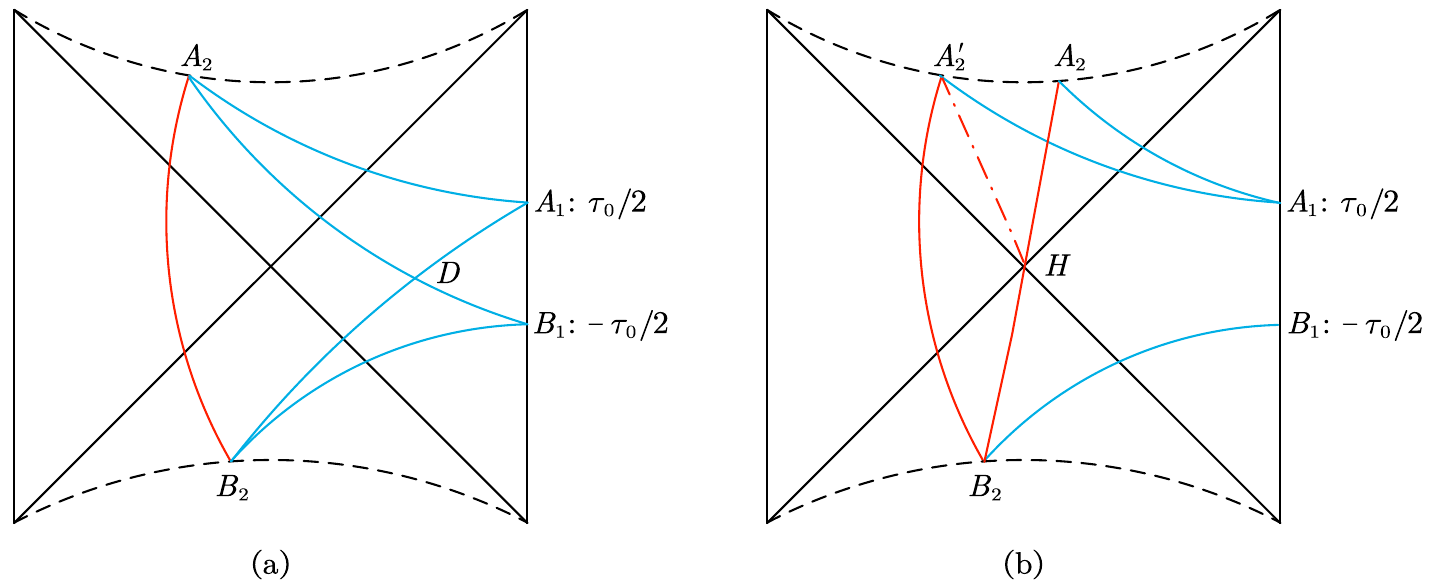}
 \end{center}
\caption{Configurations of CWES of time-like strip in SAdS black hole. Blue lines stand for space-like surface and red lines represent time-like surface.  (a) There are two types of relevant configurations of CWESs, which are $A_1A_2B_2B_1$ and $A_1B_2A_2B_1$.  (b) Comparison of time-like paths: The symmetric surface $A_2B_2$ passes through the bifurcation surface $H$ of the event horizon, whereas $A'_2B_2$ represents a generic path.}  \label{fig:CWES-AdS-Sch}
\end{figure}

With this setup, there are two relevant segment-smooth surfaces that could touch the singularity within the black hole interior, shown in Fig.~\ref{fig:CWES-AdS-Sch}(a), one of which is $A_1A_2B_2B_1$ and the other is  $A_1B_2A_2B_1$. Applying arguments analogous to the ``triangle inequality'' for spatial extremal surfaces, and defining  $\mathcal{A}=\mathscr{A}/\mathcal{V}_{d-2}$ as the complex valued area density (where $\mathcal{V}_{d-2}\equiv\int\dd^{d-2}y$ is the transverse volume of the boundary strip)\footnote{Here we work with the area density $\mathcal{A}$ to replace the area $\mathscr{A}$, because the total area is trivially infinite due to the infinite transverse volume of the boundary strip for $\boldsymbol{y}_{d-2}\in \mathbb{R}^{d-2}$.}, we have
\begin{equation}
    \begin{aligned}
         \mathcal{A}(A_1A_2)&<\mathcal{A}(A_1D)+\mathcal{A}(DA_2)\ ,\\
        \mathcal{A}(B_1B_2)&<\mathcal{A}(B_1D)+\mathcal{A}(DB_2)\ ;\\
         \mathcal{A} (A_1A_2B_2B_1)&\prec \mathcal{A}(A_1B_2A_2B_1)\ .
    \end{aligned}
\end{equation}
Here ``$\prec$'' in last line denotes the ordering relationship defined in Ref.~\cite{Li:2022tsv} to obtain the minimal CWES.

Consequently, we only need to consider the segment-smooth surface $A_1A_2B_2B_1$ of Fig.~\ref{fig:CWES-AdS-Sch}(a), which is related to two different cases, ``non-crossing'' configuration $A_1A'_2B_2B_1$ and ``crossing'' configuration $A_1A_2B_2B_1$ shown in Fig.~\ref{fig:CWES-AdS-Sch}(b). The red solid straight line $A_2B_2$ stands for the time-like surface passing through the bifurcation surface $H$ of event horizon, while $A'_2B_2$ represents a general time-like surface connecting the points of past and future singularity. Let $A'_2 H$ denote the time-like segment connecting  $A'_2$ and $H$. Since time-like surfaces obey the ``anti-triangle inequality'', we then find
\begin{equation}
    \mathcal{A} (A_2'B_2)\succ \mathcal{A}(A_2'H)+\mathcal{A}(HB_2)\ .
\end{equation}
Furthermore, the time translation symmetry of SAdS black hole (generated by the Killing vector $\partial_t$) implies that $\mathcal{A}(A_2'H)=\mathcal{A}(A_2H)$. Combining these relations, we obtain:
\begin{equation}
     \mathcal{A}(A_1A_2B_2B_1)\prec \mathcal{A}(A_1A'_2B_2B_1)\ .
\end{equation}
This result confirms that, to find the minimal CWES and compute the TEE in the SAdS black hole, it suffices to study the configuration where the time-like segment passes through the bifurcation surface, as depicted by the segment-smooth surface $A_1A_2B_2B_1$ in Fig.~\ref{fig:CWES-AdS-Sch}(b).

\subsection{Minimal Geometric Configuration of Schwarzschild-AdS Black Hole}\label{subsec:vertical}

To describe the surfaces $A_1 A_2$ and $B_1 B_2$, which smoothly cross the event horizon and terminate at the singularity ($r=0$), the standard Schwarzschild coordinates~\eqref{eq:AdSmetric} are inadequate due to the coordinate singularity at the horizon. We therefore switch to ingoing and outgoing Eddington coordinates for segments $A_1 A_2$ and $B_1 B_2$ respectively by defining:
\be\label{eq:nullsheets}
\begin{aligned}
     v=t+r_*=t+\int_\infty^r{\frac{\dd \tilde{r}}{f\left( \tilde{r} \right)}}\ ,\\
     u=t-r_*=t-\int_\infty^r{\frac{\dd \tilde{r}}{f\left( \tilde{r} \right)}}\ .
\end{aligned}
\ee
Accordingly, we utilize the metrics adapted to these respective coordinate patches (ingoing and outgoing):
\begin{equation}\label{eq:Eddington}
\begin{aligned}
    \dd s^2 =-f\left( r \right) \dd v^2+2\dd v\dd r+r^2\left( \dd x^2+\dd \boldsymbol{y}_{d-2}^{2} \right) \ ,\\
    \dd s^2 =-f\left( r \right) \dd u^2-2\dd u\dd r+r^2\left( \dd x^2+\dd \boldsymbol{y}_{d-2}^{2} \right) \ ;
\end{aligned}
\end{equation}
and in this coordinate system, the metric behaves regularly across the horizon $r=r_h$.

\begin{figure}[htbp]
 \begin{center}
   \includegraphics[width=0.8\textwidth]{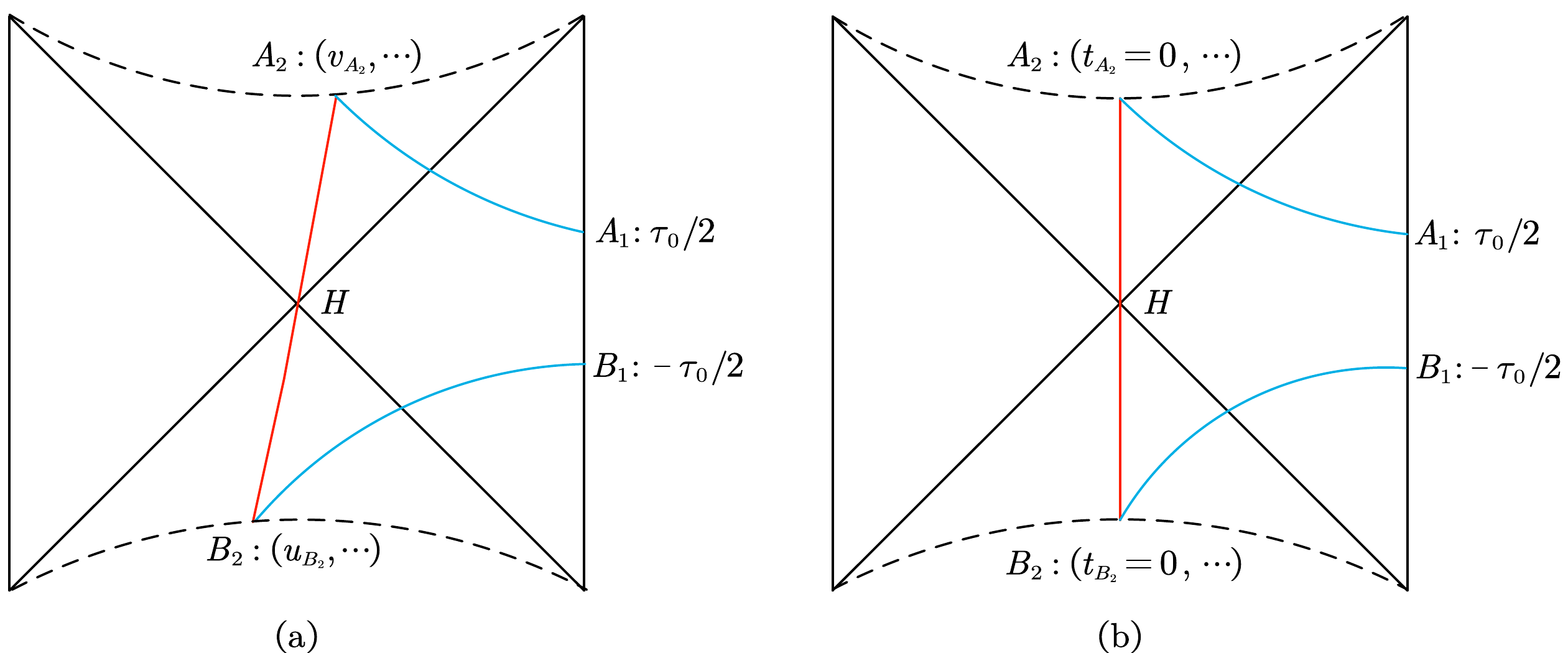}
 \end{center}
\caption{(a) A family of piece-wise configurations
parameterized by the endpoint parameter $v_{A_2}$, where varying it generates a series of candidate extremal surfaces. (b) The time-symmetric extremal configuration of CWES, corresponding to the time-symmetric case $t|_{A_2}=t|_{B_2}=0$. } \label{fig:Hamilton-Jacobi}
\end{figure}

In Eddington coordinates~\eqref{eq:Eddington}, we parameterize the endpoints at the singularity of the segment-smooth surface $A_1A_2B_2B_1$ as $A_2:(v_{A_2},\cdots)$ and $B_2:(u_{B_2},\cdots)$ respectively, as illustrated in Fig.~\ref{fig:Hamilton-Jacobi}(a).
Here the ``endpoints'' $A_2$ and $B_2$ are related to each other so that the time-like extremal surface passes through the bifurcation surface of event horizon. This leads to two important results. First, the area contribution from the time-like segment $A_2B_2$ is independent of the boundary temporal width $\tau_0$ (see Sec.~\ref{subsubsec:imagipart} for analytic result). Therefore, for the extremization problem, we treat the area of $A_2B_2$ as a constant and focus on extremizing the contribution from the space-like segments $A_1A_2$ and $B_2B_1$.
Second, constraint from $A_2B_2$ requires that the $t$-coordinates of $A_2$ and $B_2$ are the same, i.e., $t|_{A_2}=t|_{B_2}$. In other words, the time-like surface $A_2B_2$ acts as a geometric constraint, fixing the Schwarzschild coordinate time of the space-like surfaces $A_1A_2$ and $B_2B_1$ at the singularity. As a result, Eddington coordinate time $v_{A_2}$ and $u_{B_2}$ at the singularity are not independent. This reduces the variational problem to a single independent parameter, which we choose to be $v_{A_2}$ for the following analysis.
As depicted in Fig.~\ref{fig:Hamilton-Jacobi}(a), varying $v_{A_2}$ generates a family of candidate extremal surfaces. Our task is therefore reduced to find the specific $v_{A_2}$ that satisfies the joint extremality condition, i.e., the physical configuration with the minimal extremal area among all candidate configurations.

To identify this minimal configuration, our analysis proceeds in two main steps. First, we will express the real part of the CWES area $\mathcal{A}(A_1A_2B_2B_1)$ as a function of the endpoint parameter $v_{A_2}$, thereby reducing the problem to a one-variable extremization problem. Second, we will utilize the Hamilton-Jacobi formalism to derive the general condition satisfied by any extremal surface. This condition reveals an intuitive solution: the time-symmetric ``vertical'' configuration (corresponding to $t|_{A_2}=t|_{B_2}=0$, see Fig.~\ref{fig:Hamilton-Jacobi}(b)), which we subsequently demonstrate to be the unique solution. The details of each step are presented below.

(1) \textbf{Expressing the Area via a Conserved Quantity}.
Our first step is to find an expression for the area of the space-like surfaces, $\mathcal{A}_{A_1 A_2}$ and $\mathcal{A}_{B_1 B_2}$, as a function of the endpoint coordinate $v_{A_2}$.

In Eddington coordinates~\eqref{eq:Eddington}, we can parameterize the extremal surface by $\{v,r=r\left( v \right),x=0, \boldsymbol {y}_{d-2}\in \R^{d-2}\}$, subject to the boundary conditions $r(\tau_0/2)\rightarrow\infty$ and $r(v_{A_2})=0$. The area density functional $\mathcal{A}_{A_1 A_2}$ for the segment $A_1 A_2$ is then given by the integral of the Lagrangian functional $L$:
\be\label{eq:space-like area}
\begin{aligned}
     \mathcal{A}_{A_1 A_2} &= \int_{v_{A_2}}^{\tau_0/2}{\dd v\, L(r,r')}\ , \\
     L(r,r') &=r^{d-2}\left( v \right) \sqrt{ -f\left( r \right) +2r'}\ ,
\end{aligned}
\ee
where $r'\equiv\frac{\dd r}{\dd v}$.
Crucially, the SAdS metric is static, meaning the Lagrangian $L$ has no explicit dependence on the ``time'' coordinate $v$. We could find an associated conserved quantity along the extremal surface $E =L-r^\prime \frac{\partial L}{\partial r^\prime}$,
\be\label{eq:conservedH}
    \begin{aligned}
     E &= \frac{r^{d-2}\left(f(r)-r^\prime \right)}{\sqrt{-f(r)+2r^\prime}}\equiv H_{*}^{2}\ .
    \end{aligned}
\ee
Here $H_*$ is assumed to be a real constant. The equation of motion for the surface can now be solved for $r'$ in terms of $r$ and the constant $H_*$. We obtain the following equation from\footnote{Directly solving for $r^\prime$ from Eq.~\eqref{eq:conservedH} leads to $r^\prime = r^{4-2d}X\left( X\pm H_{*}^{2} \right)$. Here, only the positive branch ``$+$'' is admissible because the negative sign would imply $r' = 0$ at the horizon, which is physically invalid.} Eq.~\eqref{eq:conservedH} for surface $A_1 A_2$
\be\label{eq:r'(v)}
        r^\prime = r^{4-2d}X\left( X + H_{*}^{2} \right) \ ,
\ee
where we have introduced $X=\sqrt{H_{*}^{4} +r^{2d-4}f\left( r \right)}$.
Notice that when $f(r) < 0$ (inside the horizon) and $|f(r)|$ is large enough, term $H_*^4 + r^{2d-4}f(r)$ in the root $\sqrt{H_{*}^{4} +r^{2d-4}f\left( r \right)}$ may become negative, preventing a real square root.
This effectively limits the domain of existence for $H_*$. The condition for a real-valued solution is that $H_*$ must have a lower bound $H_*^4 \geqslant -\max\{r^{2d-4}f(r)\}$.  For SAdS case, we have
\begin{equation}\label{eq:Hstarmin}
    H_{*,\min}=G^{1/4}(r_c)\ ,
\end{equation}
where $G(r)=\omega ^{d-2}r^{d-2}-r^{2d-2}/\ell _{\AdS}^{2}$ and $r_c=[(d-2)\omega^{d-2}\ell^2_{\AdS}/(2d-2)]^{1/d}$ is the critical point where $G(r)$ reaches its extremum. The critical point $r_c$ derived here signifies that the space-like extremal surface $A_1A_2$, which connects the boundary to the singularity, approaches a critical extremal surface located at $r=r_c$. The surface $A_1A_2$ lingers at this critical extremal surface for an arbitrarily long time\footnote{The phenomenon that a critical extremal surface governs the behavior of TEE in the limit of large temporal width is analogous to that observed in the context of space-like entanglement entropy~\cite{Hartman:2013qma, Li:2022cvm} and complexity~\cite{Carmi:2017jqz}. Furthermore, Ref.~\cite{Heller:2024whi} also identified such a critical surface controlling the behavior of complex bulk extremal surfaces in the large $\tau_0$ regime in the context of TEE, following a proposal different from this paper.}, which corresponds precisely to the large boundary temporal width $\tau_0 \to \infty$, as we will demonstrate both numerically (Fig.~\ref{fig:tau-ReAofH}(a)) and analytically (Eq.~\eqref{eq:tau0Hlim}) in Sec.~\ref{subsec:asymptau}.

For the surface $A_1 A_2$, we fix the boundary time at $v=\tau_0/2$ at $A_1$. The boundary conditions for the extremal surface are thus given by
\be
    A_1:(r=+\infty ,v=\tau_0/2)\ , \quad A_2:\left( r=0,v=v_{A_2} \right) \ .
\ee
The relationship between constant $H_*$ and the time difference $\tau_0/2-v_{A_2}$ is determined by integrating with respect to $\dd  v = \dd  r/r'$
\be\label{eq:tau_0&v_A&H_*}
    \begin{aligned}
       \frac{\tau_0}{2}-v_{A_2}  &= \int_{r=0}^{r=+\infty}{\frac{\dd r}{r^{4-2d}X\left(X+H_{*}^{2} \right)}}\ .
    \end{aligned}
\ee
With the conserved quantity $H_*$ implicitly fixed by the above equation, the on-shell area functional of the surface $A_1 A_2$ can be expressed as $\mathcal{F}(\tau_0,v_{A_2})$, defined by the radial integral:
\be\label{eq:A&H_*}
    \begin{aligned}
        \mathcal{F}(\tau_0,v_{A_2})\equiv \int_{v_{A_2}}^{\tau_0/2} {\dd v\,\frac{r^{2d-4}\left( v \right)}{H_{*}^{2}}\left( f(r)-r^\prime \right)}= \int_{r=0}^{r=+\infty}{\dd r\,\frac{r^{2d-4}}{X}}\ .
    \end{aligned}
\ee
Here $X=\sqrt{H_{*}^{4} +r^{2d-4}f\left( r \right)}$ and $\mathcal{F}(\tau_0,v_{A_2})$ is the ``on-shell'' value of area integration \eqref{eq:space-like area}. Note that in this expression, $H_*$ is treated as a function of the $v_{A_2}$ and $\tau_0$ via Eq.~\eqref{eq:tau_0&v_A&H_*}.

\begin{figure}[htbp]
 \begin{center}
   \includegraphics[width=0.44\textwidth]{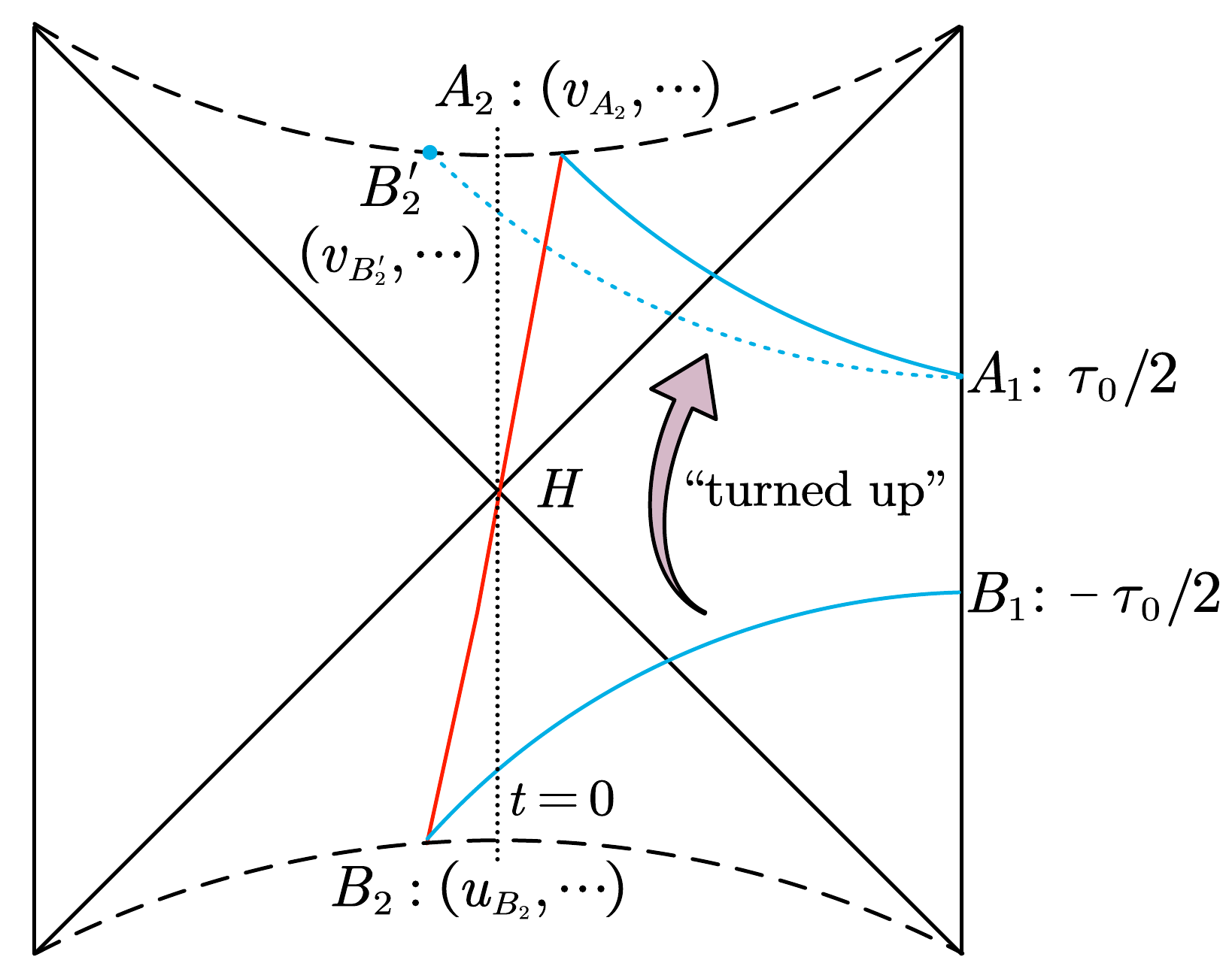}
 \end{center}
\caption{The surface $A_1 B'_2$ corresponds to the ``turned up'' time-reversed image of surface $B_1 B_2$, where geometric constraint requires $t|_{B'_2}=-t|_{B_2}=-t|_{A_2}$. } \label{fig:Hamilton-Jacobi2}
\end{figure}

For the other surface $B_1 B_2$, which originates at the boundary $B_1$ with $r(-\tau_0/2)\rightarrow\infty$ and ends at a singularity point $B_2$ with $r(u_{B_2})=0$, one could repeat the above procedure in an analogous manner. However, a full recalculation is unnecessary due to the time-reversal symmetry of the SAdS spacetime ($t \to -t$, or equivalently $u \leftrightarrow -v$ in Eddington coordinates).
Consequently, surface $B_1 B_2$ can be mapped to a time-reversed mirror image, which we denote as the ``turned up'' surface $A_1 B'_2$, as illustrated in Fig.~\ref{fig:Hamilton-Jacobi2}.
Similar to the surface $A_1 A_2$, this mirror surface $A_1 B'_2$ is parameterized by ingoing Eddington coordinates $(r,v, \cdots)$ and is anchored to the same asymptotic boundary condition $A_1: (r \to +\infty, v=\tau_0/2)$.
However, its behavior at the singularity $B'_2$ is distinct. Governed by time-reversal symmetry and the geometric constraints of the time-like segment $A_2 B_2$, the Schwarzschild coordinate time at the singularity $B'_2$ satisfies the relation $t|_{B'_2}=-t|_{B_2}$. Combined with the condition $t|_{A_2}=t|_{B_2}$, this implies $t|_{B'_2} = -t|_{A_2}$. In terms of ingoing Eddington coordinates, this imposes a condition about the endpoints of $A_1 A_2$ and $A_1 B'_2$ at the singularity
\begin{equation}\label{eq:constrain}
    v_{A_2} + v_{B'_2} = v_0\ ,
\end{equation}
where we have defined the constant $v_0 \equiv 2\int_\infty^0 \frac{\dd\tilde{r}}{f(\tilde{r})}$.
Therefore, the area density $\mathcal{A}_{B_1 B_2}(u_{B_2})$ can be identified with $\mathcal{A}_{A_1 B'2}(v_{B'_2})$, subject to the parameter shift $v_{B'_2} = v_0 - v_{A_2}$. This implies that $\mathcal{A}_{A_1 B'2}(v_{B'_2})$ is governed by the same functional form as $\mathcal{A}_{A_1 A_2}(v_{A_2})$, but evaluated for a distinct shifted boundary condition
\be
    A_1:(r=+\infty ,v=\tau_0/2)\ , \quad B'_2:\left( r=0,v=v_{B'_2}=v_0-v_{A_2} \right) \ .
\ee
Explicitly, for the surface $A_1 B'_2$, the conserved quantity $\tilde{H}_*$ is determined by:
\be\label{eq:bc2}
    \frac{\tau_0}{2}-\left(v_0-v_{A_2}\right) = \int_{r=0}^{r=+\infty}{\frac{\dd r}{r^{4-2d}\tilde{X}\left( \tilde{X}+\tilde{H}_{*}^{2} \right)}}\ ,
\ee
where $\tilde{X}=\sqrt{\tilde{H}_{*}^{4} +r^{2d-4}f\left( r \right)}$. From this equation we can solve $\tilde{H}_{*}$ for given $v_{A_2}$. The area of extremal surface then reads
\be\label{eq:bc2b}
    \mathcal{A}_{A_1 B'_2}= \int_{r=0}^{r=+\infty}{\dd r\,\frac{r^{2d-4}}{\tilde{X}}}\ .
\ee
Comparing Eqs.~\eqref{eq:bc2}, \eqref{eq:bc2b} and Eqs.~\eqref{eq:tau_0&v_A&H_*}, \eqref{eq:A&H_*}, we can find
\be\label{eq:bc2c}
    \mathcal{A}_{A_1 B'_2}=\mathcal{F}(\tau_0,v_0-v_{A_2})\ ,
\ee
and combining these results, the total real part of the CWES area is given by:
\be\label{eq:realpartSchAdS}
\begin{aligned}
     \mathcal{A}(A_1 A_2B_2 B_1) &=\mathcal{A}_{A_1 A_2}+\mathcal{A}_{A_1 B'_2}=\mathcal{F}(\tau_0,v_{A_2})+\mathcal{F}(\tau_0,v_0-v_{A_2})\ .
\end{aligned}
\ee
We have thus completed the first step: the area is expressed as a function $\mathcal{F}(x,y)$ via Eq.~\eqref{eq:A&H_*} and \eqref{eq:tau_0&v_A&H_*}. This function is determined by $\tau_0$ (or equivalently, the conserved quantity $H_*$) and $v_{A_2}$. For a fixed $\tau_0$, varying $v_{A_2}$ generates a family of candidate extremal surfaces.

(2) \textbf{Hamilton-Jacobi theory of extremal surface}.
With the area now expressed as a function of  $v_{A_2}$ (Step 1), we proceed to the second step: determining the condition that extremizes this area. For this task, the Hamilton-Jacobi formalism provides the most natural and powerful language, as it directly relates the variation of the on-shell action (here, the area density) with respect to the endpoint coordinates to the conserved quantities.

According to the Hamilton-Jacobi equation, the derivative of the action with respect to the ``time'' coordinate at the endpoint yields the negative of the conserved energy:
\begin{equation}\label{defdfdv1}
  \frac{\partial\mathcal{A}_{A_1 A_2}}{\partial v_{A_2}}=\frac{\partial\mathcal{F}(\tau_0,v_{A_2})}{\partial v_{A_2}} = -E(v_{A_2})\ .
\end{equation}
Similarly, for the second segment $A_1 B'_2$, applying the chain rule to the shifted argument $v_0 - v_{A_2}$ we will obtain
\begin{equation}\label{defdfdv2}
  \frac{\partial\mathcal{A}_{A_1 B'_2}}{\partial v_{A_2}}=\frac{\partial\mathcal{F}(\tau_0,v_0-v_{A_2})}{\partial v_{A_2}}  = E(v_0-v_{A_2})\ .
\end{equation}
For a given $\tau_0$, here the function $E$ is defined according to Eq.~\eqref{eq:conservedH}. Thus, the variation of the total area is given by:
\be
    \begin{aligned}
      \frac{\partial  \mathcal{A}_{A_1A_2B_2 B_1}}{\partial v_{A_2}} &=E(v_0-v_{A_2})-E(v_{A_2})\ .
    \end{aligned}
\ee
To determine the extremal value of total area $\mathcal{A}_{A_1A_2B_2 B_1}$, we only need to impose the following stationarity equation
\be\label{eq:HJofvA}
\frac{\partial \mathcal{A}_{A_1A_2B_2 B_1}}{\partial v_{A_2}} =0\ ,
\ee
which means a simple condition for the extremal configuration $E(v_{A_2})=E(v_{B'_2}=v_0-v_{A_2})$. A manifest solution to this condition is $v_{A_2}=v_{B'_2}=v_0-v_{A_2}$, which corresponds to the perfectly symmetric ``vertical'' configuration $t|_{A_2}=t|_{B'_2}=t|_{B_2}=0$, as illustrated in Fig.~\ref{fig:Hamilton-Jacobi}(b). In Appendix.~\ref{appen:unique}, we will show that $E(v_{A_2})$ is indeed strictly monotonic, which rigorously establishes that $v_{A_2}=v_0-v_{A_2}\Leftrightarrow t|_{A_2}=t|_{B_2}=0$ is the unique extremal solution.

Therefore, the minimal CWES configuration for a SAdS black hole is the time-symmetric ``vertical'' one shown in Fig.~\ref{fig:Hamilton-Jacobi}(b). This establishes a well-defined and unambiguous geometric dual for the TEE, allowing us to proceed with its analytical and numerical evaluation.

In summary, we can conclude that the CWES proposal is well-posed for SAdS black holes, where both the time-like and space-like extremal surfaces satisfy the extremal conditions. To obtain the TEE of SAdS case, we now only need to evaluate the area of this specific vertical configuration.

\subsection{Real and Imaginary Part}

The analysis in Sec.~\ref{subsec:vertical} rigorously established that the minimal CWES corresponds to the time-symmetric configuration ($t|_{A_2}=t|_{B_2}=0$). This provides us with a well-defined geometric quantity to compute. We now proceed to evaluate its area, which consists of a real part from the two space-like segments $A_1 A_2$, $B_1 B_2$ and an imaginary part from the time-like segment $A_2 B_2$
\begin{equation}
    \mathcal{A}\left( A_1A_2B_2B_1 \right)  =\text{Re} \mathcal{A}_{A_1A_2B_2B_1}(\tau_0) +\i\,\text{Im} \mathcal{A}_{A_1A_2B_2B_1}\ .
\end{equation}

\subsubsection{Real part}

With the condition $t|_{A_2}=t|_{B_2}=0$, the relationships we derived in Eqs.~\eqref{eq:tau_0&v_A&H_*} and~\eqref{eq:bc2} simplify significantly. The total boundary temporal width $\tau_0$ and the total real area $\text{Re}\mathcal{A}$ can be expressed as integrals dependent on the conserved quantity $H_*$:
\begin{align}
    \tau_0(H_*) &= 2\int_{r=0}^{r=+\infty}{\dd r\,\frac{r^{2d-4}}{X\left( X+H_{*}^{2} \right)}}\ , \label{eq:tau_H_final}\\
    \text{Re}\mathcal{A}(H_*) &=2 \int_{r=0}^{r=+\infty}{\dd r\,\frac{r^{2d-4}}{X}}\ , \label{eq:area_H_final}
\end{align}
where, as before, $X=\sqrt{H_*^4+r^{2d-4}f(r)}$.

Our objective is to determine the real part $\text{Re}\mathcal{A}(\tau_0)$. A direct approach would require inverting Eq.~\eqref{eq:tau_H_final} to find $H_*(\tau_0)$ analytically and then substituting it into Eq.~\eqref{eq:area_H_final}.
However, the integral in Eq.~\eqref{eq:tau_H_final} is highly non-trivial and cannot be inverted analytically in higher dimensions SAdS black hole.
To overcome this challenge, we adopt a more powerful numerical strategy: we treat $H_*$ as the fundamental independent parameter. By varying $H_*$ starting from its allowed minimum value, $H_{*,\min}$, we can numerically compute a series of corresponding pairs $(\tau_0(H_*), \text{Re}\mathcal{A}(H_*))$. Plotting these pairs directly yields the desired functional relationship $\text{Re}\mathcal{A}(\tau_0)$ without the need for explicit inversion:
\be
    \begin{dcases}
        \tau_0  = \tau_0(H_*)\ ,\\
        \text{Re}\mathcal{A}  =\text{Re}\mathcal{A}(H_*)\ ;
    \end{dcases}
        \ \Longleftrightarrow\ \text{Re}\mathcal{A}=\text{Re}\mathcal{A}(\tau_0)\ .
\ee

Before calculating the real part of area density $\mathcal{A}(A_1 A_2B_2 B_1)$, we should notice that the entanglement entropy generally has a divergence when we integrate from $r= 0$ to $r=\infty$. This is a standard feature of holographic entanglement entropy, corresponding to the infinite entanglement of short-distance modes in the dual field theory. Thus in order to evaluate the finite, physical area from the integral~\eqref{eq:area_H_final}, we employ the standard vacuum background subtraction scheme~\cite{Ryu:2006bv}, subtracting the divergent area in pure AdS spacetime. The regularized area is given by:
\be\label{eq:subtractvacuum}
\begin{aligned}
   \text{Re}\mathcal{A}_{\reg} &=
\text{Re}\mathcal{A}_\text{BH} -\text{Re}\mathcal{A}^0_\AdS  \ ,
\end{aligned}
\ee
where $\text{Re}\mathcal{A}^0_\AdS$ is the (also divergent) area in pure AdS spacetime, whose contribution is scheme-dependent but does not affect the universal features of the final result. Adopting a standard cutoff regularization, this term is given by~\cite{Doi:2023zaf}:
\begin{equation}
    \text{Re} \mathcal{A}_{\AdS}^{0}=\frac{2}{ d-2 }\ell _{\AdS}^{d-1} \epsilon _{\AdS}^{2-d}\ ,
\end{equation}
where $\epsilon$ is a UV cutoff near the AdS boundary. Crucially, since the vacuum contribution is a constant (independent of the bulk path variation), this regularization scheme does not affect the extremization analysis or the stability of the solution derived in the previous section.

\begin{figure}[htbp]
 \begin{center}
   \includegraphics[width=0.5\textwidth]{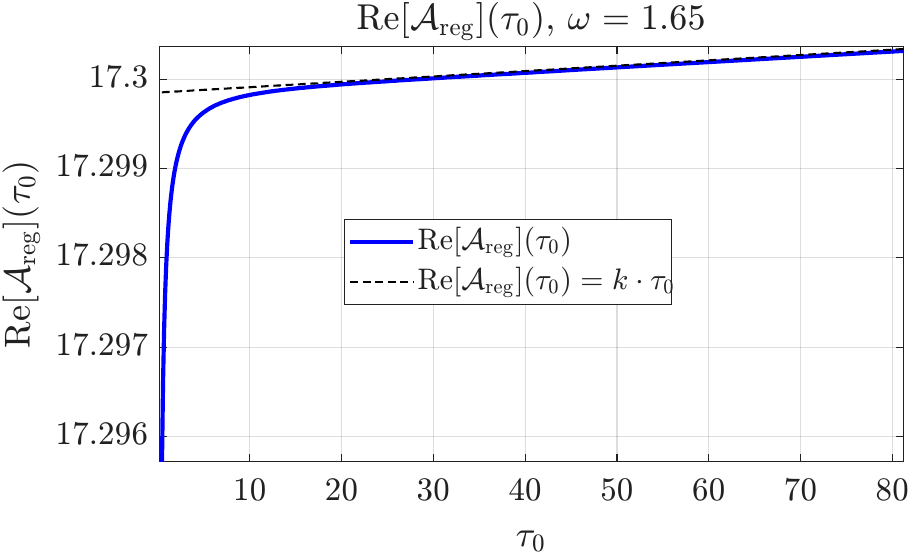}
\caption{The regularized real part of the TEE $\text{Re}\mathcal{A}_{\reg}=\text{Re}\mathcal{A}(\tau_0)$. The solid blue curve represents the full numerical result obtained by parametrically solving Eqs.~\eqref{eq:tau_H_final} and~\eqref{eq:area_H_final}. The dashed black line shows the theoretical linear asymptotic behavior for large $\tau_0$, $\text{Re}\mathcal{A}\approx k\cdot\tau_0$, as predicted in Eq.~\eqref{eq:linearReAtau}. The numerical curve is shown to converge perfectly to this theoretical line.} \label{fig:Area of tau_0}
 \end{center}
\end{figure}

Under the subtraction scheme of Eq.~\eqref{eq:subtractvacuum}, the final numerical result for the regularized real part $\text{Re}\mathcal{A}_{\reg}(\tau_0)$  of the TEE  as a function of $\tau_0$ (with $\omega=1.65$, $\ell_\AdS=1$ and $d=3$) is shown in Fig.~\ref{fig:Area of tau_0}. The cases of different $\omega$ and dimensions $d\geqslant 3$ are similar to each other.

Notice that for the specific case of $d=2$ (BTZ black hole), the exponent $2d-4$ in Eqs.~\eqref{eq:tau_H_final} and~\eqref{eq:area_H_final} vanishes. Taking the metric function as $f(r)=r^2-r^2_h$ (setting $\ell_{\AdS}=1$ and $r_h$ to represent the horizon radius), the integral simplifies to a logarithmic form. Explicitly, the regularized integration yields:
\begin{equation}\label{eq:Aofd=2}
\begin{aligned}
\left.\text{Re}\mathcal{A}\right|_{d=2} &=  2\int_{0}^{1/\epsilon} \frac{\dd r}{\sqrt{r^2 + \left(H_*^4 - r^2_h\right)}} \\
&= 2\left.\ln\left(r + \sqrt{r^2 + H_*^4 - r^2_h}\right) \right|_{0}^{1/\epsilon} \\
&\approx  2\ln \frac{2}{\epsilon} - \ln\left(H_*^4 - r^2_h\right)\ .
\end{aligned}
\end{equation}
Here the factor of 2 accounts for the two symmetric space-like branches of the extremal surface.
This result explicitly exhibits the logarithmic UV divergence $\ln(1/\epsilon)$ characteristic of CFT$_2$, which is effectively removed by the subtraction scheme Eq.~\eqref{eq:subtractvacuum}.

\subsubsection{Imaginary part}\label{subsubsec:imagipart}

The imaginary part of the TEE arises physically from the area of the time-like segment $A_2 B_2$, which connects the past and future singularities. As established in Sec.~\ref{subsec:vertical}, for the minimal configuration, this segment passes through the bifurcation surface $H$, as shown in Fig.~\ref{fig:Hamilton-Jacobi}(b). To calculate its area, we evaluate the area of a surface at constant $t$ and $x$, extending from the singularity $r=0$ to the horizon $r=r_h$ (the radius of the horizon $r_h$ has been defined in Eq.~\eqref{eq:horizon}, satisfying $r^d_h=\omega^{d-2}\ell^2_\AdS$) and back. The induced metric on this codimension-2 time-like surface is
\be
\begin{aligned}
    \dd s^2_{\text{TL}} &=-\frac{\dd r^2}{f(r)}+r^2\dd \boldsymbol{y}_{d-2}^{2}\ ,
\end{aligned}
\ee
Notice that inside the horizon ($r<r_h$), the function $f(r)$ is negative. Consequently, the area element is  $\dd\,\text{Im}\mathscr{A} = \sqrt{|h|} \,\dd^{d-1}y = r^{d-2} / \sqrt{-f(r)} \,\dd r \,\dd^{d-2}y$. The total imaginary area density $\text{Im} \mathcal{A}\left( A_1A_2B_2B_1 \right) \equiv \text{Im}\mathscr{A}/\mathcal{V} _{d-2}$ is twice the integral from the singularity to the horizon:
\be\label{eq:imaginarypartSchAdS}
\begin{aligned}
\text{Im} \mathcal{A}\left( A_1A_2B_2B_1 \right) &=2\int_0^{r_h}{\dd r\,\frac{r^{d-2}}{\sqrt{-f(r)}}}=\frac{2\ell_\AdS r_h^{d-2}}{d} B\left( \frac{3}{2} - \frac{2}{d}, \frac{1}{2} \right) \ ,
\end{aligned}
\ee
where $B\left(\frac{3}{2} - \frac{2}{d}, \frac{1}{2}\right)$ is the Euler beta function.
As a consistency check, for $d=2$ (the BTZ black hole), this formula correctly reproduces the well-known universal result
\be\label{eq:BTZ}
\text{Im} \mathcal{A}\left( A_1B_1B_2A_2 \right) =2\times \frac{\pi\ell_{\AdS}}{2}=\pi\ell_{\AdS} \ .
\ee
It is exactly what happens in the BTZ black hole~\cite{Li:2022tsv}.

While the real part of the TEE varies with the boundary time width $\tau_0$, the imaginary part is independent of $\tau_0$. However, it crucially depends on the black hole thermodynamics (for example, temperature of the black hole).
Using the relation between the horizon radius and the Hawking temperature for planar SAdS black holes~\eqref{eq:AdSmetric},\begin{equation}
    T_\sH= \frac{d r_h}{4\pi\ell^2_\AdS}  \ ,
\end{equation}
one can show that $\text{Im} \mathcal{A}$ scales with the Hawking temperature as
\begin{equation}
\begin{aligned}
      \text{Im} \mathcal{A} &=2\ell _{\AdS}^{2d-3}\frac{\left( 4\pi \right) ^{d-2}}{d^{d-1}}B\left( \frac{3}{2}-\frac{2}{d},\frac{1}{2} \right)\times T_{\sH}^{d-2}\propto T^{d-2}_\sH\ .
\end{aligned}
\end{equation}

The scaling behavior derived above ($\text{Im} \mathcal{A} \propto T^{d-2}_\sH$) suggests that, at least within the framework of the CWES proposal, the imaginary part of TEE encodes fundamental information about the thermal properties of the black hole. Since this contribution depends explicitly on the spacetime geometry (specifically, the temperature) for $d \geqslant 3$, it distinguishes itself from a generic UV cutoff.

However, the interpretation of this imaginary component remains a subject of active discussion. Recent studies utilizing alternative definitions (e.g., Ref.~\cite{Jiang:2025pen}) have suggested that a constant imaginary contribution might be unphysical and could be absorbed into the definition of the UV cutoff, effectively yielding a real-valued TEE.
While our results for $d \geqslant 3$ (which exhibit dynamic dependence on the geometry) suggest that the imaginary part may possess a richer physical structure than previously thought, it is possible that its interpretation depends on the specific computational scheme or dimension.
Whether the imaginary part is a universal physical feature or an artifact of specific holographic proposals remains an open and intriguing question. We do not attempt to settle this debate here, but rather highlight it as a compelling direction for future investigation.

\subsection{Late-time Linear Growth of Time-like Entanglement}\label{subsec:asymptau}

Our numerical results show that for large boundary temporal width $\tau_0\to \infty$, the real part of the TEE appears to grow linearly with $\tau_0$, as shown by the black dashed line in Fig.~\ref{fig:Area of tau_0}. In this section we will verify this asymptotic behavior analytically and numerically.

\begin{figure}[htbp]
\centering
\begin{subfigure}[t]{0.44\textwidth}
\centering
\includegraphics[width=\linewidth]{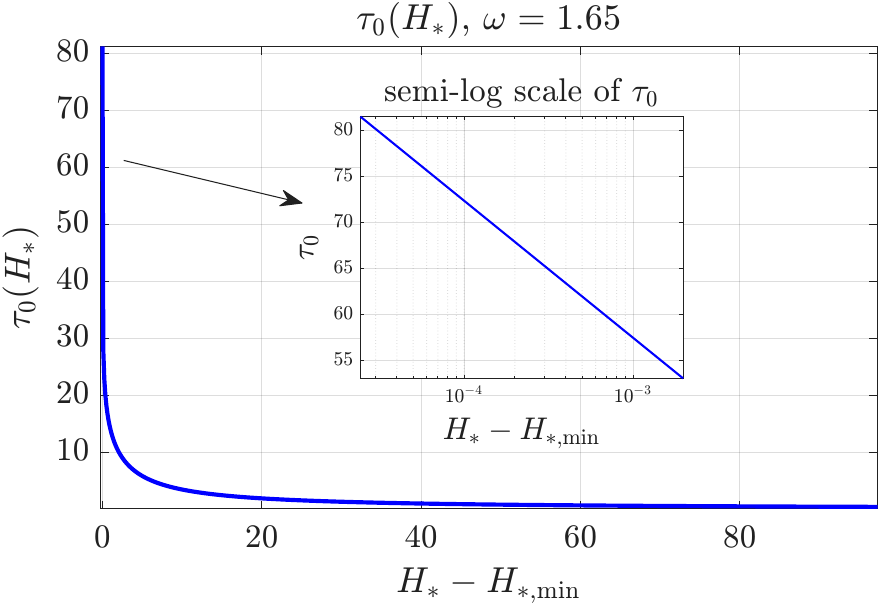}
\caption{  }
\end{subfigure}
\qquad
\begin{subfigure}[t]{0.456\textwidth}
\centering
\includegraphics[width=\linewidth]{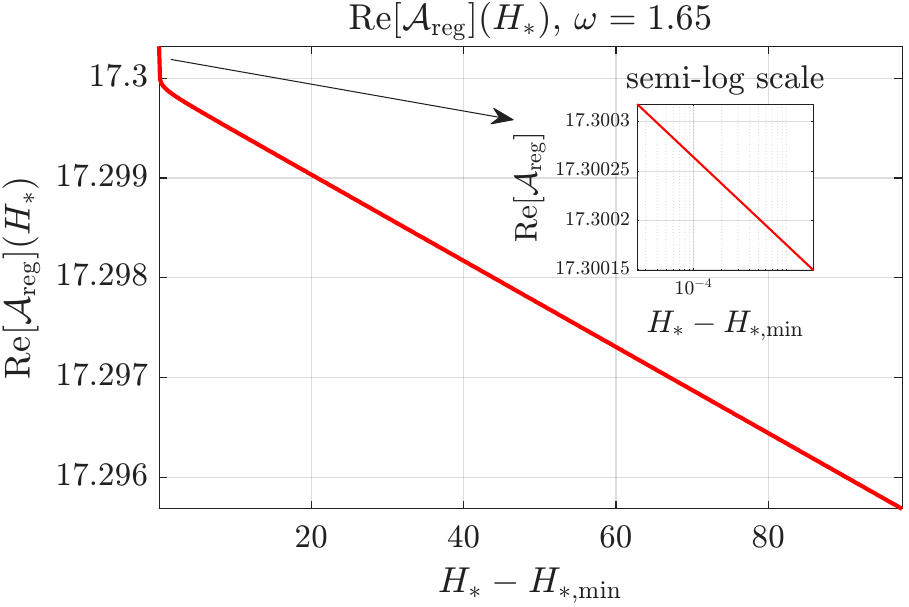}
\caption{  }
\end{subfigure}
 \caption{(a) The boundary temporal width $\tau_0(H_*)$ and (b) regularized real area $\text{Re}\mathcal{A}_\reg(H_*)$ as functions of the conserved quantity $H_*$. Both quantities decrease monotonically as $H_*$ increases. The insets show the same data plotted on a semi-logarithmic scale against the distance to the minimum value, $H_* - H_{*,\min}$. The clear linear behavior in these insets provides strong numerical evidence for the logarithmic divergence predicted by Eqs.~\eqref{eq:tau0Hlim} and~\eqref{eq:ReAHlim}.}
  \label{fig:tau-ReAofH}
\end{figure}

As suggested by the numerical data in Fig.~\ref{fig:tau-ReAofH}, the limit $\tau_0\to \infty$ corresponds to the conserved quantity $H_*$ approaching its minimum value, $H_{*,\min}$, defined by Eq.~\eqref{eq:Hstarmin}.
As $H_* \to H_{*,\min}$, the integrand in both Eqs.~\eqref{eq:tau_H_final} and~\eqref{eq:area_H_final} becomes dominated by the behavior near this critical radius $r_c$, where $r_c$ is defined in Eq.~\eqref{eq:Hstarmin}. To analyze this, we perform a Taylor expansion of the term $X^2$ around $r=r_c$. Defining $\delta \equiv H_*^4 - H_{*,\min}^4 \to 0^+$ and $u \equiv r-r_c$, we find:
\begin{equation}
X \rightarrow \sqrt{\delta +a^2 u^2}\ ,
\end{equation}
where $a^2 = -G''(r_c)/2 > 0$.

With this quadratic approximation, our integrals Eqs.~\eqref{eq:tau_H_final} and~\eqref{eq:area_H_final} simplify significantly. When $H_* \to H_{*,\min}$ , $\tau _0(H_*)$ becomes
\begin{equation}\label{eq:tau0Hlim}
    \tau_0(H_*)\approx  -\frac{2r_{c}^{2d-4}}{a H_{*,\min}^{2}}\ln \delta +\text{finite part}\ .
\end{equation}
Here we have used the fact
\[
    \int \frac{\dd u}{\sqrt{\delta+a^2u^2}}=\frac{1}{a}\text{arcsinh}\left(\frac{a u}{\sqrt{\delta}}\right)\ ,
\]
and $\text{arcsinh}(z)\approx \ln(2z)$ for $z\to \infty$. While for $\text{Re} \mathcal{A}_{A_1A_2} (H_*)$, as $H_* \to H_{*,\min}$ we also have
\begin{equation}\label{eq:ReAHlim}
    2\text{Re} \mathcal{A}_{A_1A_2} (H_*)\approx -\frac{ 2r_{c}^{2d-4}}{a}\ln \delta +\text{finite part}\ .
\end{equation}
This analysis predicts that both temporal width $\tau_0(H_*)$ and area $\text{Re} \mathcal{A}(H_*)$ should diverge logarithmically as $H_* \to H_{*,\min}$ (i.e., as $\delta\to 0$). This precise logarithmic behavior is explicitly confirmed in our numerical results. As shown in the insets of Fig.~\ref{fig:tau-ReAofH}(a) and~\ref{fig:tau-ReAofH}(b), when plotted on a semi-logarithmic scale against $H_*- H_{*,\min}$, both $\tau_0(H_*)$ and $\text{Re} \mathcal{A}(H_*)$ exhibit a clear linear relationship, which is the signature of a logarithmic divergence.

\begin{figure}[htbp]
 \begin{center}
   \includegraphics[width=0.55\textwidth]{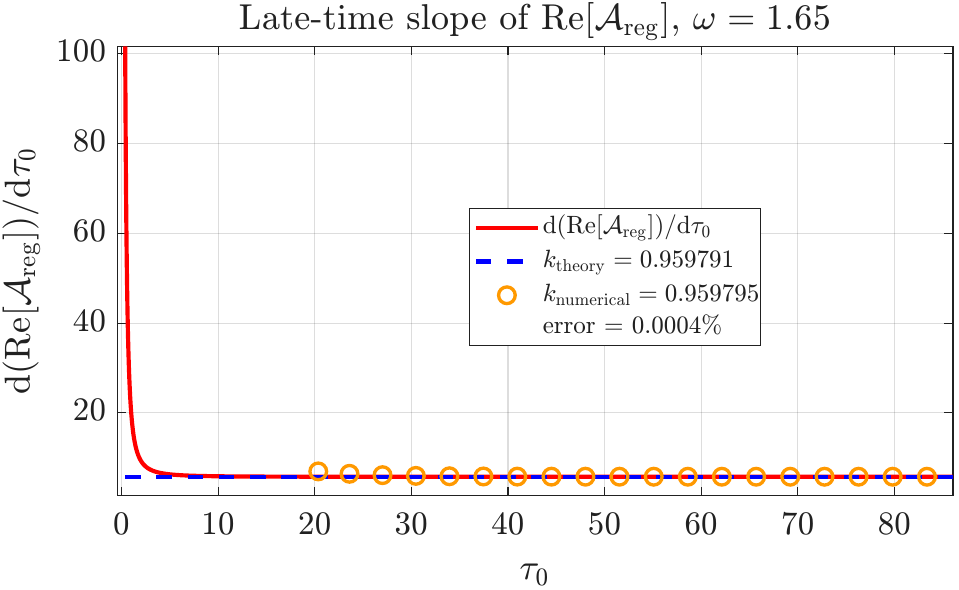}
 \end{center}
\caption{Comparison between the numerical and theoretical growth rates of TEE. The solid red curve denotes the numerical derivative $\dd(\text{Re}\mathcal{A})/\dd\tau_0$. The dashed black line represents the theoretical asymptotic slope $k_{\text{theory}}$ predicted by Eq.~\eqref{eq:linearReAtau}. The yellow circles highlight the numerical data in the large $\tau_0$ regime, indicating convergence to the theoretical limit.}
\label{fig:dReA-dtau}
\end{figure}

By eliminating the $\ln \delta$ term between these two asymptotic analysis expressions Eq.~\eqref{eq:tau0Hlim} and Eq.~\eqref{eq:ReAHlim} for large $\tau_0\to \infty$, we obtain a direct linear relationship between them:
\begin{equation}\label{eq:linearReAtau}
    \text{Re} \mathcal{A}_{A_1A_2B_2B_1}(\tau _0 ) \approx  H_{*,\min}^{2} \,\tau_0+\text{finite part}\ , \quad \text{as} \quad \tau_0 \to \infty\ ,
\end{equation}
which means we can theoretically calculate the slope $k_{\text{theory}}= H_{*,\min}^{2}$ of linear growth $\text{Re} \mathcal{A}(\tau _0 )=k_{\text{theory}}\cdot\tau_0$ for large $\tau_0 \to \infty$.
This linear relationship implies that the growth rate $k_{\text{theory}}$ of TEE should asymptotically approach a constant value. To verify this prediction, we numerically compute the derivative $\dd(\text{Re}\mathcal{A})/\dd\tau_0$ and compare it against this theoretical slope. As illustrated in Fig.~\ref{fig:dReA-dtau}, the numerical derivative converges precisely to the predicted constant in the large $\tau_0 \to \infty$ limit. This agreement confirms the validity of our asymptotic analysis.

Geometrically, as the boundary temporal width increases ($\tau_0 \to \infty$), the extremal surface $A_1 A_2 B_2 B_1$ approaches and lingers at the critical radius $r_c$ (defined in Eq.~\eqref{eq:Hstarmin}) for an arbitrarily long width. In this regime, the surface accumulates area at a constant rate $k_{\text{theory}}= H_{*,\min}^{2}$. This critical extremal surface therefore governs the asymptotic behavior of TEE in the limit of large temporal width, serving as the holographic dual to the linear growth of entanglement entropy.

\section{Time-like Entanglement in Hairy Black Holes}\label{sec:Type-II}

In the previous section, we established a robust framework for calculating the TEE and applied it to the SAdS black hole. The discussion in Sec.~\ref{sec:SAdS} relied on a specific geometric feature: the causal structure of the singularity permits space-like extremal surfaces with non-zero area to connect the boundary to the interior.
However, this causal structure is not universal to all black holes. In particular, the geometric properties of simple vacuum solutions like the SAdS and BTZ black holes are not universal. Moving beyond simple vacuum solutions, one encounters spacetimes where matter backreaction fundamentally alters the interior causal structure.

\begin{figure}[htbp]
\centering
   \includegraphics[width=0.95\textwidth]{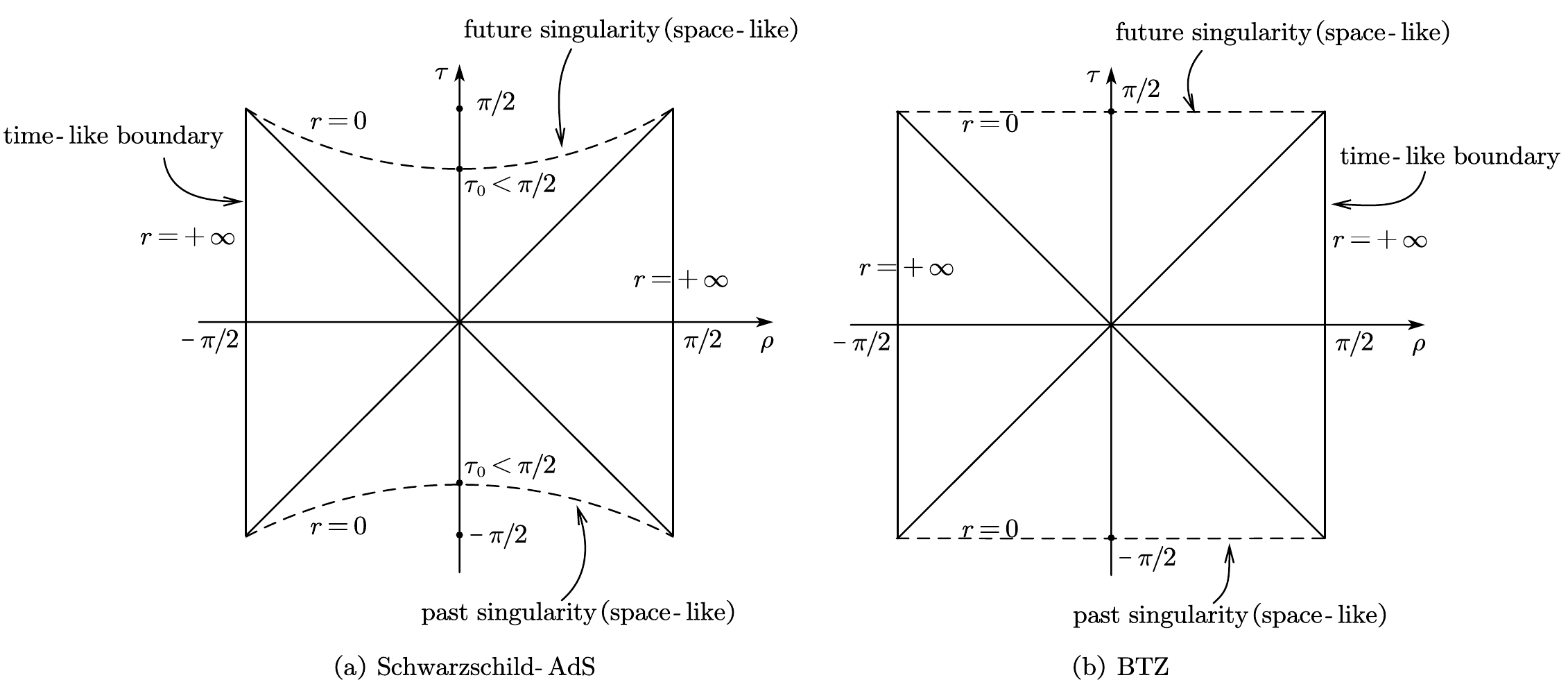}
\caption{(a) Penrose diagram of AdS-Schwarzschild black hole, where the space-like singularity behaves as ``concave''. (b) Penrose diagram of the BTZ black hole, where the space-like singularity is represented as a horizontal straight line.  } \label{fig:Sch-AdS-BTZ}
\end{figure}

In fact, the interior causal structure of spacetime will constrain the behavior of both space-like and time-like extremal surfaces. This structure is clearly reflected in the geometry or ``shape'' of the Penrose diagram.
For instance, for a SAdS black hole, the space-like singularity is depicted as a ``concave'' curve, as shown in the Penrose diagram in  Fig.~\ref{fig:Sch-AdS-BTZ}(a).
Specifically, the $d=2$  BTZ black hole is similar to SAdS case, while with the slight difference that its space-like singularity is drawn as a parallel direction of space-like straight line(as illustrated in Fig.~\ref{fig:Sch-AdS-BTZ}(b)), and the overall Penrose diagram shows a square shape.
In contrast, for black holes with scalar hair, the singularity can become ``convex''~\cite{An:2022lvo, Caceres:2022smh}. We will demonstrate that this geometric shift forces the TEE of small boundary temporal width to enter a novel ``time-like phase'', where the time-like entanglement entropy originates solely from the time-like extremal surface, with the space-like contribution reduces to a UV regulator effect. A detailed discussion will be provided in Sec.~\ref{subsec:phasetrasTEE}.

To investigate this phenomenon, we turn our attention to a well-understood, non-vacuum solution: the hairy black hole that arises in the context of holographic superconductivity. This system serves as an ideal laboratory to classify these interior structures and to study the emergence of this time-like entanglement.

\subsection{Classifying Black Hole Interiors} \label{subsec:holoType-II}

\subsubsection{Type-I/II spacetimes}

Consider a static black hole possessing a single event horizon with non-zero surface gravity; in such cases, the singularity is typically space-like. We begin by introducing a criterion based on the causal nature of the singularity to classify such spacetimes as either {\it Type-I} or {\it Type-II}, and proceed to confirm that the SAdS black hole in Sec.~\ref{sec:SAdS} and BTZ black hole belong to the Type-I category.

The idea of classifying black hole interiors based on their causal ``shape'' is not new. Pioneering work used the behavior of ingoing null geodesics to describe the ``shape'' of Penrose diagram relative to the boundary~\cite{Fidkowski:2003nf}.
More recently, this classification has been explored in the context of the Complexity=Action (CA) conjecture, where the geometry of the Wheeler-DeWitt patch led to a distinction between ``Type-D'' (down) and ``Type-U'' (up) spacetimes~\cite{An:2022lvo, Caceres:2022smh, Auzzi:2022bfd}.
These classifications describe the same underlying physical dichotomy; our Type-I (concave) spacetimes correspond to the Type-D (down) cases, while our Type-II (convex) spacetimes correspond to the Type-U (up) cases. We adopt the Type-I/II nomenclature in this work. These findings suggest that such a classification reflects a robust feature of black hole interior geometry.

\begin{figure}[htbp]
 \begin{center}
   \includegraphics[width=0.95\textwidth]{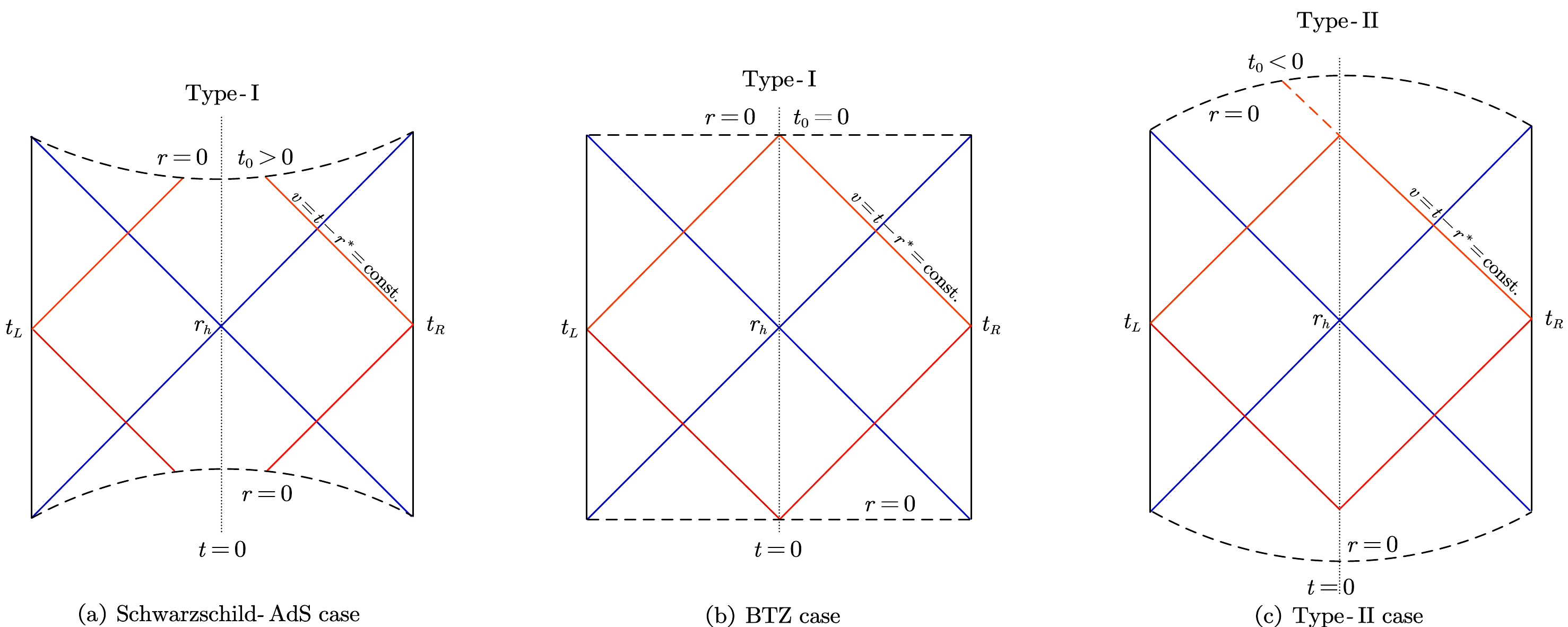}
 \end{center}
\caption{Red lines denote the null sheets that start from boundary and end at the singularities.  (a) In SAdS case, the time $t_0$ is greater than zero, which shows that the space-like singularity behaves as ``concave''.  (b)  In BTZ case, the time $t_0$ is exactly zero, which shows that the ``shape of spacetime'' is ``square''. They all belong to Type-I spacetime. (c) In Type-II case, the time $t_0$ is less than zero, which shows that the space-like singularity behaves as ``upward convex''.}\label{fig:Type-I-II}
\end{figure}

To define these two kinds of spacetimes, we consider the in-falling null sheet that starts from the opposite boundary at coordinate time $t=0$. Such a null sheet will meet the singularity at coordinate time $t_0$. In order to find the value of $t_0$, we transform into the Eddington coordinates defined in Eq.~\eqref{eq:Eddington}. In-falling light rays follow paths of constant $v =t+r_*=\text{const}$, which are indicated by the red lines in Fig.~\ref{fig:Type-I-II}. For the null sheet starting from boundary with $t=0$, we have $v=t+r_*=0+r_*(\infty)=0$.
When the null sheets reach the singularity, the time $t_0$ is given by the equation
\be\label{eq:reachessingularity1}
   t=t_0 \equiv-\int^{0}_{\infty}{\frac{\dd \tilde{r}}{f\left( \tilde{r} \right)}}\ .
\ee
Based on the sign of $t_0$, which characterizes the ``degree of concavity (or convexity)'' of the space-like singularity of spacetime, we classify spacetimes into two families (see Fig.~\ref{fig:Type-I-II}):
\begin{itemize}
    \item \textbf{Type-I (Concave/Flat Singularity):}  $t_0 \geqslant 0$, the singularity is ``concave'' (shown in Fig.~\ref{fig:Type-I-II}(a)) or ``flat''(shown in Fig.~\ref{fig:Type-I-II}(b)). Two null sheets coming from $t_L = t_R = 0$ will first meet the space-like singularity at  $r=0$ before their coordinate time $t$ decreases into negative. Two typical examples are SAdS ($t_0>0$) and BTZ ($t_0=0$) black holes\footnote{While other classification schemes (e.g., Ref.~\cite{Caceres:2022smh}) may treat the $t_0=0$ case as a distinct critical scenario, our classification is based on the behavior of the Time-like Entanglement Entropy (TEE). From the perspective of TEE, the crucial distinction is simply whether a non-zero ``time-like entanglement gap'' $\tau_c$ is required for its definition, as we will discuss in Sec.~\ref{subsec:phasetrasTEE}. Since for both $t_0>0$ and $t_0=0$ this gap is absent ($\tau_c=0$), we group them together as Type-I.}. In this scenario, the CWES and the behavior of TEE will be similar to what we have found in SAdS case in Sec.~\ref{sec:SAdS}.
    \item \textbf{Type-II (Convex Singularity):} If $t_0 < 0$, the singularity is ``convex'', as shown in Fig.~\ref{fig:Type-I-II}(c). The light rays will meet each other before they arrive at the singularity. As we will argue in Sec.~\ref{subsec:causal}, this seemingly minor change in causal structure has dramatic consequences: for small boundary temporal width, the TEE comes only from a time-like entanglement contribution, with the space-like contribution reduces to a UV regulator effect.
\end{itemize}
This classification, being rooted purely in the causal structure deep within the horizon, defines two distinct causal phases of the black hole interior, which we label as \textit{Type-I} and \textit{Type-II}.

To apply this criterion, we must evaluate the integral for $t_0$. This presents a technical challenge: taking the SAdS black hole as an example, the integrand $1/f(r)$ leads to a logarithmic divergence at the event horizon $r=r_h$, where $f(r_h)=0$. However, the logarithmic divergence encountered when integrating from the outside ($r>r_h$) to the horizon is precisely canceled by the divergence from the inside ($r<r_h$). To see this, notice that $r=r_h$ is a simple pole of the integral, then we could separate the principal value part at the horizon from the integral, leading to a numerically stable form
\be\label{eq:poles2}
    \int^r_{+\infty}{\frac{\dd r}{f\left( r \right)}}=\int^r_{+\infty}{\dd r\left( \frac{1}{f\left( r \right)}-\frac{2r_h}{f^{\prime}\left( r_h \right) \left( r^2-r_{h}^{2} \right)} \right)}+f^{\prime}\left( r_h \right) ^{-1}\ln \left| \frac{r-r_h}{r+r_h} \right|\ .
\ee
Notice that when on the boundary $r\to \infty$ and  at the singularity $r\to 0$, the term $f^{\prime}\left( r_h \right) ^{-1}\ln \left|(r-r_h)/(r+r_h)\right|$ in Eq.~\eqref{eq:poles2} actually does not contribute to integral, which means in following calculations we can drop this term. We then obtain
\be\label{eq:reachessingularity2}
\begin{aligned}
   t_0 &=-\int^{ 0}_{+\infty}{\dd  r\left( \frac{1}{f\left( r \right)}-\frac{2r_h}{f^{\prime}\left( r_h \right) \left( r^2-r_{h}^{2} \right)} \right)}\ .
\end{aligned}
\ee
From this formula one can verify that the SAdS black hole has $t_0>0$ and BTZ black hole has $t_0=0$, and so both belong to Type-I case.

Having established our criterion and validated it, we now generalize it for general static, planar-symmetric black hole spacetimes that will be the focus of the rest of our investigation. Consider a general metric of the form ($z_h$ is the position of horizon, i.e., $g(z_h)=0$)
\begin{equation}\label{eq:generalAdS}
    \dd s^2 = \frac{1}{z^2}\left[-g(z)\e^{-\chi(z)}\dd t^2+g^{-1}(z)\dd z^2+\dd x^2+\dd \boldsymbol{y}_{d-2}^{2}\right]\ .
\end{equation}
Introducing the in-falling null coordinate
\begin{equation}
    \begin{aligned}
        v &=t-z^*(z)\ ,\\
        z^*(z) &=\int^z_0\dd z\, g^{-1}\e^{\chi/2}\ ,
    \end{aligned}
\end{equation}
the in-falling null sheets are given by the equation
\begin{equation}
     v =t-z^*(z)=\text{const.}\ .
\end{equation}
In this case the same logic applies, and the regularized time $t_0$ of in-going null sheets from boundary meeting singularity of metric~\eqref{eq:generalAdS} is determined by the equation:
\begin{equation}\label{eq:reachessingularity3}
    t_0 =\int^{\infty}_0 \dd z\left(g^{-1}(z)\e^{\chi(z)/2}-\frac{2\e^{\chi(z_h)/2}z_h}{g'(z_h)(z^2-z^2_h)}\right)\ .
\end{equation}
The sign of this quantity will serve as our guide in exploring novel phenomena in more complex spacetimes, such as hairy black holes.

\subsubsection{The Type-II interior from holography}\label{subsec:TypeIIex}

To provide a concrete example of a spacetime where the space-like singularity behaves convexly (i.e., Type-II interior), we now turn to the well-studied $D=(d+1)$-dimensional Einstein-Maxwell-scalar model, in which gravity is coupled with a gauge field $A_\mu$ and a
charged scalar field $\Phi$, which provides the standard holographic description of a superconductor in $d$-dimensional boundary field theory~\cite{Hartnoll:2008kx,Cai:2015cya}
\be\label{eq:actionofRN}
I=\frac{1}{16\pi}\int \dd^D x\,\sqrt{-g}\left(R+\frac{(D-2)(D-1)}{\ell_{\AdS}^2}-\frac{1}{4}F^2 - \left|D \Phi\right|^2-m^2\left|\Phi\right|^2\right)\ ,
\ee
where $R$ is the Ricci scalar of spacetime,  $F_{\mu\nu}=\left(\dd A\right)_{\mu\nu}$ is the field strength tensor, and $D_\mu =\nabla_\mu-\i q A_\mu$ is the gauge covariant derivative with charge $q$ and mass $m$ for the charged scalar field. When the temperature is high enough, the solution is a planar Reissner-Nordstr\"{o}m-AdS black hole, which possesses a Cauchy horizon and a time-like singularity. When temperature is lower than a critical value $T_c$, the superconducting phase transition happens, the Cauchy horizon disappears and the singularity will become space-like. With  decreasing temperature, the backreaction of matter fields can sufficiently alter the interior geometry to make the singularity ``convex'' ($t_0 < 0$)~\cite{An:2022lvo}.
After the superconducting phase transition happens, planar-symmetric hairy black hole and matter fields ansatz can be written in the general form:
\be\label{eq:ansatzofRN}
\begin{aligned}
   \dd s^2 &=\frac{1}{z^2}\left(-g(z)\e^{-\chi(z)}\dd t^2+g^{-1}(z)\dd z^2+ \dd x^2+\dd \boldsymbol{y}_{d-2}^{2}\right)\ ,\\
   A_\mu &=A_t(z)(\dd t)_\mu\ ,\\
   \Phi &=\phi(z)\ .
\end{aligned}
\ee
Here the metric functions $g(z)$ and $\chi(z)$ now encode the effects of both the black hole's mass and charge, and the profile of the scalar hair.
In this coordinate system,  the spacetime boundary is located at $z = 0$ and the singularity of black hole lies at $z \to \infty$. With this ansatz, the Hawking temperature $T$ could be expressed as
\be
T= -\frac{\e^{-\chi(z_h)/2}g'(z_h)}{4\pi} \ .
\ee
The solutions of these functions can be obtained numerically~\cite{Hartnoll:2008kx,Cai:2015cya}.

\begin{figure}[htbp]
 \begin{center}
   \includegraphics[width=0.55\textwidth]{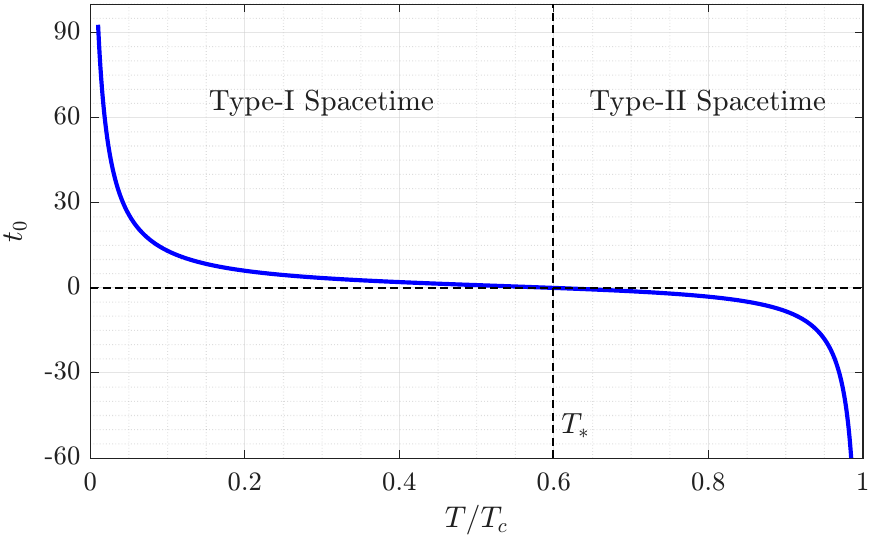}
 \end{center}
\caption{When $d=3$, $m^2=-2$, $q=1$, the result of $t_0$ as the function of $T/T_c$ shows that when approaching the critical temperatures, $t_0$ becomes negative, the spacetime belongs to the Type-II class.  $T_*$ denotes the critical temperature for the phase transition from Type-II to Type-I spacetime.}  \label{fig:t0ofT}
\end{figure}

For each numerically-obtained matter field and hairy black hole metric field solution~\eqref{eq:ansatzofRN} (parameterized by its temperature $T$), we can compute the time $t_0$ that null sheets reach the singularity from the boundary using our generalized formula in Eq.~\eqref{eq:reachessingularity3}.
The result of this procedure for a standard holographic superconductor model (with $m^2=-2, q=1$ in $d=3$) is presented in Fig.~\ref{fig:t0ofT}. The plot of $t_0$ as a function of the normalized temperature $T/T_c$ shows precisely the behavior we anticipated.
At very low temperatures, $t_0$ is positive (indicating Type-I interior). However, as the temperature increases towards the critical temperature $T_c$ of the superconductor phase transition, $t_0$ decreases, crosses zero, and becomes negative. This unambiguously demonstrates that Type-II spacetimes are physically realizable within this well-understood framework.

It is important to note that the holographic superconductor model itself is not crucial in this context; it serves merely as an example that generates a Type-II interior and provides numerical results for the Type-I/II classification criteria. The phenomenon is general: Type-II behavior typically arises from specific deformations of an unstable inner Cauchy horizon, a sufficiently small perturbation can destroy the inner horizon and produce a Type-II interior~\cite{Cai:2020wrp, An:2022lvo}.

\subsection{Time-like Entanglement Phase in Type-II Spacetimes}\label{subsec:causal}

In the previous section, we found that hairy black holes can exhibit a Type-II interior, characterized by a ``convex'' singularity with $t_0 < 0$. This seemingly minor causal geometric change has a dramatic consequence for TEE: for Type-II interiors, there exists a distinct ``time-like entanglement phase'' for narrow boundary strips. In this phase, the space-like contribution to the TEE reduces to a UV regulator effect, and the TEE is determined solely by the area of time-like extremal surfaces.

\begin{figure}[htbp]
 \begin{center}
   \includegraphics[width=0.95\textwidth]{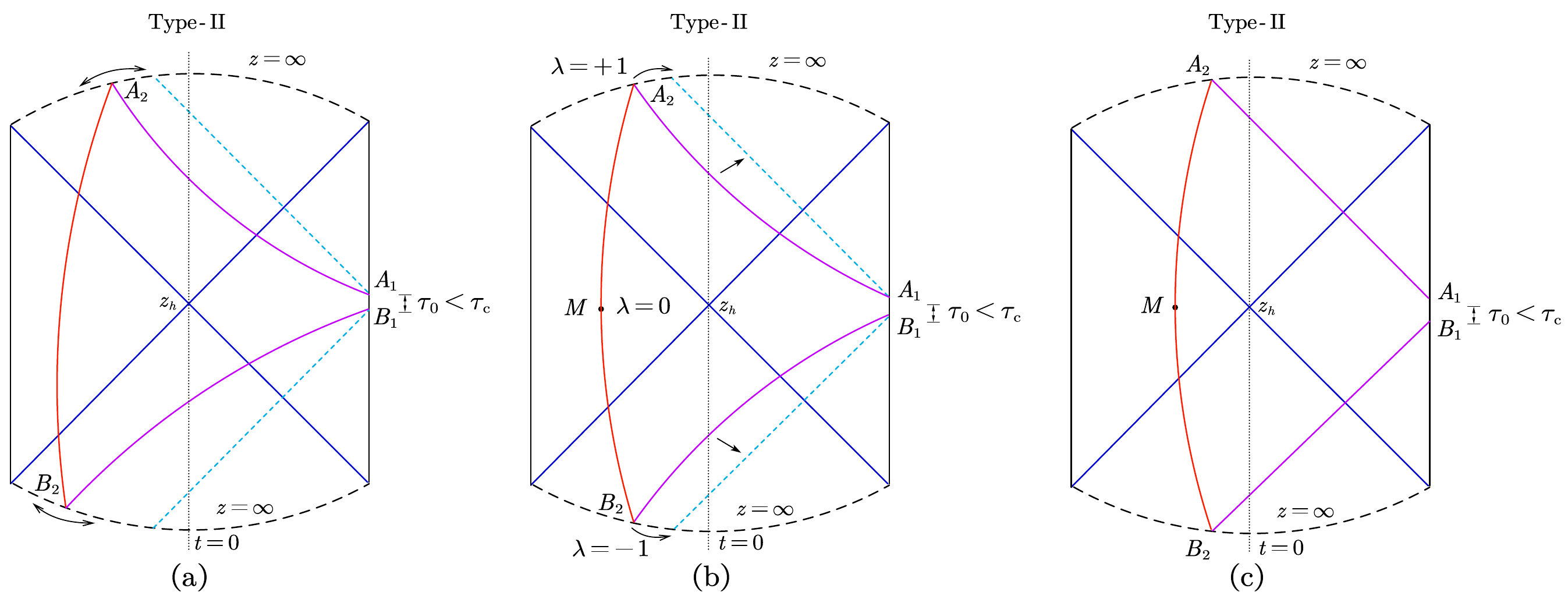}
 \end{center}
\caption{CWES configurations in the ``time-like phase'' for $\tau_0 < -2t_0$. Blue dashed lines indicate ingoing null sheets from boundary endpoints $A_1$ and $B_1$. Due to causality, space-like segments (purple lines) are confined to the left of these null sheets, as seen in (a) and (b). For non-symmetric configuration (a), one can move the ``endpoints'' $A_2$ and $B_2$ towards opposite directions without changing the area of time-like part and so obtain the configuration (b). The minimal area is achieved in the limiting configuration (c), where these space-like segments will become null surfaces with zero area, leaving only the time-like segment (red line) to contribute to the TEE. It should be noted that, though most parts of $A_1A_2$ and $B_1B_2$ are along two $45^\circ$ lines, their intersections with AdS boundary are still ``perpendicular'' (this is not exhibited in this figure, see Fig.~\ref{fig:nulllimit}). In fact, the limit where a space-like extremal surface approaches a null hypersurface involves a more subtle regulator effect, which needs to impose both  spatial and temporal UV cut-offs. We will discuss this in Sec.~\ref{subsec:phasetrasTEE}.} \label{fig:Type-II-CWES1}
\end{figure}

We now present the proof for this conclusion. Let us first consider a CWES anchored on a boundary strip of temporal width $\tau_0$. An ingoing null ray from the boundary point $A_1(t=\tau_0/2)$ reaches the singularity at a coordinate time $t|_{r=0} = \tau_0/2 + t_0$. This null sheet gives the boundary of the future ``light cone'' of $A_1$. Due to causality, any space-like segment $A_1 A_2$ anchored at $A_1$ must remain outside of this light cone, i.e., the ``endpoint'' $A_2$ should satisfy $t|_{A_2}<\tau_0/2 + t_0$. Similarly, any space-like segment $B_1 B_2$ anchored at $B_1$ must remain exterior of the past light cone of $B_1$ and the ``endpoint'' $B_2$ should satisfy $t|_{B_2}>-(\tau_0/2 + t_0)$.
If we choose a sufficiently small temporal width $\tau_0 < -2t_0$ (recalling that $t_0 < 0$ in Type-II spacetime), we find that $t|_{A_2} < 0$ and $t|_{B_2} >0$ for any valid space-like extremal surfaces $A_1A_2$ and $B_1B_2$. This implies that the entire causally accessible region for both segments, $A_1 A_2$ and $B_1 B_2$, lies strictly to the left of the vertical $t=0$, as illustrated in Fig~\ref{fig:Type-II-CWES1}(a). As a result, it is geometrically impossible to construct a time-like extremal surface connecting $A_2$ and $B_2$ that passes through the bifurcation surface $H$ in Type-II spacetime for narrow boundary strips.

In order to explain what this restriction will bring to the TEE, we start from the general setup in Fig.~\ref{fig:Type-II-CWES1}(a). By the extremality condition of TEE, we should begin by taking the time-like surface to be extremal. Suppose the time-like extremal surface has endpoints $A_2B_2$ that lie asymmetrically on the upper and lower singularities.
One can move these two endpoints along opposite directions without changing the total time-like area until the surface becomes the symmetric configuration with $t|_{A_2}=-t|_{B_2}$, and still remain to the left of the vertical dashed line $t=0$, shown in Fig.~\ref{fig:Type-II-CWES1}(b). Therefore, without loss of generality, we can restrict our search for the minimal CWES to symmetric configurations where $t|_{A_2}=-t|_{B_2}$.
The symmetric codimension-2 time-like extremal surface $A_2B_2$ can be parameterized by $\left\{t=t(\lambda),z=z\left( \lambda \right) ,x=0,\boldsymbol{y}_{d-2}\in \mathbb{R} ^{d-2} \right\}$.
For the symmetric configuration, we choose the middle point $M$ at $\lambda = 0$ with $t\left( \lambda =0 \right) =t_M=0$ and $z\left( \lambda =0 \right) =z_M$. The endpoints $A_2,B_2$ are set by $A_2:\lambda =+1$ and $B_2:\lambda =-1$ at the upper and lower singularities respectively.

The induced metric and the time-like area density functional $\mathcal{S} _{\text{TL}}$ for this time-like surface are
\begin{equation}
    \begin{aligned}
        \dd s^2 &=\frac{1}{z^2}\left[ -g(z)\e^{-\chi (z)}\dot{t}^2\dd \lambda ^2+g^{-1}(z)\dot{z}^2\dd \lambda ^2+\dd \boldsymbol{y}_{d-2}^{2} \right]\ ,\\
        \mathcal{S} _{\text{TL}} &\equiv \text{Im} \mathcal{A} =2 \int_{\lambda =0}^{\lambda =+1}{\dd \lambda \,L}\ ,
    \end{aligned}
\end{equation}
where  $L=z^{1-d}\sqrt{g(z)\e^{-\chi (z)}\dot{t}^2-g^{-1}(z)\dot{z}^2}$ is the Lagrangian functional. Notice that $L$ does not depend explicitly on generalized coordinate $t$, hence there is a conserved momentum $\Pi$
\begin{equation}
    \Pi =\frac{\partial L}{\partial \dot{t}}=\frac{\dot{t}z^{1-d}g(z)\e^{-\chi (z)}}{\sqrt{g(z)\e^{-\chi (z)}\dot{t}^2-g^{-1}(z)\dot{z}^2}}\ .
\end{equation}
At the symmetric midpoint $M$, we have $z(\lambda =0)=z_M,\dot{z}(\lambda =0)=0$. Note that $\dot{t}<0$ around $M$ since the time-like Killing vector $\partial_t$ points down around $M$, which is opposite to the direction of increasing affine parameter $\lambda$. Substituting this $\dot{t}|_{M}<0$ and $\dot{z}|_{M}=0$ back into the expression for the conserved momentum $\Pi$, we will get
\begin{equation}
    \begin{aligned}
        \Pi (z_M) &=-\frac{\sqrt{g(z_M)\e^{-\chi (z_M)}}}{z_{M}^{d-1}}<0\ .
    \end{aligned}
\end{equation}
Since $z_M>0$ and $g(z_M) > 0$ outside the horizon, the conserved momentum $\Pi$ is strictly negative.

According to Hamilton-Jacobi theory, the variation of the on-shell time-like area $\mathcal{S} _{\text{TL}}$ with respect to its time $t_{A_2}$ of upper endpoint is given by this conserved momentum
\begin{equation}\label{eq:pureSTL}
    \frac{\partial \mathcal{S} _{\text{TL}}}{\partial t_{A_2}}=\left. 2\Pi\right|_{A_2}<0\ .
\end{equation}
The factor of 2 arises because the constraint requirement $t|_{B_2}=-t|_{A_2}$ implies that the lower endpoint $B_2$ varies simultaneously.
Since $\Pi$ is constant along the time-like surface and $\Pi|_{M}<0$, we have rigorously proven that $\frac{\partial \mathcal{S}_{\text{TL}}}{\partial t_{A_2}} < 0$.

Eq.~\eqref{eq:pureSTL} provides the critical insight: the derivative is strictly negative, which implies that the time-like area $\mathcal{S}_\text{TL}$ monotonically decreases as the endpoint $A_2$ and $B_2$ move to the right (increasing $t_{A_2}$, but decreasing $t_{B_2}$). Consequently, the infimum of the time-like area is achieved at the maximum value of $t_{A_2}$ allowed by causality.
In other words, among all candidate extremal surfaces, the time-like extremal surface reaches its infimum in the limit where the configuration from Fig.~\ref{fig:Type-II-CWES1}(b) approaches the configuration shown in Fig.~\ref{fig:Type-II-CWES1}(c). In that limit, the space-like extremal surfaces approach the null configurations anchored at $A_1$ and $B_1$ (purple lines in Fig.~\ref{fig:Type-II-CWES1}(c)).
Since both components simultaneously approach their respective infimum at this limiting configuration, the global minimum of the total area is realized precisely in this ``null limit'' (Fig.~\ref{fig:Type-II-CWES1}(c)).

It is important to emphasize that this null limit is the ``limit'' of a family of space-like surfaces which are all ``perpendicular to'' the AdS boundary. As a result, though the surfaces will increasingly resemble null surfaces in the region far away from the AdS boundary, their UV behaviors always keep ``perpendicular to'' boundary rather than intersect with the AdS boundary at a $45^\circ$ angle.
In this null limit, the contribution from space-like terms yields a non-zero value due to this UV regularization effect, in contrast to the strict zero expected from a purely null surface. We will analyze this null limit in Sec.~\ref{subsec:phasetrasTEE} and show that it is governed by a connection between the UV cutoff on the AdS boundary and the cutoff regulating the approach to the null surface. Up to this regulator effect, the imaginary part from the time-like surface dominates the behavior of the TEE. Therefore, in this regulated sense, we can identify the existence of a ``time-like entanglement phase'' for small temporal width in Type-II interior.

\subsection{Phase Transition of Time-like Entanglement}\label{subsec:phasetrasTEE}

\begin{figure}[htbp]
 \begin{center}
   \includegraphics[width=0.95\textwidth]{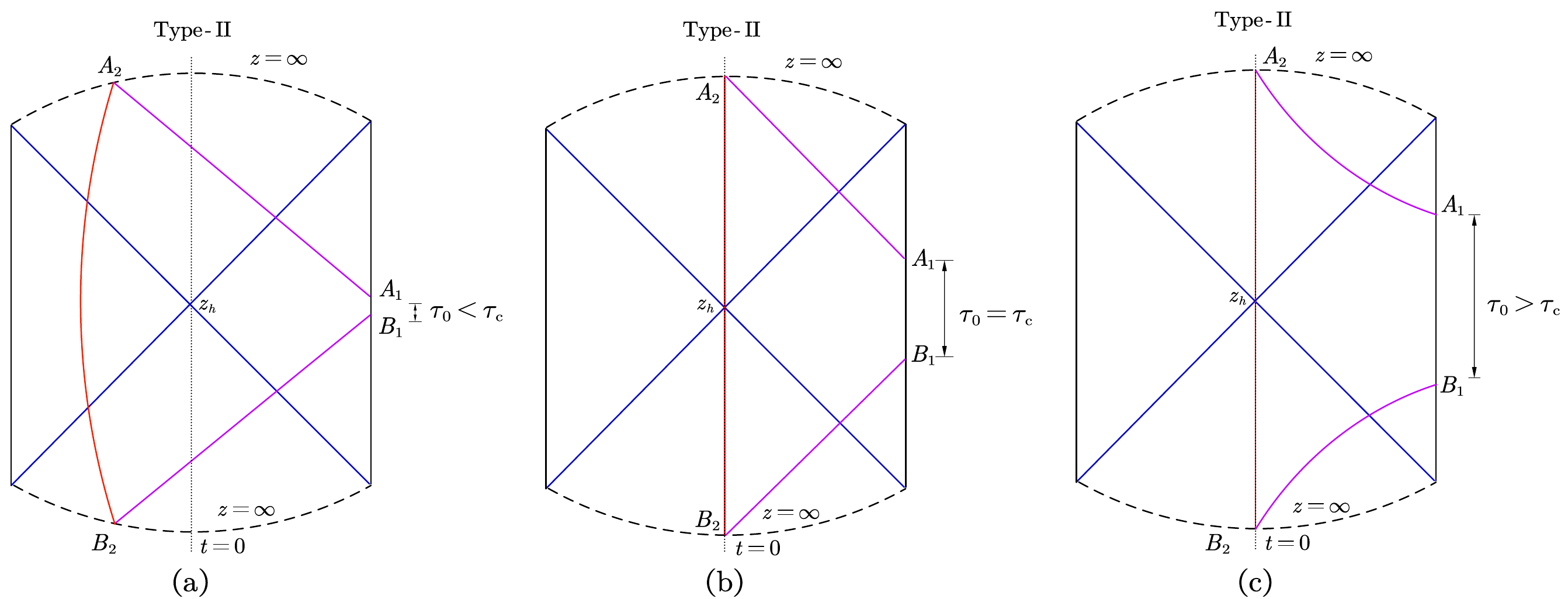}
 \end{center}
\caption{There are three possible CWESs configurations in the Type-II spacetime. (a) When $\tau_0 <\tau_c$, only time-like entanglement contributes to time-like entanglement entropy, up to a regulator effect. (b) TEE is given by a critical configuration, where time-like contribution becomes ``saturated''.  (c) In $\tau_0>\tau_c$, TEE reduces to the case of Type-I spacetime, which have been discussed in Sec.~\ref{sec:SAdS}. Once again, though most parts of $A_1A_2$ and $B_1B_2$ in (a) and (c) are along two $45^\circ$ lines, their intersections with AdS boundary are still ``perpendicular'' (this is not exhibited in these figures, see Fig.~\ref{fig:nulllimit}).} \label{fig:Type-II-CWES2}
\end{figure}

As we increase boundary temporal width $\tau_0$, the endpoint $t_{A_2} = t_0 + \tau_0/2$ is pushed towards the future. For Type-II geometries we have $t_0 <0$, this defines a critical threshold $\tau_c$
\begin{equation}
    \tau_c \equiv -2t_0 \,,
\end{equation}
where $t_0$ is defined in Eq.~\eqref{eq:reachessingularity3}. If $\tau_0<\tau_c$, both endpoints $A_2,B_2$ lie on the same side of the vertical dashed line $t=0$, as shown in Fig.~\ref{fig:Type-II-CWES2}(a). In this regime the TEE is contributed only by the time-like extremal surface up to a regulator effect, as we have discussed in Sec.~\ref{subsec:causal}. In this case, the imaginary part of TEE decreases with increasing temporal width of boundary strip, and though the extremal surfaces $A_1A_2$ and $B_1B_2$ are ``most'' null, the UV cut-off effect may yield a non-zero real part.
When $\tau_0=\tau_c$, the two endpoints $A_2$ and $B_2$ now lie on a critical line, the TEE is given by a critical configuration as illustrated in  Fig.~\ref{fig:Type-II-CWES2}(b). When $\tau_0>\tau_c$, $A_2$ and $B_2$ could lie on opposite sides of $t = 0$ and time-like extremal surface can pass through the bifurcation of event horizon (see Fig.~\ref{fig:Type-II-CWES2}(c)). In this case TEE acquires both time-like and space-like contributions and the situation reduces to the Type-I case, which have been analyzed in Sec.~\ref{sec:SAdS}. In this regime, the imaginary part of TEE becomes saturated and keeps constant, but the real part of TEE begins to increase with the temporal width $\tau_0$.

Thus, for $\tau_0 \in [0,\tau_c)$ the TEE probe measures only time-like entanglement. At $\tau_0\geqslant \tau_c$, the time-like contribution becomes ``saturated'' and the space-like contribution appears.
The window of boundary time $[0,\tau_c)$ therefore defines a time-like entanglement range, or ``time-like entanglement gap'', and the emergence of this gap is the central signature of a Type-II interior. For a Type-II metric of the form~\eqref{eq:generalAdS}, we could express the gap $\tau_c$ as follows
\be\label{eq:critlen}
\begin{aligned}
    \tau_c &= -2t_0=-2\int^{\infty}_{0} \dd z \left(g^{-1}(z)\e^{\chi(z)/2} - \frac{2\e^{\chi(z_h)/2}z_h}{g'(z_h)(z^2-z^2_h)}\right)\ .
\end{aligned}
\ee

\begin{figure}[htbp]
 \begin{center}
   \includegraphics[width=0.6\textwidth]{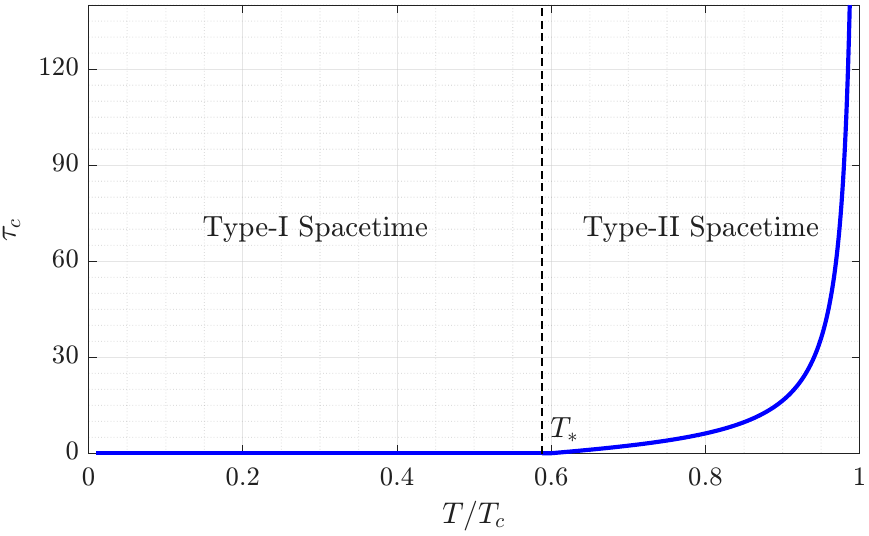}
 \end{center}
\caption{Order parameter $\tau_c$ as the function of  temperature $T/T_c$. $T_*$ is the critical temperature for the phase transition from Type-II to Type-I spacetime.  } \label{fig:taucofT}
\end{figure}

As an example, we study the behavior of TEE in Type-II spacetime based on holographic hairy black hole solutions~\eqref{eq:actionofRN}. In this model, we could numerically express the time-like entanglement gap $\tau_c$~\eqref{eq:critlen} as the function of temperature $T/T_c$, shown in Fig.~\ref{fig:taucofT}.
This reveals a causal phase transition, entirely distinct from the well-known superconducting phase transition, governed by the causal structure of the black hole interior, where the behavior of time-like entanglement entropy is completely different in Type-I and Type-II spacetimes.

Therefore, the critical temporal width $\tau_c$ acts as a bulk diagnostic and potentially a boundary ``order parameter'' for this transition. As shown in Fig.~\ref{fig:taucofT}, it cleanly distinguishes the two phases: it is identically zero in the Type-I phase ($T < T_*$) and becomes non-zero in the Type-II phase ($T > T_*$). The phase transition occurs at the critical temperature (as marked by the vertical dotted line in Fig.~\ref{fig:taucofT}), $T_*$, where this critical scale vanishes $\tau_c(T_*) = 0$.

\begin{figure}[htbp]
\centering
\begin{subfigure}[t]{0.46\textwidth}
\centering
\includegraphics[width=\linewidth]{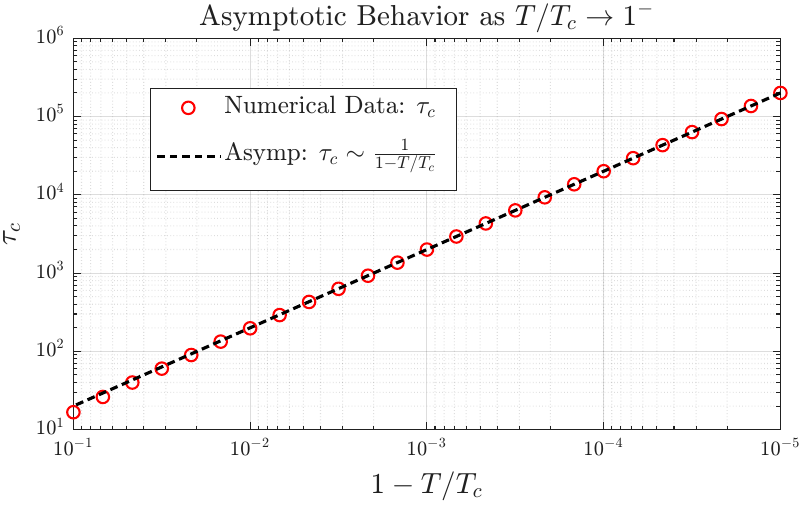}
\caption{  }
\end{subfigure}
\qquad
\begin{subfigure}[t]{0.45\textwidth}
\centering
\includegraphics[width=\linewidth]{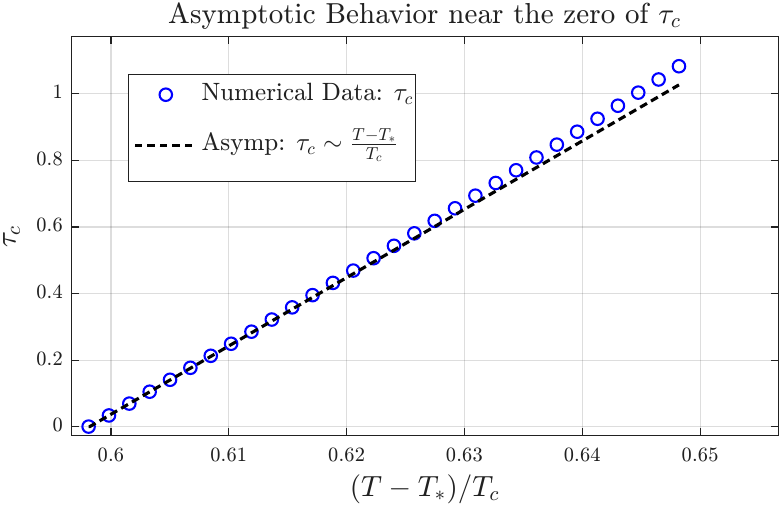}
\caption{  }
\end{subfigure}
 \caption{ (a) When $T/T_c \to 1$,the critical length $\tau_c$ is approximating a reciprocal-type divergence $\tau_c \sim (1-T/T_c)^{-1}$.  (b) Near $T_*$, the behavior of $\tau_c$ follows a linear scaling, which behaves as $\tau_c \sim (T-T_*)/T_c$.}
  \label{fig:twoasym1}
\end{figure}

Notice that it is important not to confuse this $T_*$ with the condensation critical temperature $T_c$ of the holographic superconductor discussed earlier. The size of this diagnostic gap, $\tau_c$, grows as the temperature increases above $T_*$, as quantified in the phase diagram (Fig.~\ref{fig:taucofT}) and analyzed asymptotically in Fig.~\ref{fig:twoasym1}. Fig.~\ref{fig:twoasym1}(a) shows that, as $T/T_c\to 1^{-}$, $\tau_c$ exhibits a reciprocal-type divergence, roughly $\tau_c \sim (1-T/T_c)^{-1}$. Fig.~\ref{fig:twoasym1}(b) shows that near $T_*$ (the zero point $\tau_c(T_*)=0$), the behavior of $\tau_c$ follows a linear scaling $\tau_c\sim (T-T_*)/T_c$.

The phase diagram Fig.~\ref{fig:taucofT} reveals a rich structure. At high temperatures\footnote{In this paper, we only focus on the situation of $T<T_c$, where the singularity becomes space-like and no inner horizon will appear. For the case $T>T_c$, the interior has Cauchy horizon and the singularity becomes time-like, which leads to additional difficulty in finding the correct configuration of TEE. We leave this subtle issue in future.} $T_*<T<T_c$,  the system is in the Type-II phase ($\tau_c\neq 0$). As we decrease the temperature, the system undergoes a phase transition at $T_* \approx 0.58 T_c$, entering the Type-I phase. Deeper in the Type-II phase, as $T$ approaches the superconducting critical temperature $T_c$, the required minimum temporal width $\tau_c$ grows without bound. This signals a dramatic change in the interior geometry and causes the boundary's time-like entanglement range to expand rapidly.

\begin{figure}[htbp]
\centering
\begin{subfigure}[t]{0.46\textwidth}
\centering
\includegraphics[width=\linewidth]{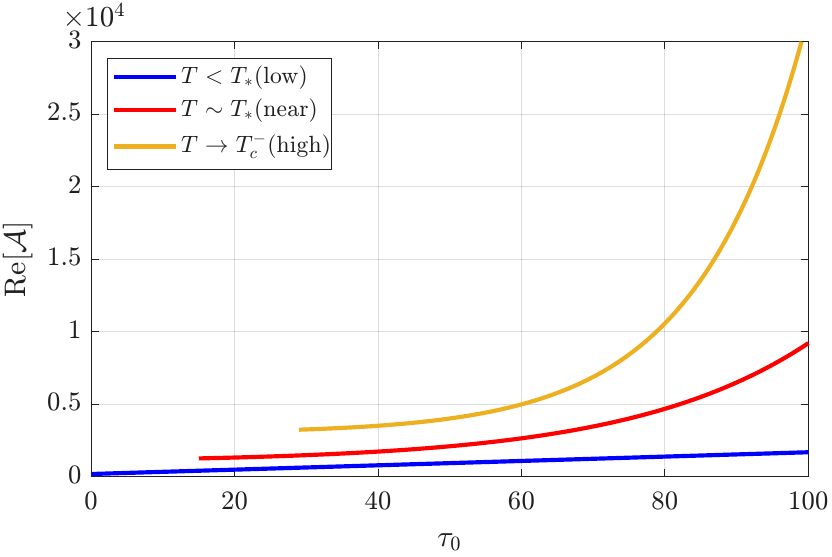}
\caption{}
\end{subfigure}
\qquad
\begin{subfigure}[t]{0.43\textwidth}
\centering
\includegraphics[width=\linewidth]{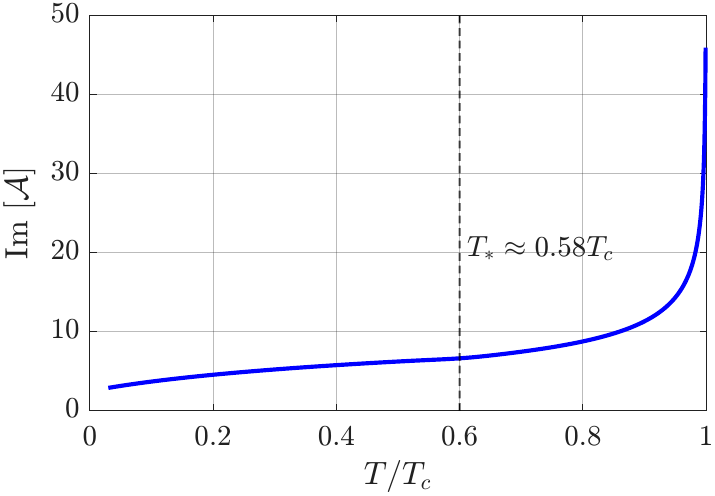}
\caption{}
\end{subfigure}
 \caption{(a) Relationship between Re$\mathcal{A}$ and $\tau_0$ for different temperatures. In the low temperature $T < T_*$ regime,  boundary temporal width $\tau_0$ can start from zero. For temperatures slightly above $T \gtrsim T_*$, a nonzero critical temporal width appears.  For the case that temperature is close to critical temperature $T\rightarrow T^{-}_c$, the real part of TEE grows rapidly and reaches much larger values. (b) When $\tau_0>\tau_c$, the imaginary part of TEE as the function of  temperature $T/T_c$. }
  \label{fig:imagpartviaT}
\end{figure}

This phase transition at $T_*$ leaves sharp, observable signatures in the behavior of the time-like entanglement itself. By applying the numerical framework for computing the TEE's real and imaginary parts, which we established and validated in Sec.~\ref{sec:SAdS}, we can now probe these signatures directly. The most direct signature is seen in the real part of the TEE $\text{Re}\mathcal{A}(\tau_0)$, shown in Fig.~\ref{fig:imagpartviaT}(a):
\begin{itemize}
    \item  For $T > T_*$ (in the Type-II phase, red curve for $T\gtrsim T_*$ and orange curve for $T\to T^-_c$), a clear ``time-like entanglement gap'' $\tau_c$ appears, which means the spacetime is in the ``gapped'' Type-II phase. The real part of TEE, $\text{Re}\mathcal{A}(\tau_0)$, can only begin at the non-zero critical value $\tau_0 = \tau_c$. This entanglement gap is a signature of the Type-II interior. For any boundary width $\tau_0<\tau_c$, the TEE is only given by the imaginary part of extremal surfaces, leading to a `` time-like entanglement'' phase.
    \item  For $T < T_*$ (in the Type-I phase, blue curve for $T<T_*$), because the critical separation is zero $\tau_c=0$, the TEE curve could start from the origin, exhibiting a ``gapless'' behavior similar to the SAdS case. There TEE contains both time-like entanglement and space-like entanglement contributions, i.e., space-like entanglement re-emerges. The imaginary part contributes a constant value independent of $\tau_0$ in this phase, because time-like contribution have ``saturated''.
\end{itemize}

\begin{figure}[htbp]
\centering
\begin{subfigure}[t]{0.43\textwidth}
\centering
\includegraphics[width=\linewidth]{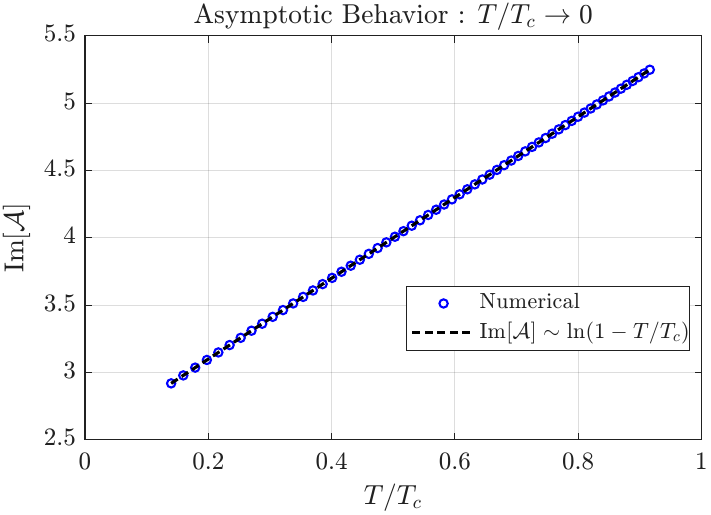}
\caption{  }
\end{subfigure}
\qquad
\begin{subfigure}[t]{0.445\textwidth}
\centering
\includegraphics[width=\linewidth]{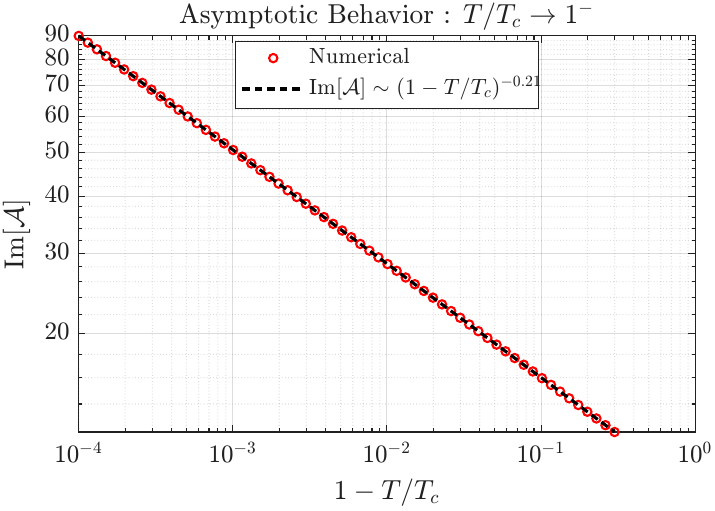}
\caption{  }
\end{subfigure}
 \caption{(a) When $T/T_c \to 0$, imaginary part of TEE $\text{Im}\mathcal{A}$ exhibits a logarithmic dependence $\text{Im}\mathcal{A} \sim \ln(1-T/T_c)$.  (b) When $T/T_c \to 1^-$, the behavior of $\text{Im}\mathcal{A}$  follows a power-law scaling $\text{Im}\mathcal{A} \sim (1-T/T_c)^{-0.21}$.}
  \label{fig:twoasym2}
\end{figure}

Besides the real part of TEE and ``entanglement gap'' $\tau_c$, the imaginary part $\text{Im}\mathcal{A}$ also serves as a probe of the causal transition from Type-I phase to Type-II phase, exhibiting distinct functional dependencies on temperature $T/T_c$ in the different phases, as shown in Fig.~\ref{fig:imagpartviaT}(b).
The imaginary part of the TEE will monotonically decrease with the decreasing temperature after the superconducting phase transition happens. In addition, we sketch its asymptotic behavior in two limits: the low temperature limit and the approach to the condensation onset (the condensation critical temperature).
As illustrated in Fig.~\ref{fig:twoasym2}(a), when low temperature limit $T/T_c \to 0$, the imaginary part of the TEE shows logarithmic asymptotics $\text{Im}\mathcal{A}\sim\ln(1-T/T_c)$ (see Fig.~\ref{fig:twoasym2}(a)).
As the system approaches the condensation critical temperature $T/T_c \to 1^-$, in other words $1-T/T_c \to 0^+$ (shown in Fig.~\ref{fig:twoasym2}(b)), the imaginary part of the TEE exhibits a clear power-law asymptotics, scaling as $\text{Im}\mathcal{A}\sim (1-T/T_c)^{-0.21}$.

\subsubsection{Null limit in Type-II interior}

Note that the real part in Type-II spacetime for $\tau_0<\tau_c$ is obtained via a null-limiting procedure (see Fig.~\ref{fig:Type-II-CWES1}). In this null limit,  we use a family of space-like surfaces which are all ``perpendicular to'' the AdS boundary to approach a null surface. As a result, these surfaces will increasingly resemble null in the region far away from the boundary but their UV behaviors always keep ``perpendicular to'' boundary rather than intersect with the boundary at a $45^\circ$ angle required by an exact null surface.
In that limit the surfaces $A_1A_2$ and $B_1B_2$ will become null far away from the AdS boundary, yet the area still produces a non-zero, and in fact unbounded, contribution due to the space-like contribution near AdS boundary. This behavior is controlled by two scales. One is the UV ``spatial'' cutoff near the AdS boundary.
The other is an emergent effective temporal scale, an UV ``temporal'' cutoff in the time difference $\tau_0/2-v_{A_2}$ between the arrival time on boundary of the space-like extremal surface and the corresponding null value of null surface on boundary; this second scale measures how closely the space-like surfaces approaches the null surfaces. Under the null limit only the divergent spatial and temporal pieces of the space-like extremal surface survive for $\tau_0<\tau_c$.

\begin{figure}[htbp]
\centering
\begin{subfigure}[t]{0.37\textwidth}
\centering
\includegraphics[width=\linewidth]{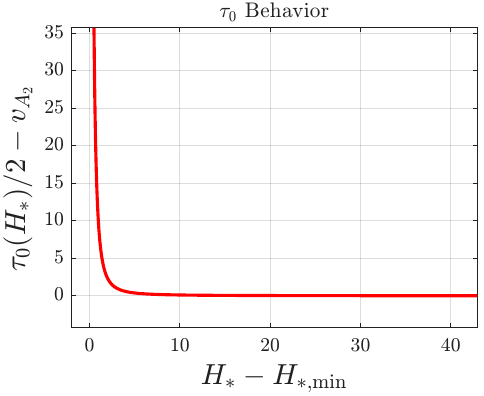}
\caption{  }
\end{subfigure}
\qquad
\begin{subfigure}[t]{0.37\textwidth}
\centering
\includegraphics[width=\linewidth]{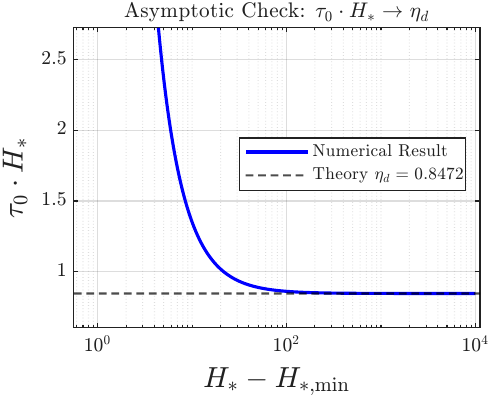}
\caption{  }
\end{subfigure}
\caption{ (a) The null limit corresponds to $H_*\to \infty$ (or $h\to \infty$), (b) giving $\tau_0/2 -v_{A_2}\sim \eta_d h^{-1}$ and driving the space-like arrival time to the null value(here we have set $d=3$ and worked in Type-II spacetime for $\tau_0 <\tau_c$). Here $\eta_d$ is a dimension-dependent constant defined in Appendix.~\ref{appen:null}. }
  \label{fig:taucofLH}
\end{figure}

To make the role of these cutoffs explicit, note that the null limit corresponds to taking the conserved quantity $H_*$ to infinity value $H_*\to \infty$. It is convenient to introduce the auxiliary variable $h^{2d-2}\equiv H_{*}^{4}$, and for $d=3$ we have $h= H_{*}$. The gap between the arrival time on boundary of the space-like extremal surface and the corresponding null value of null surface $\Delta t$ on boundary actually is $\Delta t=\tau_0/2 -v_{A_2}$. In the large-$H_*$(large-$h$) limit, time-gap or time-difference $\Delta t$ is driven to its asymptotic form $\Delta t_{\text{UV}}$. One finds (see Appendix.~\ref{appen:null} for the derivation)
\begin{equation}\label{eq:asymptoticnullform1}
    \Delta t_{\text{UV}}= \tau _0(h\to \infty)/2-v_{A_2} =\eta_d h^{-1}+\cdots\ ,
\end{equation}
where the constants $\eta_d=\frac{1}{\sqrt{\pi}}\Gamma\left(\frac{d}{2d-2}\right)\Gamma\left(\frac{2d-3}{2d-2}\right)$ are defined in Eq.~\eqref{eq:defetad1}, and the omitted terms represent the higher-order terms in $h^{-1}$ that we have neglected. As $h\to \infty$, the right-hand side behaves as $\Delta t_{\text{UV}}=\tau_0(h)/2-v_{A_2}\propto h^{-1}$. Numerically, this $h^{-1}$ power-scaling is shown in Fig.~\ref{fig:taucofLH}. Hence the null limit is physically realized by taking the conserved quantity $H_*$ to infinity value, or $h\to \infty$.

In the limit where the space-like surfaces approach the null configuration ($H_* \to \infty$), the real part of the area functional exhibits characteristic UV divergences. As carefully derived in Appendix.~\ref{appen:null}, to accommodate the broad class of hairy black hole backgrounds considered here, we adopt a generalized asymptotically AdS near-boundary expansion $g(z)\e^{-\chi(z)} \sim 1 - c_1 z^{\alpha_1} + \cdots$. Here, the leading exponent $\alpha_1$ and the coefficient $c_1$ are background-dependent constants encoding specific UV data, such as the boundary energy density and the scalar condensate.

By systematically isolating the divergent structures (see Appendix.~\ref{appen:null} for details and for the definitions of $c_1$ and $\beta _{d,\alpha _1}$), we find that the real part of the area\footnote{Here we assume $d\geqslant 3$. In the case $d=2$, a logarithmic divergency on $h$ will appear. See Eq.~\eqref{eq:d2log} in Appendix.~\ref{appen:null} for details.} can be expressed in terms of both the spatial radial cutoff $\epsilon$ and the temporal scale $\Delta t_{\rm UV}$:
\begin{equation}\label{eq:asymptoticnullform2}
   \mathcal{A} _{A_1A_2}(H_*)=\frac{c_1 \beta _{d,\alpha_1}}{2h^{\alpha_1 -d+2}}+\frac{\epsilon ^{2-d}}{d-2} - \eta_d^{d-1}\frac{\Delta t_{{\rm UV}}^{2-d}}{d-2}+\cdots\ .
\end{equation}

\begin{figure}[htbp]
 \begin{center}
   \includegraphics[width=0.57\textwidth]{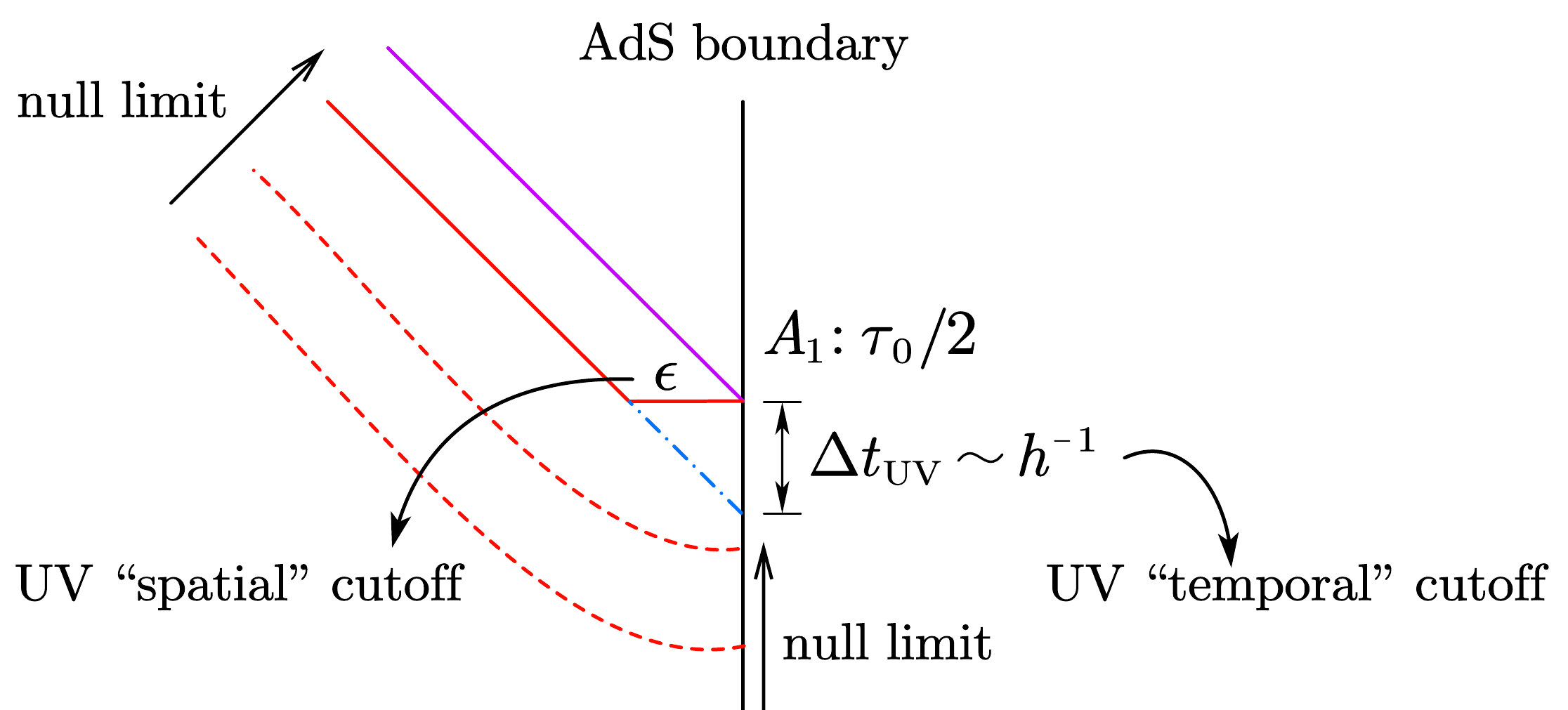}
 \end{center}
\caption{Schematic near-boundary geometry. Red dashed lines show a family of space-like extremal surfaces approaching the null surface (purple solid line). The UV spatial cutoff $\epsilon$ imposes a finite spatial ``resolution'' near the boundary; for spatial scales below $\epsilon$ the red and purple curves cannot be distinguished. As the spacelike extremal surface approaches a null configuration, we find that alongside the standard UV spatial cutoff, an additional regulator necessarily emerges. This new cutoff $\Delta t_{\text{UV}} \sim h^{-1}$ quantifies the degree of approach to the null limit or, equivalently, how closely the boundary value of the space-like surfaces (blue dashed line) approximate that of the null surface boundary. Cutoff $\Delta t_{\text{UV}}$ sets a UV ``temporal'' resolution: time differences smaller than $\mathcal{O}(\eta_d h^{-1})$ cannot be resolved.}\label{fig:nulllimit}
\end{figure}
Particularly, we find that there are two kinds of divergences $\epsilon$ and $\Delta t_{\text{UV}}$ in Eq.~\eqref{eq:asymptoticnullform2}, as depicted in Fig~\ref{fig:nulllimit}. Here $\epsilon$ is the UV ``spatial'' cutoff near the AdS boundary, and $\Delta t_{\text{UV}}$ corresponds to null limit where the space-like extremal surface approaches to null (i.e., $h\rightarrow\infty$).
The term $\epsilon ^{2-d}$ in Eq.~\eqref{eq:asymptoticnullform2} is the familiar UV contribution from the boundary region. The new term, proportional to $(\Delta t_{\text{UV}})^{2-d}$ (or equivalently $h^{d-2}$), represents a UV temporal divergence in the null limit. These UV terms $\epsilon$ and $\Delta t_{\text{UV}}$ yield a nonzero contribution even when the space-like piece becomes almost null.

The reason is, near the AdS boundary the space-like extremal surface remains nearly orthogonal to the boundary, whereas a true null surface meets the boundary at a $45^\circ$ angle. The nearly vertical piece therefore always produces a UV contribution to the area that does not vanish in the null limit, as shown in Fig.~\ref{fig:nulllimit}. As we have known, the spatial UV cut-off $\epsilon$ can be interpreted as the minimal resolution in spatial scale. We can make a similar interpretation for the emergent scale $\Delta t_{\text{UV}}$.  From the large-$H_*$ (or large-$h$) analysis in Eq.~\eqref{eq:asymptoticnullform1}, we see the finite cut-off for $\Delta t_{\text{UV}}$ means we set a minimal value for $\tau_0/2-v_{A_2}$. Thus, this is equivalent to an emergent UV ``temporal'' resolution, as shown in Fig.~\ref{fig:nulllimit}. On time scales smaller than this scale $\Delta t_{\text{UV}}$ one cannot resolve the difference between the arrival time of the space-like extremal surface (red solid curve) and that of the null surface (purple solid curve). Thus both regulators, the UV ``spatial'' cutoff $\epsilon$ near the boundary and the emergent effective temporal scale $\Delta t_{\text{UV}}$ controlled by $h$, produce residual, non-vanishing contributions to $\text{Re}\mathcal{A}$ in the Type-II spacetimes with $\tau_0<\tau_c$.

It is crucial to clarify the physical nature of $\Delta t_\text{UV}$ as a ``temporal resolution scale''. Unlike the standard spatial radial cutoff $\epsilon$, which is introduced artificially by hand to regularize the boundary CFT volume, $\Delta t_\text{UV}$ is not an externally imposed regulator. Rather, it is an emergent effective scale intrinsically associated with the null-limit family of extremal surfaces (i.e., as $H_* \to \infty$). Geometrically, a space-like extremal surface anchored at the boundary cannot coincide with a globally null hypersurface; a residual temporal gap naturally emerges, setting an intrinsic resolution limit for how closely a space-like probe can mimic a true null surface in the bulk.

To systematically extract the finite, physically meaningful TEE, we must carefully subtract these divergences. As established in the holographic renormalization framework for hairy black holes~\cite{Taylor:2016aoi}, because our holographic setup focuses on spontaneous condensation (vanishing scalar source $\phi_{(0)}=0$) on a planar symmetry black hole, all scalar- and curvature-induced counterterms identically vanish.
The divergence structure of our hairy black hole is therefore safely identical to that of pure AdS spacetime. Consequently, we define the renormalized physical area by explicitly subtracting the spatial divergent piece $\text{Re}\mathcal{A}_0(\epsilon) = \frac{\epsilon^{2-d}}{d-2}$ and the temporal divergent piece $\text{Re}\mathcal{A}_\text{temporal}(\Delta t_{\text{UV}}) = - \eta_d^{d-1}\frac{\Delta t_{\text{UV}}^{2-d}}{d-2}$, and subsequently taking both regulators to the continuum limit:
\begin{equation}\label{eq:renormalization_scheme}
    \mathcal{A}_{\text{ren}} = \lim_{\Delta t_{\text{UV}} \to 0} \lim_{\epsilon \to 0} \Big[ \text{Re}\mathcal{A}_{\text{BH}}(\epsilon, \Delta t_{\text{UV}}) - \text{Re}\mathcal{A}_{0}(\epsilon) - \text{Re}\mathcal{A}_\text{temporal}(\Delta t_{\text{UV}}) \Big]\ .
\end{equation}
In this explicit subtraction scheme, both the spatial cutoff $\epsilon$ and the emergent scale $\Delta t_{\text{UV}}$ are ultimately removed. Since divergencies are universal and independent of the details of dynamics of matters, the finite $\mathcal{A}_{\text{ren}}$ will also carry all the physical information beyond vacuum case.

\subsubsection{Cauchy Horizon and Pure Time-like Entanglement}

The double limit $T \to T^-_c$ and $\tau_c\to \infty$ show that the system exhibits the extreme limit of the Type-II phase. The divergence of $\tau_c$, shown in our phase diagram (Fig.~\ref{fig:taucofT}), together with the rapid growth of the real part of TEE (orange line in Fig.~\ref{fig:imagpartviaT}(a)), results from the fact that the space-like extremal surfaces $A_1A_2$ and $B_1B_2$ approach the null singularity when $T\rightarrow T_c^-$ and $\tau_0\rightarrow\infty$.

\begin{figure}[htbp]
 \begin{center}
   \includegraphics[width=0.8\textwidth]{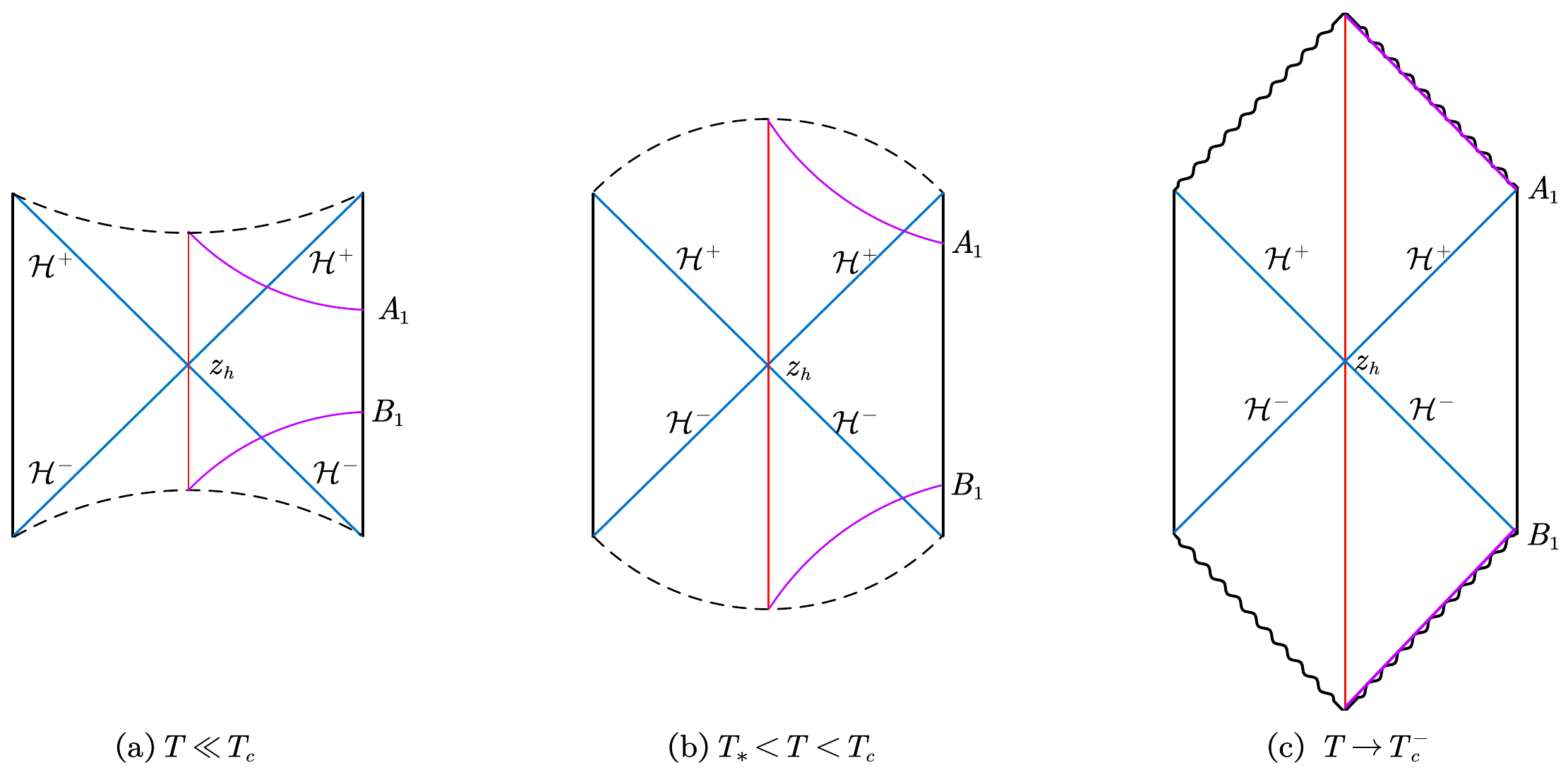}
 \end{center}
\caption{As the temperature increases to the critical condensation temperature $T/T_c\to 1$, the space-like singularity converges to the null singularity, with the interior geometry transforming from Type-I to Type-II.}\label{fig:Cauchy}
\end{figure}

Physically, this reflects a drastic change of the interior geometry: as the temperature increases to $T_c$, the space-like singularity disappears, the Cauchy horizon and time-like singularity will recover. We can illustrate this process with a schematic diagram. Fig.~\ref{fig:Cauchy}(a) shows the original space-like singularity of a hairy black hole.
As the temperature increases, the time $t_0$~\eqref{eq:reachessingularity3} that fall from the boundary to the singularity shifts from positive to negative, causing the space-like singularity to ``bulge'' as seen in Fig.~\ref{fig:Cauchy}(b).
At the critical condensation temperature $T\to T^-_c$, as the condensation becomes infinitesimally small, the spacetime singularity converges to the null singularity shown in Fig.~\ref{fig:Cauchy}(c).
At this point the boundary's time-like entanglement window grows unbounded (see Fig.~\ref{fig:taucofT}): for any finite boundary time spacing $\tau_0$, the TEE is dominated by the time-like contribution (up to the aforementioned regulator effect) because the entanglement gap becomes effectively infinite.
The space-like singularity and the Cauchy horizon merge into a null singularity.
This again highlights the pathological nature of the Cauchy horizon and null singularities, as discussed in the context of the cosmic censorship hypothesis. We therefore conjecture: the presence of an inner Cauchy horizon is associated with a nonzero entanglement gap $\tau_c$ in the boundary theory. Further, an infinitesimal perturbation that turns the Cauchy horizon into a null singularity drives $\tau_c\to \infty$, so that TEE is entering a ``pure time-like entanglement phase'', up to a regulator effect.

The nature of the zero-temperature ground state is, however, of key theoretical importance. It is crucial to note that the infrared (IR) geometry of this ground state in holographic superconductors is not universal and depends sensitively on the model parameters $m^2$ and $q$, see Ref.~\cite{Horowitz:2009ij}.
Here we present results for the holographic superconductor model with parameters $m^2=-2,q=1$. A full analysis of how the TEE behaves in the qualitatively different ground states that arise in other parameter regimes, such as those with emergent conformal (AdS), Lifshitz geometry or Poincar\'e symmetry discussed in Ref.~\cite{Horowitz:2009ij}, is beyond the scope of our current analysis, but presents an intriguing question.

In conclusion, there is causal phase transition between the type II and Type I spacetimes after the superconducting phase transition happens. Our analysis in this section has uncovered the causal phase transition at a critical temperature $T_*$ within the black hole interior, diagnosed entirely by the time-like entanglement entropy.
Within the CWES geometric framework, the behavior of this time-like quantity serves as a remarkably sharp, single-boundary diagnostic of the hidden interior geometry. It therefore provides a compelling holographic bridge, directly linking a boundary quantum informational measure to a fundamental change in the bulk's causal structure.

\section{Time-like Entanglement First Law for the Real Part}\label{sec:TEEfirstlaw}

The first law of entanglement provides a universal relationship between quantum information and energy~\cite{Bhattacharya:2012mi, Blanco:2013joa}. For a small spatial subsystem, a low-energy excitation of the vacuum satisfies a thermodynamics-like relation, $T_\text{ent} \Delta S = \Delta E$, where $\Delta S$ and $\Delta E$ are the variations of entanglement entropy and energy respectively, and $T_\text{ent}$ is the effective entanglement temperature. Extending this framework to time-like subregions of the boundary CFT poses unique challenges.
 
Motivated by the profound connection between entanglement and gravity, recent works, e.g., Refs.~\cite{Li:2025tud, Xiao:2026nvu}, established a time-like analogue of the entanglement first law via double Wick rotations for hyperbolic subregions, demonstrating equivalence with linearized Einstein equations and even generalizing to arbitrary-order Lovelock gravity in the bulk. However, they also revealed that direct Lorentzian calculations for hyperbolic boundaries using piece-wise smooth extremal surfaces encounter severe subtleties and apparent inconsistencies.

Here, we adopt an alternative perspective by working directly in Lorentzian signature with piece-wise smooth extremal surfaces~\cite{Doi:2022iyj, Li:2022tsv, Doi:2023zaf, Bohra:2025mhb}. In this section, we demonstrate that for the time-like \textit{strip} subsystem ($|t| < \tau_0/2$) defined in Eq.~\eqref{eq:time-like strip}, the CWES proposal~\cite{Li:2022tsv} operating directly in Lorentzian signature yields a consistent time-like entanglement first-law-like relation for its real part. We establish the holographic dictionary between the variation of the real part of TEE and the boundary energy. In the low-temperature perturbative regime, we confirm the relation $T_\text{ent}^{(\text{Re})} \Delta \text{Re}\,\mathcal{S}_{\mathcal{T}} = \Delta \mathcal{E}_{\mathcal{T}}$ and derive an effective entanglement temperature inversely proportional to the temporal width, $T_\text{ent}^{(\text{Re})} \propto \tau_0^{-1}$. Furthermore, we explore the late-time asymptotic limit (large temporal width) to analyze the thermodynamic behavior of this effective entanglement temperature. Alongside this, we highlight the special role of the imaginary part, which, rather than obeying a linear perturbative first law, characterizes the non-analytic change in the global causal structure from the AdS vacuum to a black hole interior. 

Because the CWES area is complex, it is essential to distinguish the variation of its real part from the behavior of the full complex quantity. Throughout this section, we study only the real-part variation $\Delta\text{Re}\,\mathcal{A}$ between the vacuum and a weakly excited thermal state at fixed strip width $\tau_0$. While our analysis demonstrates a robust first-law-like relation for this real component, the formulation of a generalized first law for the full complex TEE in Lorentzian spacetime, potentially involving a modular Hamiltonian as discussed in Refs.~\cite{Li:2025tud, Xiao:2026nvu}, remains a compelling open question.

\subsection{Perturbative Verification of the First Law in Time-like Strip Case}

We begin with the metric ansatz
\begin{equation}
    \dd s^2 = \frac{1}{z^2}\left(-g(z)\e^{-\chi(z)}\dd t^2+g^{-1}(z)\dd z^2+ \dd x^2+\dd \boldsymbol{y}_{d-2}^{2}\right)\ .
\end{equation}
For a Schwarzschild-AdS (SAdS) spacetime in $d+1$ dimensions, the blackening factors are $g(z)=1-(z/z_h)^d$ and $\e^{-\chi(z)}=1$ (setting $\ell_\AdS =1$).
As shown in Sec.~\ref{sec:SAdS}, the real part of the TEE for a boundary time-like strip of width $\tau_0$ is governed by the conserved quantity $H_*$:
\begin{align}
    \tau_0(H_*) &= 2\int_0^\infty \frac{\dd z}{X \left(X + H_{*}^{2}z^{d-1}\right)}\ , \\
    \text{Re}\mathcal{A}(H_*) &= 2\int_0^\infty \dd z\,\frac{z^{1-d}}{X}\ ,
\end{align}
where we have introduced $X(H_*) = \sqrt{H_{*}^{4}z^{2d-2}+g(z)}$.

The imaginary part of the TEE originates from the segment of the time-like extremal surface completely localized inside the black hole horizon ($z_h < z < \infty$):
\begin{equation}
    \text{Im} \mathcal{A} = 2\int_{z_h}^{\infty}{\dd z\,\frac{z^{1-d}}{\sqrt{-g(z)}}} = 2\frac{(4\pi)^{d-2}}{d^{d-1}}B\left( \frac{3}{2}-\frac{2}{d},\frac{1}{2} \right)\times T_{\sH}^{d-2}\ .
\end{equation}
Here, the Hawking temperature is given by $T_{\sH} = -\frac{g'(z_h)}{4\pi} = \frac{d}{4\pi z_h}$.

We next clarify the non-analytic behavior of the imaginary part before turning to the perturbative real-part correction.
For $d>2$, the imaginary part vanishes in the strict zero-temperature limit $T_{\sH}\to 0$. However, this does not imply that it can be omitted from the variation between the vacuum and a weakly excited thermal state. Using the perturbative parameter $m = z_h^{-d} = (4\pi T_{\sH}/d)^d$, the thermal variation of the imaginary part is given by:
\begin{equation}
    \Delta\text{Im}\,\mathcal{A} = \frac{2}{d} B\left(\frac{3}{2}-\frac{2}{d},\frac{1}{2}\right) m^{\frac{d-2}{d}}\ .
\end{equation}
Taking the derivative with respect to the perturbation parameter yields $\frac{\partial}{\partial m} \Delta\text{Im}\,\mathcal{A} \propto m^{-2/d}$. Since $d>2$, this derivative diverges as $m \to 0^+$. This divergent behavior is indeed inherent to the imaginary part of TEE and physically explains why the imaginary part does not obey a simple linear first-law-like variation. Physically, while the variation of the real part $\Delta\text{Re}\,\mathcal{A}$ captures the local geometric response to small energy excitations, the emergence of the imaginary part reflects a non-analytic change in the singular structure of the spacetime.  Specifically, for $d > 2$, the excitation introduces a physical curvature singularity and an event horizon in the deep bulk interior. The transition from a smooth AdS vacuum to a spacetime containing a physical curvature singularity is topologically and geometrically non-continuous. Since the CWES is known to probe the deep interior of the black hole, if a physical observable computed from it is sensitive to this sudden appearance of a singularity, it is highly likely that its behavior  need not admit a smooth analytic expansion at the moment this singularity is born. This non-analytic onset associated with the qualitative change from ``no singularity'' to ``having a singularity'' naturally breaks any smooth, perturbative expansion near the vacuum limit ($m \to 0^+$), explaining why the imaginary part violates linear perturbation theory. 

Interestingly, this divergence of the imaginary part disappears in the case of the BTZ black hole ($d = 2$). The reason is that the BTZ black hole is locally AdS$_3$ and its local bulk geometry remains smooth and free of curvature singularities. Consequently, the TEE does not encounter any local geometric singularity, rendering the transition from AdS$_3$ to the BTZ black hole smooth and leaving the imaginary part free of any non-perturbative divergence in three dimensions.

Therefore, to analytically compute the thermal corrections to the real part, we introduce the perturbative parameter $m$:
\begin{equation}
    g(z) = 1-m z^{d},\qquad m \ll 1\ .
\end{equation} 
In the SAdS case, $m= z_h^{-d}=\left( 4\pi T_{\sH}/d\right)^d$.

In the pure AdS background (corresponding to vacuum CFT limit $m\to 0$), $X$ reduces to $X_0(z)=\sqrt{1+\left(H_{*}^{\left( 0 \right)}\right)^4 z^{2d-2}}$. Introducing the dimensionless variable $u=\left(H_{*}^{\left( 0 \right)}\right)^{\frac{2}{d-1}}z$, we have
\begin{equation}
    \tau_0^{(0)}(H_{*}^{\left( 0 \right)}) = 2\left(H_{*}^{\left( 0 \right)}\right)^{-\frac{2}{d-1}}\int_0^\infty \dd u\left(1-\frac{u^{d-1}}{\sqrt{1+u^{2d-2}}}\right)\ .
\end{equation}
Defining the numerical constant $K_d$:
\begin{equation}
    \begin{aligned}
    K_d &\equiv \int_0^\infty \dd u \left(1-\frac{u^{d-1}}{\sqrt{1+u^{2d-2}}}\right) \\
    &= \frac{2^{-\frac{1}{d-1}}}{d-1}B\!\left(\frac{2d-3}{2d-2},\frac{1}{d-1}\right)\ ,
    \end{aligned}
\end{equation}
we obtain $\tau_0^{(0)}=2K_d\,\left(H_{*}^{\left( 0 \right)}\right)^{-\frac{2}{d-1}}$, which implies the scaling relation $H_{*}^{\left( 0 \right)} \propto \tau_0^{-\frac{d-1}{2}}$.

Similarly, the area integral in pure AdS is:
\begin{equation}
    \text{Re}\mathcal{A}^{(0)} = 2\int_0^\infty \dd z\, \frac{z^{1-d}}{\sqrt{1+\left(H_{*}^{\left( 0 \right)}\right)^4 z^{2d-2}}}\ .
\end{equation}
Using the same rescaling $u=\left(H_{*}^{\left( 0 \right)}\right)^{\frac{2}{d-1}}z$, and separating the UV divergence (as discussed in Eq.~\eqref{eq:subtractvacuum}), we get:
\begin{equation}
    \text{Re}\mathcal{A}^{(0)} = \frac{2}{d-2}\,\epsilon^{2-d} - 2c_d\,\left(H_{*}^{\left( 0 \right)}\right)^{\frac{2(d-2)}{d-1}}\ ,
\end{equation}
where the finite integration constant is defined as:
\begin{equation}
    \begin{aligned}
    c_d &\equiv \int_0^\infty \dd u \left( \frac{1}{u^{d-1}} - \frac{1}{u^{d-1}\sqrt{1+u^{2d-2}}} \right)\\
    &= \frac{1}{(d-2)\sqrt{\pi}} \Gamma\!\left(\frac{d}{2d-2}\right) \Gamma\!\left(\frac{2d-3}{2d-2}\right)\ .
    \end{aligned}
\end{equation}
Expressing this in terms of the temporal width $\tau_0$, the pure AdS real part reads:
\begin{equation}
    \text{Re}\mathcal{A}^{(0)}(\tau_0) = \frac{2}{d-2}\,\epsilon^{2-d} - \frac{A_0}{\tau_0^{d-2}}\ ,
\end{equation}
with the constant $A_0=2c_d(2K_d)^{d-2}$.

Now we evaluate the variation of the area $\text{Re}\mathcal{A}(m,H_*)$ (and temporal width $\tau_0(m,H_*)$) when the small thermal perturbation $m$ is turned on. Because the boundary temporal width $\tau_0$ defines the subsystem and must be held fixed, the total variation is composed of two parts:
\begin{equation}
    \begin{aligned}
        \Delta \text{Re}\mathcal{A}(m,H_*) &=\left( \frac{\partial \text{Re}\mathcal{A}}{\partial m} \right) _{H_*=H_{*}^{\left( 0 \right)}}m+\left( \frac{\partial \text{Re}\mathcal{A}}{\partial H_*} \right) _{m=0}\delta H_*+\mathcal{O} \left( m^2 \right) \\
        &=\left. \Delta \text{Re}\mathcal{A} \right|_{H_*=H_{*}^{\left( 0 \right)}}+\left. \Delta \text{Re}\mathcal{A} \right|_{m=0}\ .
    \end{aligned}
\end{equation}

For the first term, when a small temperature is introduced ($m \neq 0$), the integrand expands as $\frac{1}{X} = \frac{1}{X_0} + \frac{m z^d}{2X_0^3} + \mathcal{O}(m^2)$. Applying the rescaling $u=H_*^{\frac{2}{d-1}}z$, the explicit $\mathcal{O}(m)$ correction to the area integral is:
\begin{equation}
    \left. \Delta \text{Re}\mathcal{A} \right|_{H_*=H_{*}^{\left( 0 \right)}} = m\,H_*^{-\frac{4}{d-1}}\,\mathcal{I}_d + \mathcal{O}(m^2)\ ,
\end{equation}
where the coefficient $\mathcal{I}_d$ evaluates to:
\begin{equation}
    \begin{aligned}
    \mathcal{I}_d &= \int_0^\infty \dd u\,\frac{u}{\left(1+u^{2d-2}\right)^{3/2}}\\\
    &= \frac{1}{2d-2} B\!\left(\frac{1}{d-1},\frac{3}{2}-\frac{1}{d-1}\right)\ .
    \end{aligned}
\end{equation}

To compute the second term $\left. \Delta \text{Re}\mathcal{A} \right|_{m=0}$, we must account for the shift in the conserved quantity $H_* = H_{*}^{\left( 0 \right)} + \delta H_*$ required to keep $\tau_0$ fixed. Defining a dimensionless combination $\mu\equiv mH_*^{-2d/(d-1)}$, the temporal width expands as:\begin{equation}
    \begin{aligned}
        \tau_0(m,H_*) &= 2H_*^{-2/(d-1)}\int_0^{\infty}\dd u\,\frac{1}{X(X+u^{d-1})}\\
        &= H_*^{-\frac{2}{d-1}}\left( \mathcal{C}_0 + \mu\,\mathcal{C}_1 + \mathcal{O}(\mu^2) \right)\ ,
    \end{aligned}
\end{equation}
where $X = \sqrt{1+u^{2d-2}-\mu u^d}$. The zeroth-order term is $\mathcal{C}_0 = 2K_d$, and evaluating the derivative with respect to $\mu$ at $\mu=0$ gives:
\begin{equation}
    \mathcal{C}_1 = \int_0^\infty \dd u\, \frac{u^d(2X_0+u^{d-1})}{X_0^3(X_0+u^{d-1})^2}\ ,
\end{equation}
with $X_0 = \sqrt{1+u^{2d-2}}$. Imposing the constraint $\delta\tau_0 = 0$, the necessary geometric shift is:
\begin{equation}
    \delta H_* = \frac{d-1}{4K_d}\,\mathcal{C}_1\,m\,\left(H_{*}^{\left( 0 \right)}\right)^{-\frac{d+1}{d-1}}\ .
\end{equation}
This shift feeds back into the area variation, yielding the implicit correction:
\begin{equation}
    \begin{aligned}
       \left. \Delta \text{Re}\mathcal{A} \right|_{m=0} &= \left.\frac{\partial}{\partial H_*}\left(\text{Re}\mathcal{A}^{(0)}\right)\right|_{m=0}\delta H_* \\
       &= -c_d\,\frac{d-2}{K_d}\,\mathcal{C}_1\, m\, \left(H_{*}^{\left( 0 \right)}\right)^{-\frac{4}{d-1}}\ .
    \end{aligned}
\end{equation}

Combining the two contributions, the total thermal correction to the real part of the TEE is:
\begin{equation}
     \Delta \text{Re}\mathcal{A} = m\,\left(H_*^{(0)}\right)^{-\frac{4}{d-1}} \left[ \mathcal{I}_d - c_d\,\frac{d-2}{K_d}\,\mathcal{C}_1 \right] + \mathcal{O}(m^2)\ .
\end{equation}
Substituting $H_*^{(0)}\propto \tau_0^{-\frac{d-1}{2}}$, we observe that the thermal correction scales exactly as $\Delta\text{Re}\mathcal{A} \sim m\,\tau_0^2 \sim T_{\sH}^d\,\tau_0^2$.
Defining the constant $\Xi_d \equiv \frac{1}{4K_d^2} \left[ \mathcal{I}_d - \frac{c_d(d-2)}{K_d}\,\mathcal{C}_1 \right]$, the complete expression for the real part of the TEE in the low-energy limit $T_{\sH} \ll 1$ becomes:
\begin{equation}
    \text{Re}\mathcal{A}(\tau_0) = \frac{2}{d-2}\,\epsilon^{2-d} - \frac{A_0}{\tau_0^{d-2}} + m\,\tau_0^2\,\Xi_d + \mathcal{O}(m^2)\ .
\end{equation}

To relate this variation to the boundary energy, we utilize the standard Fefferman-Graham (FG) expansion~\cite{Fefferman:2007rka}
\begin{equation}
    \begin{aligned}
       \dd s^2 &= \frac{1}{\xi^2} \left( \dd\xi^2 + g_{\mu\nu}(x, \xi) \dd x^\mu \dd x^\nu \right)\ ,\\
        g_{\mu\nu}(x, \xi) &= g^{(0)}_{\mu\nu}(x) + \xi^2\, g^{(2)}_{\mu\nu}(x) + \cdots + \xi^d\, g^{(d)}_{\mu\nu}(x) + \cdots\ .
    \end{aligned}
\end{equation}
Near the boundary $\xi \to 0$, our perturbative metric takes the form:
\begin{equation}
        \dd s^2 = \frac{1}{\xi^2}\left[-\left(1 + \frac{1-d}{d}\cdot\frac{\xi^d}{z_h^d}\right)\dd t^2 + \dd\xi^2 + \left(1 + \frac{\xi^d}{d\,z_h^d}\right)\left(\dd x^2+\dd \boldsymbol{y}_{d-2}^{2}\right)\right]\ .
\end{equation}
The boundary stress tensor is given by $\langle T_{\mu\nu} \rangle = \frac{d}{16\pi G_\text{N}} \, g^{(d)}_{\mu\nu}$, and its $tt$-component reads:
\[
    \langle T_{tt} \rangle = \frac{d}{16\pi G_\text{N}} \cdot g^{(d)}_{tt} = \frac{d-1}{16\pi G_\text{N} z_h^d}\ .
\]
Thus, the increased amount of energy density $\Delta\mathcal{E}_\mathcal{T}=\Delta E_\mathcal{T}/\mathcal{V}_{d-2}$ for the time-like subsystem $\mathcal{T}$ with temporal width $\tau_0$ is~\cite{Li:2025tud}:
\begin{equation}
    \begin{aligned}
        \Delta\mathcal{E}_\mathcal{T}&= \int_{\mathcal{T}} \dd t \,\langle T_{tt} \rangle = \frac{d-1}{16\pi G_\text{N} z_h^d} \tau_0\ .
    \end{aligned}
\end{equation} 

Correspondingly, the change in the real part of the TEE density $\Delta\text{Re}\,\mathcal{S}_\mathcal{T} = \Delta\text{Re}\,\mathcal{A}/(4G_\text{N})$ is given by
\begin{equation}
   \Delta\text{Re}\,\mathcal{S}_\mathcal{T} = \frac{\Delta \text{Re}\mathcal{A}}{4G_\text{N}} = \frac{z_h^{-d}\tau_0^2 \Xi_d}{4G_\text{N}}\ .
\end{equation}
Therefore, the effective entanglement temperature satisfies the expected universal inverse-size relationship:
\begin{equation}\label{eq:T_ent}
    \begin{aligned}
        T_\text{ent}^{(\text{Re})} &= \frac{\Delta \mathcal{E}_\mathcal{T}}{\Delta \text{Re}\,\mathcal{S}_\mathcal{T}} \\
        &= \frac{d-1}{4\pi \,\Xi _d}\cdot\tau _{0}^{-1} \propto \tau _{0}^{-1}\ .
    \end{aligned}
\end{equation}
This relation allows us to define a time-like entanglement first-law-like relation for the real part, $T_\text{ent}^{(\text{Re})} \Delta \text{Re}\,\mathcal{S}_{\mathcal{T}} = \Delta \mathcal{E}_{\mathcal{T}}$ with $T_\text{ent}^{(\text{Re})}$ as the effective entanglement temperature. This derivation confirms that a first-law-like relation holds for the real part of strip subregions within the CWES framework. The validity of a generalized time-like entanglement first law~\cite{Li:2025tud, Xiao:2026nvu} for the full complex TEE of time-like strip system in Lorentzian spacetimes remains an open question.
 
\subsection{Late-Time Asymptotics and Effective Temperature}

The entanglement temperature defined in Eq.~\eqref{eq:T_ent} loses its universality in the large-size limit ($\tau_0 \to \infty$) of time-like subregions. In this regime, the entanglement properties become sensitive to the deep infrared (IR) geometry details of the excitations.

Consider an AdS black hole at Hawking temperature $T_\sH$. In the late-time limit $\tau_0 \to \infty$, $T_\text{ent}$ approaches a constant multiple of $T_\sH$. As shown in Sec.~\ref{subsec:asymptau}, the real part of the TEE exhibits a universal linear growth at late times:
\begin{equation}
    \begin{aligned}
        \text{Re}\mathcal{A}_{A_1A_2B_2B_1}(\tau _0) &\approx H_{*,\min}^{2}\tau_0+\text{finite part}\ , \quad \text{as} \quad \tau_0 \to \infty\ , \\
        \lim_{\tau_0 \to \infty}\frac{\dd \mathcal{A}_{A_1 A_2B_2 B_1}(\tau _0)}{\dd \tau_0} &= H_{*,\min}^{2}= \sqrt{G(z_c)}\ .
    \end{aligned}
\end{equation}
Here, the function $G(z) \equiv -z^{2-2d}g(z)$ dictates the internal geometry, and the critical location $z_c$ is implicitly determined by the geometric extremum condition $G'(z_c)=0$. For the planar SAdS black hole with $g(z)=1-(z/z_h)^d$, solving this equation yields the critical point $z_c=z_h\left(\frac{2(d-1)}{d-2}\right)^{1/d}$ for $d>2$. Evaluating the geometric minimum yields:
\begin{equation}
    H_{*,\min}^2 = \left(\frac{4\pi T_\sH}{d}\right)^{d-1}\sqrt{\frac{d}{d-2}}\left(\frac{2(d-1)}{d-2}\right)^{\frac{1-d}{d}} = \nu_d\,T_\sH^{\,d-1}\ ,
\end{equation}
where $\nu_d=\left(\frac{4\pi}{d}\right)^{d-1}\sqrt{\frac{d}{d-2}}\left(\frac{2(d-1)}{d-2}\right)^{\frac{1-d}{d}}$.

For $d=2$ (the planar BTZ case), the late-time behavior must be treated carefully. With $g(z)=1-(z/z_h)^2$, the function $G(z) = \frac{1}{z_h^2}-\frac{1}{z^2}$ is monotonically increasing and reaches its supremum only asymptotically as $z\to\infty$.
Consequently, there is no finite interior critical point $z_c$. Instead, $G'(z_c)=0$ is replaced by the asymptotic saturation condition $H_{*,\min}^2=\lim_{z\to\infty} \sqrt{G(z)}= z^{-1}_h=2\pi T_\sH$, where the BTZ Hawking temperature is $T_\sH=\frac{1}{2\pi z_h}$.

We can define the late-time entanglement temperature $T_\text{ent}^\text{late}$ via the geometric growth rate ratio:
\begin{equation}
    T_\text{ent}^\text{late}\equiv \frac{\Delta\mathcal{E}_{\mathcal T}}{\Delta\mathcal{S}_{\mathcal T}^\text{late}}\ .
\end{equation}
For $d>2$, this evaluates to a constant multiple of the Hawking temperature:
\begin{equation}
        T_\text{ent}^\text{late} =\frac{d-1}{4\pi}\,\frac{z_h^{-d}}{H_{*,\min}^{2}} =\kappa_d T_\sH\ ,
\end{equation}
where $\kappa_d=\frac{d-1}{d}\sqrt{\frac{d-2}{d}}\left(\frac{2(d-1)}{d-2}\right)^{\frac{d-1}{d}}$.
For $d=2$, this geometric ratio yields exactly $T_\text{ent}^\text{late}=\frac{1}{4\pi z_h}=\frac{T_\sH}{2}$, which is same as the result obtained in Ref.~\cite{Li:2025tud}.  

Following Ref.~\cite{Li:2025tud}, we can employ a thermodynamic definition for the late-time entanglement temperature $\widetilde{T}_\text{ent}^\text{late}$ by taking the derivative with respect to the background state (temperature) while keeping the subsystem size $\tau_0$ fixed:
\begin{equation}
    \widetilde{T}_\text{ent}^\text{late} \equiv \left.\frac{\partial \Delta\mathcal{E}_\mathcal{T} / \partial T_\sH}{\partial \Delta\mathcal{S}_\mathcal{T} / \partial T_\sH} \right|_{\tau_0}\ .
\end{equation}
In $d=2$ BTZ case, where $\Delta\mathcal{S}_\mathcal{T} = \frac{\pi}{2G_\text{N}} T_\sH \tau_0$ and $\Delta\mathcal{E}_\mathcal{T} = \frac{\pi}{4G_\text{N}} T_\sH^2 \tau_0$, this thermodynamic derivative exactly recovers the Hawking temperature~\cite{Li:2025tud}:
\begin{equation}
    \widetilde{T}_\text{ent}^\text{late} = \frac{\frac{\pi}{2G_\text{N}} T_\sH \tau_0}{\frac{\pi}{2G_\text{N}} \tau_0} = T_\sH\ .
\end{equation}
For $d>2$, applying this thermal variation yields:
\begin{equation}
    \widetilde{T}_\text{ent}^\text{late} = \frac{d}{d-1} \frac{\Delta\mathcal{E}_\mathcal{T}}{\Delta\mathcal{S}_\mathcal{T}} = \frac{d}{d-1} \kappa_d T_\sH\ .
\end{equation} 
It is important to note that while the CWES framework precisely recovers the exact Hawking temperature for the $d=2$ BTZ black hole, it introduces a dimension-dependent distortion factor ($\widetilde{T}_\text{ent}^\text{late} = \frac{d\kappa_d}{d-2} T_\sH$) in higher dimensions ($d>2$). Geometrically, within the CWES framework, this distortion arises because the late-time extremal surface does not hug the event horizon, but rather penetrates it and is governed by a deeper interior critical locus $z_c > z_h$, as discussed in Sec.~\ref{subsec:asymptau}.
 
We note that recent double-Wick-rotation approaches~\cite{Li:2025tud, Xiao:2026nvu} have successfully addressed hyperbolic subregions, where the underlying boundary field theory is well understood. However, developing a rigorous modular Hamiltonian formulation in Lorentzian spacetime for time-like strips in higher dimensions remains an outstanding open challenge. our specific $d>2$ bulk result currently lacks an independent boundary CFT validation. This dimension-dependent factor therefore constitutes an unresolved limitation of the present CWES geometric prescription. The apparent inconsistency between the outcomes of these distinct proposals raises a highly intriguing question. Further investigating the physical differences between these approaches may provide insights into determining which bulk geometric prescription serves as the most appropriate and universal holographic dual for time-like entanglement entropy, a pressing issue we plan to systematically explore in future work. 

\section{Discussion}\label{sec:discussion}

Understanding the internal structure of black holes remains a central challenge in holography and gravitational physics. In this work, we explore this problem from a new single-boundary perspective based on holographic time-like entanglement entropy (TEE), using a bulk geometric prescription. We investigate this question within the concrete and tractable CWES geometric prescription.
We first applied this framework to the Schwarzschild-AdS geometry as a controlled baseline. Our analysis demonstrated that the TEE exhibits linear growth in the  limit of large temporal width $\tau_0 \to \infty$, governed by a critical extremal surface. Furthermore, we argued that the imaginary part of the TEE is not merely a regularization artifact; instead, it encodes physical information about the thermal state that cannot be absorbed into a standard UV cutoff. We explicitly verified that for strip subsystems, the real part of this geometric formulation obeys a time-like entanglement first-law-like relation, yielding a consistent thermodynamic interpretation with an effective entanglement temperature inversely proportional to the subsystem size. The imaginary part serves as a geometric signature of the non-analytic causal shift from the pure AdS vacuum to a black hole with a horizon. 

Building on this foundation, we uncovered a previously hidden causal phase transition inside hairy black holes, which remains invisible to traditional space-like holographic probes. We identified two distinct interior phases, Type-I and Type-II, distinguished by the causal structures of their space-like singularities. This transition is marked by a robust bulk diagnostic and potentially a new order parameter: the critical temporal width $\tau_c$. This quantity manifests on the boundary as a ``time-like entanglement gap''. Inside this gap ($\tau_0 < \tau_c$), the TEE is dominated purely by time-like contributions, up to a regulator effect corresponding to finite ``spatial'' and ``temporal'' resolutions.
Crucially, this framework provides a sharp diagnostic for the stability of the black hole interior. We found that the presence of an inner Cauchy horizon drives the critical width $\tau_c$ to infinity, resulting in pure time-like entanglement. Unlike two-sided ``strong probes'' that require information from both boundaries to access the singularity, TEE is a single-boundary observable. It therefore enriches the holographic dictionary by directly relating a boundary information measure to the deep dynamical features of the bulk spacetime.

This work lays the foundation for several critical future directions. First, our classification in Sec.~\ref{sec:Type-II} relies on space-like singularities. Extending TEE to black holes with inner horizons and time-like singularities remains a daunting but essential challenge.
The most pressing example is the Reissner-Nordstr\"{o}m (RN) black hole, whose interior terminates at a time-like singularity.
The key difficulty is to determine if well-defined CWES configurations exist for such geometries, or if a completely new prescription is needed.
If such extremal surfaces exist, developing a robust method to compute TEE for RN-AdS black holes would be a major step towards a universal understanding of entanglement in realistic black holes, potentially offering a new holographic tool to probe the instability of Cauchy horizons and the strong cosmic censorship conjecture.

Furthermore, our discovery that the interior causal type is linked to the existence of TEE suggests a broader conjecture. In models like holographic superconductors with scalar or vector hair, matter backreaction can largely alter the asymptotic near space-like singularity geometry and produce Kasner-type chaotic behavior~\cite{Hartnoll:2020fhc, Cai:2020wrp,Cai:2023igv,Cai:2024ltu,Zhang:2025hkb,Zhang:2025tsa}. It is important to consider the non-trivial question of whether the TEE remains well-defined in such complex internal geometries, or equivalently, what constraints (e.g., on the Kasner exponents) the requirement of a well-defined TEE imposes on the Kasner spacetime.
Furthermore, this gives a concrete setting to study how TEE responds to detailed features of the near-singularity geometry and how to detect the characteristic anisotropic oscillations (Kasner epochs) as it approaches the singularity. Beyond space-like singularities, chaotic behavior can also emerge near time-like singularities (time-like BKL interior)~\cite{Shaghoulian:2016umj,Bhowmick:2016hph,Ren:2016xhb}.
Exploring how TEE responds in such extreme scenarios represents a further test for our framework. It requires not only overcoming the challenges of Cauchy horizons but also understanding how TEE is affected by these chaotic interior Kasner-geometries near time-like singularity.

Beyond the classification of interior phases, our results for the Schwarzschild-AdS baseline in Sec.~\ref{sec:SAdS} hint at another layer of universality. Previous studies on two-sided black holes have established that the HEE's growth rate tends toward a universal constant~\cite{Hartman:2013qma,Li:2022cvm}. Strikingly, we observed a parallel phenomenology for TEE in the single-boundary static case. As shown in Sec.~\ref{subsec:asymptau}, the real part of TEE scales linearly at large boundary temporal width, $\tau_0 \to \infty$.
This behavior is precisely controlled by a space-like critical surface that lingers at a specific bulk radius $r_c$. This compelling analogy invites a natural follow-up question: does the growth rate of single-sided TEE also obey universal upper bounds analogous to those constraining HEE? Furthermore, how do fundamental physical constraints, such as energy conditions, imprint themselves on this potential bound? We leave these interesting issues for future studies.

While we confirmed a first-law-like relation for the real part of TEE in strip geometries, recent studies~\cite{Li:2025tud, Xiao:2026nvu} highlight severe subtleties when applying direct Lorentzian calculations to both strip and hyperbolic subregions. A major theoretical hurdle is that performing a double Wick rotation on a higher-dimensional time-like strip yields a boundary spatial slab. Unlike a spherical ball, it remains a challenging open question in boundary CFT how to define a well-behaved modular Hamiltonian for a spatial slab in higher dimensions, making it a fundamentally open boundary field theory question how to define a modular Hamiltonian-based time-like entanglement first law (i.e., $\Delta S = \Delta \langle H \rangle$) for such systems. Furthermore, we speculate that apparent inconsistencies in the literature are deeply tied to the boundary topology itself. Investigating whether the CWES geometric prescription requires adaptation for hyperbolic boundaries, or if the Lorentzian first law assumes a modified form in these cases, remains a pressing conceptual puzzle. Resolving these issues will provide crucial consistency tests for future formulations of holographic TEE. 

Finally, we wish to highlight a potential connection between TEE and Renormalization Group (RG) flows. Previous studies, such as Ref.~\cite{Caceres:2022smh}, investigated ``Trans-IR'' flows that extend into the black hole interior to probe the singularity. Since our work establishes TEE as a sensitive probe of interior structure, it offers a natural candidate to quantify these deep RG flows. Recent works~\cite{Grieninger:2023knz, Roychowdhury:2025ukl, Giataganas:2025ize, Afrasiar:2025eam} have explored various relationships between boundary entanglement, RG flows, and bulk geometry. It would be fascinating to further investigate the precise mapping between TEE as a quantum information measure and the renormalization group flow towards the spacetime singularity.

\begin{acknowledgments}
We would like to thank Profs. Haitang Yang, Wu-zhong Guo, and Keun-Young Kim for valuable discussions. We are also grateful to Jaydeep Kumar Basak and Kuntal Pal for helpful discussions. This work is supported by the Natural Science Foundation of China under Grant No. 12375051 and Tianjin University Self-Innovation Fund Extreme Basic Research Project Grant No. 2025XJ22-0014 and 2025XJ21-0007.
\end{acknowledgments}

\appendix

\section{Uniqueness of the Extremal Area Configuration}\label{appen:unique}

In Sec.~\ref{subsec:vertical}, we established that the extremal CWES configuration for the holographic dual of TEE must satisfy the Hamilton-Jacobi condition $E(v_{A_2}) - E(v_{B'_2}=v_0-v_{A_2}) = 0$. While $v_{A_2}=v_{B'_2}=v_0-v_{A_2}$ is a manifest solution, we present here the proof of its uniqueness.

Our strategy is to prove that $E(v_{A_2})$ (or equivalently, $H_*(v_{A_2})$ where $E=H_*^2$, defined in Eq.~\eqref{eq:conservedH}) is a strictly monotonic function of $v_{A_2}$. If $E(v_{A_2})$ is strictly monotonic with its argument $v_{A_2}$, then the condition $E(v_{A_2}) = E(v_{B'_2}=v_0-v_{A_2})$ and Eq.~\eqref{eq:HJofvA} can only be satisfied if $v_{A_2}=v_{B'_2}=v_0-v_{A_2}$. To show this, we will examine the derivative $\dd E/\dd v_{A_2}$:
\be\label{eq:monotonicE}
    \frac{\dd E(v_{A_2})}{\dd v_{A_2}}=2H_*(v_{A_2})\frac{\dd H_*(v_{A_2})}{\dd v_{A_2}}\ .
\ee
We will prove that when $H_* > 0$ we have $\frac{\dd H_*(v_{A_2})}{\dd v_{A_2}} >0$; $H_* < 0$ we have $\frac{\dd H_*(v_{A_2})}{\dd v_{A_2}} <0$, in other words we always have $\frac{\dd E(v_{A_2})}{\dd v_{A_2}}>0$.
From the relation Eq.~\eqref{eq:tau_0&v_A&H_*}, we can compute $\dd v_{A_2}/\dd H_*$
\be
    \begin{aligned}
        \frac{\dd v_{A_2}}{\dd H_*} &=\int_{r=0}^{r=\infty}{\dd r\frac{1}{D(r;H_*)^2}\frac{\partial D(r;H_*)}{\partial H_*}}\ ,
    \end{aligned}
\ee
where we have introduced
\be
\begin{aligned}
  D(r;H_*) &=r^{4-2d}X\left( X + H_{*}^{2} \right)\ ,
\end{aligned}
\ee
and have used the fact that the integrand is smooth everywhere, then we could interchange differentiation and integration.
A direct calculation yields
\be
\begin{aligned}
    \frac{\partial D}{\partial H_*}&=\frac{4H_{*}^{3}}{r^{2d-4}}+\frac{2H_*}{r^{2d-4}}X+\frac{2H_{*}^{5}}{r^{2d-4}X}\ .
\end{aligned}
\ee
We observe that each term in $\frac{\partial D}{\partial H_*}$ is strictly positive for $r>0$, if we assume that $H_* > 0$, we will get $\frac{\partial D(r;H_*)}{\partial H_*}>0$;  if we assume that $H_* < 0$, we will get $\frac{\partial D(r;H_*)}{\partial H_*}<0$.  Consequently,
\be
\begin{aligned}
   \frac{\dd v_{A_2}}{\dd H_*} &=\int_{r=0}^{r=\infty}{\dd r\frac{1}{D(r;H_*)^2}\frac{\partial D(r;H_*)}{\partial H_*}} >0\ ,\quad \text{for}\quad  H_* > 0\ ;\\
    \frac{\dd v_{A_2}}{\dd H_*} &=\int_{r=0}^{r=\infty}{\dd r\frac{1}{D(r;H_*)^2}\frac{\partial D(r;H_*)}{\partial H_*}} <0\ , \quad \text{for}\quad  H_* < 0\ .
\end{aligned}
\ee
This shows $H_*(\dd H_*/\dd v_{A_2})>0$  and then Eq.~\eqref{eq:monotonicE} implies $E(v_{A_2})$ is strictly increasing with respect to $v_{A_2}$. We then find the unique solution for $E(v_{A_2}) = E(v_{B'_2}=v_0-v_{A_2})$ to be $v_{A_2}=v_{B'_2}=v_0-v_{A_2}$, which means a time-symmetric ``vertical'' configuration $t|_{A_2}=t|_{B_2}=0$, as illustrated in Fig.~\ref{fig:Hamilton-Jacobi}(b).

\begin{figure}[htbp]
 \begin{center}
   \includegraphics[width=0.55\textwidth]{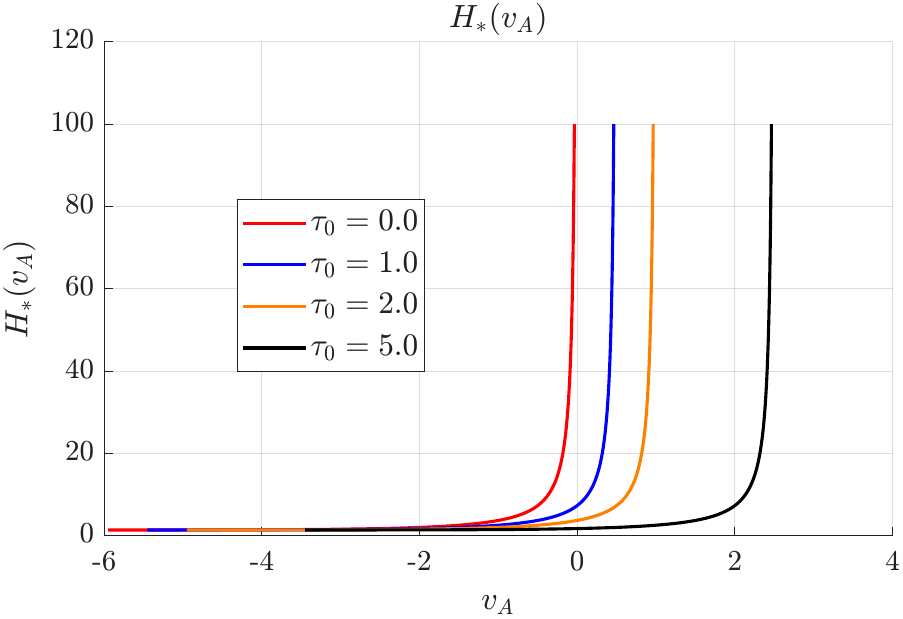}
 \end{center}
\caption{The numerical result suggests that $H_{*}(v_{A_2})$ (thus its inverse $v_{A_2}(H_{*})$) is a monotonic function, where $H_*$ varies from the lower bound~\eqref{eq:Hstarmin} $H_{*,\min}\simeq 0.979689$ to $100$. } \label{HStar-v_A.pdf}
\end{figure}

It rigorously confirms that $v_{A_2}=v_{B'_2}=v_0-v_{A_2}$ is not just \textit{an} extremum, but the \textit{unique} one.
As a double check, we also verify numerically\footnote{Here in the numerical integration of Eq.~\eqref{eq:tau_0&v_A&H_*}, we subtract the vacuum AdS background as in Eq.~\eqref{eq:subtractvacuum}.} that $v_{A_2}(H_{*})$, thus its inverse $H_{*}(v_{A_2})$, is a monotonic function, as shown in Fig.~\ref{HStar-v_A.pdf}.

\section{Asymptotic Details of Null Limit}\label{appen:null}

In this appendix, we provide a detailed derivation of the asymptotic behavior given by Eqs.~\eqref{eq:asymptoticnullform1} and~\eqref{eq:asymptoticnullform2} in the null limit, where the space-like surfaces $A_1 A_2$ and $B_1B_2$ approach null surfaces, as illustrated in Fig.~\ref{fig:nulllimit}.

For the time-difference integral with metric~\eqref{eq:generalAdS}
\begin{equation}
   \tau_0(H_*)/2 -v_{A_2} = \int_0^{\infty}{\frac{\dd z}{\e^{\chi /2}X(X+H_{*}^{2}z^{d-1})}} \ ,
\end{equation}
where $X=\sqrt{H_{*}^{4}z^{2d-2}+g\e^{-\chi}}$, we could set $h^{2d-2}\equiv H_{*}^{4}$ and perform the change of variables $z=\frac{\rho}{h},\dd z=h^{-1}\dd \rho$. This leads to the expression given in
\begin{equation}
    \begin{aligned}
        \tau _0(H_*)/2-v_{A_2} &= \int_0^{\infty}{\frac{\dd z}{\e^{\chi /2}\sqrt{H_{*}^{4}z^{2d-2}+g\e^{-\chi}}\left( \sqrt{H_{*}^{4}z^{2d-2}+g\e^{-\chi}}+H_{*}^{2}z^{d-1} \right)}}\\
        &= h^{-1} \underset{\equiv \eta _d}{\underbrace{\int_0^{\infty}{ \frac{\dd \rho}{\e^{\chi (\rho/h) /2}\sqrt{\rho ^{2d-2}+g(\rho/h) \e^{-\chi (\rho/h)}}\left( \sqrt{\rho ^{2d-2}+g(\rho/h) \e^{-\chi (\rho/h)}}+\rho ^{d-1} \right)} }}}\\
        & = \eta_d h^{-1}\ ,
    \end{aligned}
\end{equation}
which tells us the limit $H_* \to \infty$ corresponds to the space-like surfaces $A_1 A_2$ and $B_1B_2$ approaching a null configuration. 

We note that above integration will be dominated by behavior of metric near AdS boundary region $z =\rho/h\to 0$ for large enough $h$. To accommodate the broad class of hairy black hole backgrounds considered in the main text, we adopt a generalized near-boundary expansion. In our holographic superconductor setup, the source-free condition $\phi_{(0)}=0$ implies that the scalar field is dominated by the normalizable mode $\Phi \sim z^\Delta$. The backreaction of this scalar condensate (scaling as $z^{2\Delta}$), together with the black hole mass term (scaling as $z^d$), determines the explicit metric expansion near the AdS boundary ($z \to 0$). Therefore, we write out the leading and sub-leading terms explicitly:
\begin{equation}\label{eq:asymgz0}
     \begin{aligned} 
         g(z) \e^{-\chi(z)} &\sim 1 - c_1 z^{\alpha_1} - c_2 z^{\alpha_2} - \dots, \qquad \chi(z) = \mathcal{O}\left(z^{\alpha_1}\right) \ ; \\
         g(\rho/h) \e^{-\chi(\rho/h)} &\sim 1 - c_1 (\rho/h)^{\alpha_1} - c_2 (\rho/h)^{\alpha_2} - \dots \ ,
     \end{aligned}
\end{equation}
where the exponents depend on the specific background parameters, satisfying $\alpha_2 > \alpha_1 > 0$. The leading exponent is given by $\alpha_1 = \min(d, 2\Delta)$. The constants $c_1$ and $c_2$ encode the boundary energy density and the scalar condensate, respectively.

We then find for $d\geqslant 2$
\begin{equation}\label{eq:defetad1}
     \begin{aligned}
        \eta_d &=\int_0^{\infty} \frac{\dd \rho}{\sqrt{\rho ^{2d-2}+1}\left( \sqrt{\rho ^{2d-2}+1}+\rho ^{d-1} \right)}\\
        &=\int_0^{\infty}\dd \rho \left(1-\frac{\rho^{d-1}}{\sqrt{\rho ^{2d-2}+1}}\right)\\
        &= \frac{1}{\sqrt{\pi}}\Gamma\left(\frac{d}{2d-2}\right)\Gamma\left(\frac{2d-3}{2d-2}\right)\ .
     \end{aligned}
\end{equation}
The individual term $\int_0^{\infty}1\,\dd\rho$ in second line diverges, but the combination above is convergent for suitable $d\geqslant 2$ due to cancellation at large $\rho$. Thus we find
\begin{equation}\label{deltatuv}
  \Delta t_{{\rm UV}}=\eta_dh^{-1}+\cdots\,.
\end{equation}
This gives us Eq.~\eqref{eq:asymptoticnullform1}.

For the area integral with metric~\eqref{eq:generalAdS}
\begin{equation}
    \text{Re}\mathcal{A}(H_*)= \int_0^{\infty}{\dd z\,\frac{z^{1-d}}{\e^{\chi /2}X}}\ ,
\end{equation}
we introduce a small cut-off parameter $\epsilon$ to regulate the near AdS boundary (UV) region $z \to 0$. We first isolate the UV-divergent contribution, splitting the integral into sub-integrals $I_1(H_*)$ and $I_2(H_*)$ as follow
\[
    \begin{aligned}
        \mathcal{A} _{A_1A_2}(H_*) &=\int_0^{\infty}{\dd z\,\e^{-\chi /2}\frac{z^{1-d}}{\sqrt{H_{*}^{4}z^{2d-2}+g\e^{-\chi}}}}
\\
&=\int^{\infty}_{\epsilon}{\dd z\,\e^{-\chi /2}\left( \frac{z^{1-d}}{\sqrt{H_{*}^{4}z^{2d-2}+g\e^{-\chi}}}-\frac{z^{1-d}}{\sqrt{H_{*}^{4}z^{2d-2}+1}}+\frac{z^{1-d}}{\sqrt{H_{*}^{4}z^{2d-2}+1}} \right)}
\\
&=\underset{=I_1\left( H_* \right)}{\underbrace{\int^{\infty}_{\epsilon}{\dd z\,\e^{-\chi /2}\frac{z^{1-d}}{\sqrt{H_{*}^{4}z^{2d-2}+1}}}}}+\underset{=I_2\left( H_* \right)}{\underbrace{\int^{\infty}_{\epsilon}{\dd z\,\e^{-\chi /2}\left( \frac{z^{1-d}}{\sqrt{H_{*}^{4}z^{2d-2}+g\e^{-\chi}}}-\frac{z^{1-d}}{\sqrt{H_{*}^{4}z^{2d-2}+1}} \right)}}}\ .
    \end{aligned}
\]
Applying the same scaling $h^{2d-2} \equiv H_{*}^{4}$ and variable change $z = \rho / h, \dd z = h^{-1}\dd \rho$ yields the form of sub-integral $I_1\left( H_* \right)$ for $d\geqslant 3$ with limit $h\to \infty$
\[
    \begin{aligned}
        I_1\left( H_* \right) &=\int^{\infty}_{\epsilon}{\dd z\,\frac{z^{1-d}}{\sqrt{\left( hz \right) ^{2d-2}+1}}}\\
        &=h^{d-2}\int^{\infty}_{\epsilon h}{\dd \rho \,\frac{\rho ^{1-d}}{\sqrt{\rho ^{2d-2}+1}}}\\
        &=\left.-\frac{h^{d-2}}{d-2} \,{}_2F_1\left(\frac{1}{2},\frac{2-d}{2 d-2},\frac{d}{2 d-2},-\rho^{2 d-2}\right) \rho^{-d+2}\right|^{\infty}_{\epsilon h}\ .
    \end{aligned}
\]
For brevity, we define the parameters of the hypergeometric function as $\alpha = 1/2$, $\beta = \frac{2-d}{2d-2}$, and $\gamma = \frac{d}{2d-2}$. We must evaluate the expression at the boundaries $\rho \to \epsilon h$ and $\rho \to \infty$. At the lower boundary $\rho = \epsilon h$, the argument of the hypergeometric function $Z = -\rho^{2d-2}$ approaches zero for $d\geqslant 3$. Using the expansion ${}_2F_1(\alpha, \beta, \gamma, Z) = 1 + \mathcal{O}(Z)$ as $Z \to 0$, the contribution from the lower limit is:
\[
    \begin{aligned}
        \left.I_1(H_*)\right|_{\epsilon h} &=-\left( -\frac{h^{d-2}}{d-2} \left(\epsilon h\right)^{2-d} \left[ 1 + \mathcal{O}\left( \left(\epsilon h\right)^{2d-2} \right) \right] \right)\\
        &\approx \frac{\epsilon^{2-d}}{d-2}\ .
    \end{aligned}
\]
Here we have supposed that  $1 \gg 1/h \gg \epsilon$.

At the upper limit, the argument $Z = -\rho^{2d-2}$ diverges. To evaluate the limit, we utilize the linear transformation formula that maps $|Z| > 1$ to the region near zero:
\begin{equation}\label{eq:linearinver}
    \begin{aligned}
        {}_2F_1(\alpha, \beta, \gamma, Z) &= \frac{\Gamma(\gamma)\Gamma(\beta-\alpha)}{\Gamma(\gamma-\alpha)\Gamma(\beta)} (-Z)^{-\alpha} \, {}_2F_1\left(\alpha, \alpha-\gamma+1, \alpha-\beta+1, \frac{1}{Z}\right) \\
        &+ \frac{\Gamma(\gamma)\Gamma(\alpha-\beta)}{\Gamma(\gamma-\beta)\Gamma(\alpha)} (-Z)^{-\beta} \, {}_2F_1\left(\beta, \beta-\gamma+1, \beta-\alpha+1, \frac{1}{Z}\right)\ .
    \end{aligned}
\end{equation}
Substituting this back into the integral expression, which involves a prefactor $\rho^{2-d}$, we analyze the asymptotic behavior of the two terms in Eq.~\eqref{eq:linearinver} as $\rho \to \infty$. For the first term in Eq.~\eqref{eq:linearinver}, the scaling is determined by $\rho^{2-d} (-Z)^{-\alpha} \sim \rho^{2-d} (\rho^{2d-2})^{-1/2} = \rho^{3-2d}$. For $d \geqslant 3$, the exponent $3-2d$ is negative, so this term vanishes as $\rho \to \infty$. For the second term in Eq.~\eqref{eq:linearinver}, the scaling is $\rho^{2-d} (-Z)^{-\beta} \sim \rho^{2-d} (\rho^{2d-2})^{-\frac{2-d}{2d-2}} = \rho^{2-d} \rho^{d-2} = \rho^0$. This term yields a finite constant. Consequently, only the second term in Eq.~\eqref{eq:linearinver} contributes at infinity. The coefficient is calculated using $\alpha - \beta = \frac{2d-3}{2d-2}, \gamma - \beta = 1$. Thus, the finite contribution from the upper limit is:
\[
    \left.I_1(H_*)\right|_\infty= -\frac{h^{d-2}}{d-2} \frac{\Gamma\left(\frac{d}{2d-2}\right)\Gamma\left(\frac{2d-3}{2d-2}\right)}{\Gamma(1)\Gamma\left(\frac{1}{2}\right)} = -\frac{\Gamma\left(\frac{d}{2d-2}\right)\Gamma\left(\frac{2d-3}{2d-2}\right)}{\sqrt{\pi}(d-2)}h^{d-2}\ .
\]
Combining the contributions from the upper and lower limits (and the result of $\eta_d$ in Eq.~\eqref{eq:defetad1} for the time-difference integral), the final result for the sub-integral $I_1(H_*)$ (or $I_1(h)$) with $d\geqslant 3$ is
\begin{equation}
     I_1(h) = \frac{\epsilon^{2-d}}{d-2} - h^{d-2}\frac{\eta_d}{d-2}\ .
\end{equation}
For $d=2$, there will appear a logarithmic divergency on $h$:
\[
    \begin{aligned}
        I_1\left( H_* \right) &=\int_{\epsilon}^{\infty}{\mathrm{d}z\,\frac{z^{-1}}{\sqrt{\left( hz \right) ^2+1}}}
\\
&= \int_{\epsilon h}^{\infty}{\mathrm{d}\rho \,\frac{1}{\rho \sqrt{\rho ^2+1}}}
\\
&=\left.-\ln \left( \frac{\sqrt{\rho ^2+1}+1}{\rho} \right) \right|_{\epsilon h}^{\infty}
    \end{aligned}
\]
At the upper limit, we have
\[
    \left.I_1(H_*)\right|_\infty= -\ln(1) = 0\ .
\]
The contribution from the lower limit is
\[
    \left.I_1(H_*)\right|_{\epsilon h} = -\ln \left( \frac{\sqrt{(\epsilon h)^2+1} + 1}{\epsilon h} \right)  = -\ln \left( \frac{2}{\epsilon h} \right)\ .
\]
Combining the contributions from the upper and lower limits, the final result for $d=2$ is
\begin{equation}\label{eq:d2log}
     I_1\left( h \right) =  \ln \left( \frac{2}{\epsilon h} \right) + \mathcal{O}\left( \epsilon^2 h^2 \right) \ .
\end{equation}

Sub-integral $I_2\left( H_* \right)$ is treated similarly:
\begin{equation}\label{eq:zhrho}
  \begin{aligned}
        I_2\left( H_* \right) &=\int_0^{\infty}{\dd z\,\e^{-\chi \left( z \right) /2}\left( \frac{z^{1-d}}{\sqrt{H_{*}^{4}z^{2d-2}+g\left( z \right) \e^{-\chi \left( z \right)}}}-\frac{z^{1-d}}{\sqrt{H_{*}^{4}z^{2d-2}+1}} \right)}\\
        &=\int_0^{\infty}{\dd z\,\e^{-\chi \left( z \right) /2}\left( \frac{z^{1-d}}{\sqrt{\left( hz \right) ^{2d-2}+g\e^{-\chi}}}-\frac{z^{1-d}}{\sqrt{\left( hz \right) ^{2d-2}+1}} \right)}\\ 
        &=h^{d-2}\int_0^{\infty}{\dd \rho \,\e^{-\chi (\rho/h) /2}\left( \frac{\rho ^{1-d}}{\sqrt{\rho ^{2d-2}+g(\rho/h) \e^{-\chi (\rho/h)}}}-\frac{\rho ^{1-d}}{\sqrt{\rho ^{2d-2}+1}} \right)}\\
        &=h^{d-2}\int_0^{\infty}{\dd \rho \,\e^{-\chi (\rho/h) /2}\left( \frac{\rho ^{1-d}}{\sqrt{\rho ^{2d-2}+g(\rho/h) \e^{-\chi (\rho/h)}}}-\frac{\rho ^{1-d}}{\sqrt{\rho ^{2d-2}+1}} \right)}\ .
    \end{aligned}
\end{equation}
Recall $z=\rho/h$, then at the limit $h\rightarrow\infty$, the integration in the last line of Eq.~\eqref{eq:zhrho} will be dominated by the behavior of $z=\rho/h\ll 1$. In the near AdS boundary region $z \to 0$, we substitute the explicit expansion Eq.~\eqref{eq:asymgz0} into the integrand. For large $h$, we first consider the contribution from the leading-order expansion term $\alpha_1$ of sub-integral $I_2\left( H_* \right)$:
\[
    \begin{aligned}
        I_{2}(h) &= h^{d-2}\int_0^{\infty}{\dd\rho \,\mathrm{e}^{-\chi (\rho /h)/2}\left( \frac{\rho ^{1-d}}{\sqrt{1+\rho ^{2d-2}-c_1 \rho ^{\alpha_1}h^{-\alpha_1}}}-\frac{\rho ^{1-d}}{\sqrt{\rho ^{2d-2}+1}} \right)} \\
        &\approx h^{d-2}\int_0^{\infty}{\dd\rho \,\frac{\rho ^{1-d}}{\sqrt{1+\rho ^{2d-2}}}\left( \frac{c_1 \rho ^{\alpha_1}h^{-\alpha_1}}{2\left( 1+\rho ^{2d-2} \right)} \right)} \\
        &=\frac{c_1 h^{d-2-\alpha_1}}{2}\underset{\equiv \beta _{d,\alpha_1}}{\underbrace{\int_0^{\infty}{\dd\rho \,\frac{\rho ^{\alpha_1 +1-d}}{\left( 1+\rho ^{2d-2} \right) ^{3/2}}}}} \\
        &=\frac{c_1 \beta _{d,\alpha_1}}{2h^{\alpha_1 -d+2}}\ .
\end{aligned}
\]
Here, we take $\e^{-\chi(\rho/h)/2} \to 1$ in the large-$h$ limit to read off the numerical constant $\beta_{d,\alpha_1}$ as
\begin{equation}\label{eq:beta-d}
    \begin{aligned}
	\beta _{d,\alpha_1} &=\int_0^{\infty}{\dd\rho \,\frac{\rho ^{\alpha_1 +1-d}}{\left( 1+\rho ^{2d-2} \right) ^{3/2}}}\\
	&=\frac{\Gamma \left( \frac{\alpha_1 -d+2}{2d-2} \right) \Gamma \left( \frac{4d-5-\alpha_1}{2d-2} \right)}{\sqrt{\pi}\,\left( d-1 \right)}\ .
    \end{aligned}
\end{equation}

Crucially, the unitarity bound of the dual CFT dictates $\Delta > (d-2)/2$, which guarantees $2\Delta > d-2$. Since the leading exponent is $\alpha_1 = \min(d, 2\Delta)$ and $d > d-2$, we always have $\alpha_1 > d-2$. This ensures that the power of $h$ in the denominator is positive: $\alpha_1 - d + 2 > 0$. Consequently, this leading contribution naturally decays to zero as $h \to \infty$.
Following the same logic, if we consider the next-to-leading order term $\alpha_2$ (and higher-order terms) from Eq.~\eqref{eq:asymgz0}, its contribution to the integral would scale as $\mathcal{O}(h^{-(\alpha_2 - d + 2)})$. Because $\alpha_2 > \alpha_1$, these higher-order terms carry even higher negative powers of $h$ and therefore decay significantly faster in the $h \to \infty$ limit. Even if we take all higher-order contributions into account, they will all smoothly vanish at infinity. 
Therefore, considering only the leading-order term $\alpha_1$ is fully sufficient to capture the most precise dominant correction to the area in the null limit. The final asymptotic form of $I_2(h)$ is simply governed by this leading contribution:
\begin{equation}
    I_2(h \to \infty) = \frac{c_1 \beta _{d,\alpha_1}}{2h^{\alpha_1 -d+2}}\ .
\end{equation}

Combining the results from sub-integrals $I_1(H_*)$ and $I_2(H_*)$, we obtain the complete asymptotic behavior for the area in the null limit for $d\geqslant 3$:
\begin{equation}
    \mathcal{A} _{A_1A_2}(H_*)=\frac{c_1 \beta _{d,\alpha_1}}{2h^{\alpha_1 -d+2}}+\frac{\epsilon ^{2-d}}{d-2}-h^{d-2}\frac{\eta _d}{d-2}+\cdots\ .
\end{equation}
Combine it with Eq.~\eqref{deltatuv} and we then find Eq.~\eqref{eq:asymptoticnullform2}.

\bibliographystyle{JHEP}

\bibliography{TEE-singularity}

@article{Lifshitz:1963ps,
    author = "Lifshitz, E. M. and Khalatnikov, I. M.",
    title = "{Investigations in relativistic cosmology}",
    doi = "10.1080/00018736300101283",
    journal = "Adv. Phys.",
    volume = "12",
    pages = "185--249",
    year = "1963"
}

@article{Belinsky:1970ew,
    author = "Belinsky, V. A. and Khalatnikov, I. M. and Lifshitz, E. M.",
    title = "{Oscillatory approach to a singular point in the relativistic cosmology}",
    doi = "10.1080/00018737000101171",
    journal = "Adv. Phys.",
    volume = "19",
    pages = "525--573",
    year = "1970"
}

@article{Belinski:1973zz,
    author = "Belinski, V. A. and Khalatnikov, I. M.",
    title = "{Effect of Scalar and Vector Fields on the Nature of the Cosmological Singularity}",
    journal = "Sov. Phys. JETP",
    volume = "36",
    pages = "591",
    year = "1973"
}

@article{Hartnoll:2020fhc,
    author = "Hartnoll, Sean A. and Horowitz, Gary T. and Kruthoff, Jorrit and Santos, Jorge E.",
    title = "{Diving into a holographic superconductor}",
    eprint = "2008.12786",
    archivePrefix = "arXiv",
    primaryClass = "hep-th",
    doi = "10.21468/SciPostPhys.10.1.009",
    journal = "SciPost Phys.",
    volume = "10",
    number = "1",
    pages = "009",
    year = "2021"
}

@article{Cai:2020wrp,
    author = "Cai, Rong-Gen and Li, Li and Yang, Run-Qiu",
    title = "{No Inner-Horizon Theorem for Black Holes with Charged Scalar Hairs}",
    eprint = "2009.05520",
    archivePrefix = "arXiv",
    primaryClass = "gr-qc",
    doi = "10.1007/JHEP03(2021)263",
    journal = "JHEP",
    volume = "03",
    pages = "263",
    year = "2021"
}

@article{Maldacena:1997re,
    author = "Maldacena, Juan Martin",
    title = "{The Large $N$ limit of superconformal field theories and supergravity}",
    eprint = "hep-th/9711200",
    archivePrefix = "arXiv",
    reportNumber = "HUTP-97-A097, HUTP-98-A097",
    doi = "10.4310/ATMP.1998.v2.n2.a1",
    journal = "Adv. Theor. Math. Phys.",
    volume = "2",
    pages = "231--252",
    year = "1998"
}

@article{Gubser:1998bc,
    author = "Gubser, S. S. and Klebanov, Igor R. and Polyakov, Alexander M.",
    title = "{Gauge theory correlators from noncritical string theory}",
    eprint = "hep-th/9802109",
    archivePrefix = "arXiv",
    reportNumber = "PUPT-1767",
    doi = "10.1016/S0370-2693(98)00377-3",
    journal = "Phys. Lett. B",
    volume = "428",
    pages = "105--114",
    year = "1998"
}

@article{Witten:1998qj,
    author = "Witten, Edward",
    title = "{Anti de Sitter space and holography}",
    eprint = "hep-th/9802150",
    archivePrefix = "arXiv",
    reportNumber = "IASSNS-HEP-98-15",
    doi = "10.4310/ATMP.1998.v2.n2.a2",
    journal = "Adv. Theor. Math. Phys.",
    volume = "2",
    pages = "253--291",
    year = "1998"
}

@article{Maldacena:2001kr,
    author = "Maldacena, Juan Martin",
    title = "{Eternal black holes in anti-de Sitter}",
    eprint = "hep-th/0106112",
    archivePrefix = "arXiv",
    reportNumber = "NSF-ITP-01-59",
    doi = "10.1088/1126-6708/2003/04/021",
    journal = "JHEP",
    volume = "04",
    pages = "021",
    year = "2003"
}

@article{Maldacena:2013xja,
    author = "Maldacena, Juan and Susskind, Leonard",
    title = "{Cool horizons for entangled black holes}",
    eprint = "1306.0533",
    archivePrefix = "arXiv",
    primaryClass = "hep-th",
    doi = "10.1002/prop.201300020",
    journal = "Fortsch. Phys.",
    volume = "61",
    pages = "781--811",
    year = "2013"
}

@article{Ryu:2006bv,
    author = "Ryu, Shinsei and Takayanagi, Tadashi",
    title = "{Holographic derivation of entanglement entropy from AdS/CFT}",
    eprint = "hep-th/0603001",
    archivePrefix = "arXiv",
    reportNumber = "NSF-KITP-06-11",
    doi = "10.1103/PhysRevLett.96.181602",
    journal = "Phys. Rev. Lett.",
    volume = "96",
    pages = "181602",
    year = "2006"
}

@article{Hubeny:2007xt,
    author = "Hubeny, Veronika E. and Rangamani, Mukund and Takayanagi, Tadashi",
    title = "{A Covariant holographic entanglement entropy proposal}",
    eprint = "0705.0016",
    archivePrefix = "arXiv",
    primaryClass = "hep-th",
    reportNumber = "DCPT-07-13, KUNS-2069",
    doi = "10.1088/1126-6708/2007/07/062",
    journal = "JHEP",
    volume = "07",
    pages = "062",
    year = "2007"
}

@article{Stanford:2014jda,
    author = "Stanford, Douglas and Susskind, Leonard",
    title = "{Complexity and Shock Wave Geometries}",
    eprint = "1406.2678",
    archivePrefix = "arXiv",
    primaryClass = "hep-th",
    doi = "10.1103/PhysRevD.90.126007",
    journal = "Phys. Rev. D",
    volume = "90",
    number = "12",
    pages = "126007",
    year = "2014"
}

@article{Brown:2015bva,
    author = "Brown, Adam R. and Roberts, Daniel A. and Susskind, Leonard and Swingle, Brian and Zhao, Ying",
    title = "{Holographic Complexity Equals Bulk Action?}",
    eprint = "1509.07876",
    archivePrefix = "arXiv",
    primaryClass = "hep-th",
    doi = "10.1103/PhysRevLett.116.191301",
    journal = "Phys. Rev. Lett.",
    volume = "116",
    number = "19",
    pages = "191301",
    year = "2016"
}

@article{Lehner:2016vdi,
    author = "Lehner, Luis and Myers, Robert C. and Poisson, Eric and Sorkin, Rafael D.",
    title = "{Gravitational action with null boundaries}",
    eprint = "1609.00207",
    archivePrefix = "arXiv",
    primaryClass = "hep-th",
    doi = "10.1103/PhysRevD.94.084046",
    journal = "Phys. Rev. D",
    volume = "94",
    number = "8",
    pages = "084046",
    year = "2016"
}

@article{Hartman:2013qma,
    author = "Hartman, Thomas and Maldacena, Juan",
    title = "{Time Evolution of Entanglement Entropy from Black Hole Interiors}",
    eprint = "1303.1080",
    archivePrefix = "arXiv",
    primaryClass = "hep-th",
    doi = "10.1007/JHEP05(2013)014",
    journal = "JHEP",
    volume = "05",
    pages = "014",
    year = "2013"
}

@article{Fidkowski:2003nf,
    author = "Fidkowski, Lukasz and Hubeny, Veronika and Kleban, Matthew and Shenker, Stephen",
    title = "{The Black hole singularity in AdS / CFT}",
    eprint = "hep-th/0306170",
    archivePrefix = "arXiv",
    reportNumber = "SU-ITP-03-16",
    doi = "10.1088/1126-6708/2004/02/014",
    journal = "JHEP",
    volume = "02",
    pages = "014",
    year = "2004"
}

@article{Caceres:2022smh,
    author = "Caceres, Elena and Kundu, Arnab and Patra, Ayan K. and Shashi, Sanjit",
    title = "{Trans-IR flows to black hole singularities}",
    eprint = "2201.06579",
    archivePrefix = "arXiv",
    primaryClass = "hep-th",
    reportNumber = "UTTG-30-2022",
    doi = "10.1103/PhysRevD.106.046005",
    journal = "Phys. Rev. D",
    volume = "106",
    number = "4",
    pages = "046005",
    year = "2022"
}

@article{Doi:2022iyj,
    author = "Doi, Kazuki and Harper, Jonathan and Mollabashi, Ali and Takayanagi, Tadashi and Taki, Yusuke",
    title = "{Pseudoentropy in dS/CFT and Timelike Entanglement Entropy}",
    eprint = "2210.09457",
    archivePrefix = "arXiv",
    primaryClass = "hep-th",
    reportNumber = "YITP-22-121, IMPU22-0052",
    doi = "10.1103/PhysRevLett.130.031601",
    journal = "Phys. Rev. Lett.",
    volume = "130",
    number = "3",
    pages = "031601",
    year = "2023"
}

@article{Wang:2018jva,
    author = "Wang, Peng and Wu, Houwen and Yang, Haitang",
    title = "{Fix the dual geometries of $T\bar{T}$ deformed CFT$_2$ and highly excited states of CFT$_2$}",
    eprint = "1811.07758",
    archivePrefix = "arXiv",
    primaryClass = "hep-th",
    reportNumber = "CTP-SCU/2018007",
    doi = "10.1140/epjc/s10052-020-08680-7",
    journal = "Eur. Phys. J. C",
    volume = "80",
    number = "12",
    pages = "1117",
    year = "2020"
}

@article{Chu:2023zah,
    author = "Chu, Chong-Sun and Parihar, Himanshu",
    title = "{Time-like entanglement entropy in AdS/BCFT}",
    eprint = "2304.10907",
    archivePrefix = "arXiv",
    primaryClass = "hep-th",
    doi = "10.1007/JHEP06(2023)173",
    journal = "JHEP",
    volume = "06",
    pages = "173",
    year = "2023"
}

@article{Chu:2025sjv,
    author = "Chu, Chong-Sun and Parihar, Himanshu",
    title = "{Timelike entanglement entropy with gravitational anomalies}",
    eprint = "2504.19694",
    archivePrefix = "arXiv",
    primaryClass = "hep-th",
    doi = "10.1007/JHEP08(2025)038",
    journal = "JHEP",
    volume = "08",
    pages = "038",
    year = "2025"
}

@article{Jiang:2023ffu,
    author = "Jiang, Xin and Wang, Peng and Wu, Houwen and Yang, Haitang",
    title = "{Timelike entanglement entropy and $T\bar{T}$ deformation}",
    eprint = "2302.13872",
    archivePrefix = "arXiv",
    primaryClass = "hep-th",
    reportNumber = "CTP-SCU/2023004",
    doi = "10.1103/PhysRevD.108.046004",
    journal = "Phys. Rev. D",
    volume = "108",
    number = "4",
    pages = "046004",
    year = "2023"
}

@article{Narayan:2022afv,
    author = "Narayan, K.",
    title = "{de Sitter space, extremal surfaces, and time entanglement}",
    eprint = "2210.12963",
    archivePrefix = "arXiv",
    primaryClass = "hep-th",
    doi = "10.1103/PhysRevD.107.126004",
    journal = "Phys. Rev. D",
    volume = "107",
    number = "12",
    pages = "126004",
    year = "2023"
}

@article{Li:2022tsv,
    author = "Li, Ze and Xiao, Zi-Qing and Yang, Run-Qiu",
    title = "{On holographic time-like entanglement entropy}",
    eprint = "2211.14883",
    archivePrefix = "arXiv",
    primaryClass = "hep-th",
    doi = "10.1007/JHEP04(2023)004",
    journal = "JHEP",
    volume = "04",
    pages = "004",
    year = "2023"
}

@article{Doi:2023zaf,
    author = "Doi, Kazuki and Harper, Jonathan and Mollabashi, Ali and Takayanagi, Tadashi and Taki, Yusuke",
    title = "{Timelike entanglement entropy}",
    eprint = "2302.11695",
    archivePrefix = "arXiv",
    primaryClass = "hep-th",
    reportNumber = "YITP-23-22",
    doi = "10.1007/JHEP05(2023)052",
    journal = "JHEP",
    volume = "05",
    pages = "052",
    year = "2023"
}

@article{Narayan:2023zen,
    author = "Narayan, K.",
    title = "{Further remarks on de Sitter space, extremal surfaces, and time entanglement}",
    eprint = "2310.00320",
    archivePrefix = "arXiv",
    primaryClass = "hep-th",
    doi = "10.1103/PhysRevD.109.086009",
    journal = "Phys. Rev. D",
    volume = "109",
    number = "8",
    pages = "086009",
    year = "2024"
}

@article{Jiang:2025pen,
    author = "Jiang, Xin and Wu, Houwen and Yang, Haitang",
    title = "{Timelike entanglement entropy Revisited}",
    eprint = "2503.19342",
    archivePrefix = "arXiv",
    primaryClass = "hep-th",
    reportNumber = "CTP-SCU/2025004",
    month = "3",
    year = "2025"
}

@article{Milekhin:2025ycm,
    author = "Milekhin, Alexey and Adamska, Zofia and Preskill, John",
    title = "{Observable and computable entanglement in time}",
    eprint = "2502.12240",
    archivePrefix = "arXiv",
    primaryClass = "quant-ph",
    month = "2",
    year = "2025"
}

@article{Gong:2025pnu,
    author = "Gong, XiangKun and Guo, Wu-zhong and Xu, Jin",
    title = "{Entanglement measures for causally connected subregions and holography}",
    eprint = "2508.05158",
    archivePrefix = "arXiv",
    primaryClass = "hep-th",
    month = "8",
    year = "2025"
}

@article{Guo:2025ase,
    author = "Guo, Wu-zhong and He, Song and Liu, Tao",
    title = "{Entanglement of General Subregions in Time-Dependent States}",
    eprint = "2512.19955",
    archivePrefix = "arXiv",
    primaryClass = "hep-th",
    month = "12",
    year = "2025"
}

@article{Jiang:2025jnk,
    author = "Jiang, Xin and Yang, Haitang",
    title = "{Entanglement entropy of conformal field theory in all dimensions}",
    eprint = "2506.02786",
    archivePrefix = "arXiv",
    primaryClass = "hep-th",
    reportNumber = "CTP-SCU/2025011",
    doi = "10.1007/JHEP01(2026)015",
    journal = "JHEP",
    volume = "01",
    pages = "015",
    year = "2026"
}

@article{Guo:2024lrr,
    author = "Guo, Wu-zhong and He, Song and Zhang, Yu-Xuan",
    title = "{Relation between time- and spacelike entanglement entropy}",
    eprint = "2402.00268",
    archivePrefix = "arXiv",
    primaryClass = "hep-th",
    doi = "10.1103/gmkp-lrh3",
    journal = "Phys. Rev. D",
    volume = "112",
    number = "8",
    pages = "086020",
    year = "2025"
}

@article{Xu:2024yvf,
    author = "Xu, Jin and Guo, Wu-zhong",
    title = "{Imaginary part of timelike entanglement entropy}",
    eprint = "2410.22684",
    archivePrefix = "arXiv",
    primaryClass = "hep-th",
    doi = "10.1007/JHEP02(2025)094",
    journal = "JHEP",
    volume = "02",
    pages = "094",
    year = "2025"
}

@article{Harper:2025lav,
    author = "Harper, Jonathan and Kawamoto, Taishi and Maeda, Ryota and Nakamura, Nanami and Takayanagi, Tadashi",
    title = "{Non-hermitian Density Matrices from Time-like Entanglement and Wormholes}",
    eprint = "2512.13800",
    archivePrefix = "arXiv",
    primaryClass = "hep-th",
    reportNumber = "YITP-25-186",
    month = "12",
    year = "2025"
}

@article{He:2023ubi,
    author = "He, Peng-Zhang and Zhang, Hai-Qing",
    title = "{Holographic timelike entanglement entropy from Rindler method*}",
    eprint = "2307.09803",
    archivePrefix = "arXiv",
    primaryClass = "hep-th",
    doi = "10.1088/1674-1137/ad57a8",
    journal = "Chin. Phys. C",
    volume = "48",
    number = "11",
    pages = "115113",
    year = "2024"
}

@article{Wen:2024yny,
    author = "Wen, Qiang and Xu, Mingshuai and Zhong, Haocheng",
    title = "{Timelike and gravitational anomalous entanglement from the inner horizon}",
    eprint = "2412.21058",
    archivePrefix = "arXiv",
    primaryClass = "hep-th",
    doi = "10.21468/SciPostPhys.18.6.204",
    journal = "SciPost Phys.",
    volume = "18",
    number = "6",
    pages = "204",
    year = "2025"
}

@article{Nunez:2025gxq,
    author = "Nunez, Carlos and Roychowdhury, Dibakar",
    title = "{Timelike entanglement entropy: A top-down approach}",
    eprint = "2505.20388",
    archivePrefix = "arXiv",
    primaryClass = "hep-th",
    doi = "10.1103/vjyt-xc15",
    journal = "Phys. Rev. D",
    volume = "112",
    number = "2",
    pages = "026030",
    year = "2025"
}

@article{Nunez:2025puk,
    author = "Nunez, Carlos and Roychowdhury, Dibakar",
    title = "{Holographic timelike entanglement across dimensions}",
    eprint = "2508.13266",
    archivePrefix = "arXiv",
    primaryClass = "hep-th",
    doi = "10.1007/JHEP11(2025)100",
    journal = "JHEP",
    volume = "11",
    pages = "100",
    year = "2025"
}

@article{Afrasiar:2024lsi,
    author = "Afrasiar, Mir and Basak, Jaydeep Kumar and Giataganas, Dimitrios",
    title = "{Timelike entanglement entropy and phase transitions in non-conformal theories}",
    eprint = "2404.01393",
    archivePrefix = "arXiv",
    primaryClass = "hep-th",
    doi = "10.1007/JHEP07(2024)243",
    journal = "JHEP",
    volume = "07",
    pages = "243",
    year = "2024"
}

@article{Afrasiar:2024ldn,
    author = "Afrasiar, Mir and Basak, Jaydeep Kumar and Giataganas, Dimitrios",
    title = "{Holographic timelike entanglement entropy in non-relativistic theories}",
    eprint = "2411.18514",
    archivePrefix = "arXiv",
    primaryClass = "hep-th",
    doi = "10.1007/JHEP05(2025)205",
    journal = "JHEP",
    volume = "05",
    pages = "205",
    year = "2025"
}

@article{Bohra:2025mhb,
    author = "Bohra, Hardik and Sivaramakrishnan, Allic",
    title = "{Composite AdS geodesics for CFT correlators and timelike entanglement entropy}",
    eprint = "2511.22168",
    archivePrefix = "arXiv",
    primaryClass = "hep-th",
    reportNumber = "CALT-TH 2025-037",
    month = "11",
    year = "2025"
}

@article{Heller:2024whi,
    author = "Heller, Michal P. and Ori, Fabio and Serantes, Alexandre",
    title = "{Geometric Interpretation of Timelike Entanglement Entropy}",
    eprint = "2408.15752",
    archivePrefix = "arXiv",
    primaryClass = "hep-th",
    doi = "10.1103/PhysRevLett.134.131601",
    journal = "Phys. Rev. Lett.",
    volume = "134",
    number = "13",
    pages = "131601",
    year = "2025"
}

@article{Heller:2025kvp,
    author = "Heller, Michal P. and Ori, Fabio and Serantes, Alexandre",
    title = "{Temporal Entanglement from Holographic Entanglement Entropy}",
    eprint = "2507.17847",
    archivePrefix = "arXiv",
    primaryClass = "hep-th",
    doi = "10.1103/qlsv-gp22",
    journal = "Phys. Rev. X",
    volume = "15",
    number = "4",
    pages = "041022",
    year = "2025"
}

@article{Guo:2025pru,
    author = "Guo, Wu-zhong and Xu, Jin",
    title = "{Duality of Ryu-Takayanagi surfaces inside and outside the horizon}",
    eprint = "2502.16774",
    archivePrefix = "arXiv",
    primaryClass = "hep-th",
    doi = "10.1103/xndj-9ftm",
    journal = "Phys. Rev. D",
    volume = "112",
    number = "10",
    pages = "L101901",
    year = "2025"
}

@article{Guo:2025mwp,
    author = "Guo, Wu-zhong",
    title = "{Measuring the black hole interior from the exterior}",
    eprint = "2505.09878",
    archivePrefix = "arXiv",
    primaryClass = "hep-th",
    doi = "10.1142/S0218271825440067",
    journal = "Int. J. Mod. Phys. D",
    volume = "34",
    number = "16",
    pages = "2544006",
    year = "2025"
}

@article{Anegawa:2024kdj,
    author = "Anegawa, Takanori and Tamaoka, Kotaro",
    title = "{Black hole singularity and timelike entanglement}",
    eprint = "2406.10968",
    archivePrefix = "arXiv",
    primaryClass = "hep-th",
    doi = "10.1007/JHEP10(2024)182",
    journal = "JHEP",
    volume = "10",
    pages = "182",
    year = "2024"
}

@article{An:2022lvo,
    author = "An, Yu-Sen and Li, Li and Yang, Fu-Guo and Yang, Run-Qiu",
    title = "{Interior structure and complexity growth rate of holographic superconductor from M-theory}",
    eprint = "2205.02442",
    archivePrefix = "arXiv",
    primaryClass = "hep-th",
    doi = "10.1007/JHEP08(2022)133",
    journal = "JHEP",
    volume = "08",
    pages = "133",
    year = "2022"
}

@article{Auzzi:2022bfd,
    author = "Auzzi, Roberto and Bolognesi, Stefano and Rabinovici, Eliezer and Schaposnik Massolo, Fidel I. and Tallarita, Gianni",
    title = "{On the time dependence of holographic complexity for charged AdS black holes with scalar hair}",
    eprint = "2205.03365",
    archivePrefix = "arXiv",
    primaryClass = "hep-th",
    doi = "10.1007/JHEP08(2022)235",
    journal = "JHEP",
    volume = "08",
    pages = "235",
    year = "2022"
}

@article{Carmi:2017jqz,
    author = "Carmi, Dean and Chapman, Shira and Marrochio, Hugo and Myers, Robert C. and Sugishita, Sotaro",
    title = "{On the Time Dependence of Holographic Complexity}",
    eprint = "1709.10184",
    archivePrefix = "arXiv",
    primaryClass = "hep-th",
    doi = "10.1007/JHEP11(2017)188",
    journal = "JHEP",
    volume = "11",
    pages = "188",
    year = "2017"
}

@article{Li:2022cvm,
    author = "Li, Ze and Yang, Run-Qiu",
    title = "{Upper bounds of holographic entanglement entropy growth rate for thermofield double states}",
    eprint = "2205.15154",
    archivePrefix = "arXiv",
    primaryClass = "hep-th",
    doi = "10.1007/JHEP10(2022)072",
    journal = "JHEP",
    volume = "10",
    pages = "072",
    year = "2022"
}

@article{Hartnoll:2008kx,
    author = "Hartnoll, Sean A. and Herzog, Christopher P. and Horowitz, Gary T.",
    title = "{Holographic Superconductors}",
    eprint = "0810.1563",
    archivePrefix = "arXiv",
    primaryClass = "hep-th",
    doi = "10.1088/1126-6708/2008/12/015",
    journal = "JHEP",
    volume = "12",
    pages = "015",
    year = "2008"
}

@article{Cai:2015cya,
    author = "Cai, Rong-Gen and Li, Li and Li, Li-Fang and Yang, Run-Qiu",
    title = "{Introduction to Holographic Superconductor Models}",
    eprint = "1502.00437",
    archivePrefix = "arXiv",
    primaryClass = "hep-th",
    reportNumber = "CCTP-2015-06, CCQCN-2015-65",
    doi = "10.1007/s11433-015-5676-5",
    journal = "Sci. China Phys. Mech. Astron.",
    volume = "58",
    number = "6",
    pages = "060401",
    year = "2015"
}

@article{Horowitz:2009ij,
    author = "Horowitz, Gary T. and Roberts, Matthew M.",
    title = "{Zero Temperature Limit of Holographic Superconductors}",
    eprint = "0908.3677",
    archivePrefix = "arXiv",
    primaryClass = "hep-th",
    doi = "10.1088/1126-6708/2009/11/015",
    journal = "JHEP",
    volume = "11",
    pages = "015",
    year = "2009"
}

@article{Cai:2023igv,
    author = "Cai, Rong-Gen and Duan, Mei-Ning and Li, Li and Yang, Fu-Guo",
    title = "{Towards classifying the interior dynamics of charged black holes with scalar hair}",
    eprint = "2312.11131",
    archivePrefix = "arXiv",
    primaryClass = "gr-qc",
    doi = "10.1007/JHEP02(2024)169",
    journal = "JHEP",
    volume = "02",
    pages = "169",
    year = "2024"
}

@article{Cai:2024ltu,
    author = "Cai, Rong-Gen and Duan, Mei-Ning and Li, Li and Yang, Fu-Guo",
    title = "{Clarifying Kasner dynamics inside anisotropic black hole with vector hair}",
    eprint = "2408.06122",
    archivePrefix = "arXiv",
    primaryClass = "gr-qc",
    doi = "10.1007/JHEP04(2025)179",
    journal = "JHEP",
    volume = "04",
    pages = "179",
    year = "2025"
}

@article{Zhang:2025hkb,
    author = "Zhang, Xing-Kun and Zhao, Xin and Nie, Zhang-Yu and Hu, Ya-Peng and An, Yu-Sen",
    title = "{Diving into a holographic multi-band superconductor}",
    eprint = "2411.07693",
    archivePrefix = "arXiv",
    primaryClass = "hep-th",
    doi = "10.1016/j.physletb.2025.139684",
    journal = "Phys. Lett. B",
    volume = "868",
    pages = "139684",
    year = "2025"
}

@article{Zhang:2025tsa,
    author = "Zhang, Xing-Kun and Zhao, Xin and Nie, Zhang-Yu and Hu, Ya-Peng and An, Yu-Sen",
    title = "{Interior structure of the holographic s+p superconductor and chaotic-stable transition near the black hole singularity}",
    eprint = "2506.19419",
    archivePrefix = "arXiv",
    primaryClass = "hep-th",
    doi = "10.1016/j.physletb.2025.140110",
    journal = "Phys. Lett. B",
    volume = "872",
    pages = "140110",
    year = "2026"
}

@article{Shaghoulian:2016umj,
    author = "Shaghoulian, Edgar and Wang, Huajia",
    title = "{Timelike BKL singularities and chaos in AdS/CFT}",
    eprint = "1601.02599",
    archivePrefix = "arXiv",
    primaryClass = "hep-th",
    doi = "10.1088/0264-9381/33/12/125020",
    journal = "Class. Quant. Grav.",
    volume = "33",
    number = "12",
    pages = "125020",
    year = "2016"
}

@article{Ren:2016xhb,
    author = "Ren, Jie",
    title = "{Asymptotically AdS spacetimes with a timelike Kasner singularity}",
    eprint = "1603.08004",
    archivePrefix = "arXiv",
    primaryClass = "hep-th",
    doi = "10.1007/JHEP07(2016)112",
    journal = "JHEP",
    volume = "07",
    pages = "112",
    year = "2016"
}

@article{Bhowmick:2016hph,
    author = "Bhowmick, Samrat and Chatterjee, Soumyabrata",
    title = "{Towards Timelike Singularity via AdS Dual}",
    eprint = "1610.05484",
    archivePrefix = "arXiv",
    primaryClass = "hep-th",
    doi = "10.1142/S0217751X17501226",
    journal = "Int. J. Mod. Phys. A",
    volume = "32",
    number = "21",
    pages = "1750122",
    year = "2017"
}

@article{Grieninger:2023knz,
    author = "Grieninger, Sebastian and Ikeda, Kazuki and Kharzeev, Dmitri E.",
    title = "{Temporal entanglement entropy as a probe of renormalization group flow}",
    eprint = "2312.08534",
    archivePrefix = "arXiv",
    primaryClass = "hep-th",
    doi = "10.1007/JHEP05(2024)030",
    journal = "JHEP",
    volume = "05",
    pages = "030",
    year = "2024"
}

@article{Roychowdhury:2025ukl,
    author = "Roychowdhury, Dibakar",
    title = "{Holographic timelike entanglement and c theorem for supersymmetric QFTs in (0 + 1)d}",
    eprint = "2502.10797",
    archivePrefix = "arXiv",
    primaryClass = "hep-th",
    doi = "10.1007/JHEP06(2025)003",
    journal = "JHEP",
    volume = "06",
    pages = "003",
    year = "2025"
}

@article{Giataganas:2025ize,
    author = "Giataganas, Dimitrios",
    title = "{Timelike Entanglement Entropy and Renormalization Group Flow Irreversibility}",
    eprint = "2512.16499",
    archivePrefix = "arXiv",
    primaryClass = "hep-th",
    month = "12",
    year = "2025"
}

@article{Afrasiar:2025eam,
    author = "Afrasiar, Mir and Basak, Jaydeep Kumar and Kim, Keun-Young",
    title = "{Aspects of holographic timelike entanglement entropy in black hole backgrounds}",
    eprint = "2512.21327",
    archivePrefix = "arXiv",
    primaryClass = "hep-th",
    month = "12",
    year = "2025"
}

@article{Li:2025tud,
    author = "Li, Guo-Ying and Xiao, Mei-Hui and He, Song and Sun, Jia-Rui",
    title = "{Entanglement first law for timelike entanglement entropy and linearized Einstein's equation}",
    eprint = "2511.17098",
    archivePrefix = "arXiv",
    primaryClass = "hep-th",
    month = "11",
    year = "2025"
}

@article{Fefferman:2007rka,
    author = "Fefferman, Charles and Graham, C. Robin",
    title = "{The ambient metric}",
    eprint = "0710.0919",
    archivePrefix = "arXiv",
    primaryClass = "math.DG",
    journal = "Ann. Math. Stud.",
    volume = "178",
    pages = "1--128",
    year = "2011"
}

@article{Bhattacharya:2012mi,
    author = "Bhattacharya, Jyotirmoy and Nozaki, Masahiro and Takayanagi, Tadashi and Ugajin, Tomonori",
    title = "{Thermodynamical Property of Entanglement Entropy for Excited States}",
    eprint = "1212.1164",
    archivePrefix = "arXiv",
    primaryClass = "hep-th",
    reportNumber = "IPMU12-0220, YITP-12-99, IPMU12-0220; YITP-12-99",
    doi = "10.1103/PhysRevLett.110.091602",
    journal = "Phys. Rev. Lett.",
    volume = "110",
    number = "9",
    pages = "091602",
    year = "2013"
}

@article{Blanco:2013joa,
    author = "Blanco, David D. and Casini, Horacio and Hung, Ling-Yan and Myers, Robert C.",
    title = "{Relative Entropy and Holography}",
    eprint = "1305.3182",
    archivePrefix = "arXiv",
    primaryClass = "hep-th",
    doi = "10.1007/JHEP08(2013)060",
    journal = "JHEP",
    volume = "08",
    pages = "060",
    year = "2013"
}

@article{Taylor:2016aoi,
    author = "Taylor, Marika and Woodhead, William",
    title = "{Renormalized entanglement entropy}",
    eprint = "1604.06808",
    archivePrefix = "arXiv",
    primaryClass = "hep-th",
    doi = "10.1007/JHEP08(2016)165",
    journal = "JHEP",
    volume = "08",
    pages = "165",
    year = "2016"
}

@article{Xiao:2026nvu,
    author = "Xiao, Mei-Hui and Li, Guo-Ying and He, Song and Sun, Jia-Rui",
    title = "{Timelike Entanglement First Law and Linearized Field Equations in Higher Curvature Gravity}",
    eprint = "2607.12834",
    archivePrefix = "arXiv",
    primaryClass = "hep-th",
    month = "7",
    year = "2026"
}

@article{Bernamonti:2026pxo,
    author = "Bernamonti, Alice and Galli, Federico and Heller, Michal P. and Ori, Fabio and Serantes, Alexandre",
    title = "{Temporal Entanglement from Twist Correlators in 2d Conformal Field Theory and Holography}",
    eprint = "2607.14012",
    archivePrefix = "arXiv",
    primaryClass = "hep-th",
    month = "7",
    year = "2026"
}

\end{document}